%% file: Tese_Fisica_JA.tex
\documentclass[11pt,twoside,english,brazil]{report}

\usepackage[T1]{fontenc}
\usepackage[latin9]{inputenc}
\usepackage{xcolor}

\usepackage{geometry}

\usepackage{fancyhdr}
\usepackage{babel}
\usepackage{calc}
\usepackage{setspace}
\usepackage[unicode=true,pdfusetitle,
 bookmarks=true,bookmarksnumbered=false,bookmarksopen=false,
 breaklinks=false,pdfborder={0 0 1},backref=false,colorlinks=false]
 {hyperref}

\makeatletter

\usepackage{nextpage}
\newcommand\myclearpage{\cleartooddpage[\thispagestyle{empty}]}

\usepackage[fit]{truncate}
\makeatother

\usepackage{graphics}
\usepackage{graphicx}
\usepackage{ae,aecompl}
\usepackage{graphicx}
\usepackage[T1]{fontenc}
\usepackage{amsmath, amsthm, amssymb, amsfonts}
\usepackage{bm}
\usepackage{enumerate}
\usepackage[all]{xy}
\def\kb{\bm\kappa}
\def\vp{\bm\varphi}

\newtheorem{teo}{Teorema}[chapter]

\newtheorem{define}[teo]{Definição}
\newtheorem{obs}[teo]{Observação}

\usepackage{upgreek}

\def\cep{\ep} 
\def\al{\alpha}
\def\be{\beta}
\def\ga{\gamma}

\def\ep{\epsilon}

\def\et{\eta}

\def\ka{\kappa}
\def\la{\lambda}

\def\si{\sigma}

\def\vp{\varphi}
\def\ch{\chi}
\def\ps{\psi}
\def\om{\omega}
\def\Ga{\Gamma}

\def\La{\Lambda}

\def\cl{{\cal L}}

\def\mn{{\mu\nu}}

\def\half{{\textstyle{1\over 2}}}
\def\prt{\partial}
 \def\lrDmu{\stackrel{\leftrightarrow}{D_\mu}}
 \def\pt#1{\phantom{#1}}
 \def\e{$e_\mu$}
 \def\f{$f_\mu$}
\newcommand{\ba}{\begin{eqnarray}}
\newcommand{\ea}{\end{eqnarray}}
\newcommand{\bege}{\begin{equation}}
\newcommand{\enge}{\end{equation}}
\newcommand{\benu}{\begin{enumerate}}
\newcommand{\enu}{\end{enumerate}}

\newcommand{\bbbbox}{\mathop{\Box\kern -5pt\raisebox{.8pt}{$|$}}}

\newcommand{\RR}{\mathbb{R}}

\newcommand{\w}{\wedge} 
\newcommand{\vv}{{\bf v}}

\newcommand{\OO}{\mathbb{O}}

\def\nsc#1#2#3{\om_{#1}^{{\pt{#1}}#2#3}}
\def\lsc#1#2#3{\om_{#1#2#3}}
\def\lulsc#1#2#3{\om_{#1\pt{#2}#3}^{{\pt{#1}}#2}}
 \def\tor#1#2#3{T^{#1}_{{\pt{#1}}#2#3}}

\def\vb#1#2{e_{#1}^{{\pt{#1}}#2}}
 \def\ivb#1#2{e^{#1}_{{\pt{#1}}#2}}
 \def\uvb#1#2{e^{#1#2}}
 \def\lvb#1#2{e_{#1#2}}
 
\newcommand{\gdualn}[1]{\overset{\:{}^{{}^{\boldsymbol{\neg}}}}{\smash[t]{#1}}} 

\def\0{\mbox{\boldmath$\displaystyle\mathbb{O}$}}

\def\vp{\mbox{\boldmath$\displaystyle\boldsymbol{\varphi}$}}

\def\kb{\mbox{\boldmath$\displaystyle\boldsymbol{\kappa}$}}

\def\s{\mbox{\boldmath$\displaystyle\boldsymbol{\sigma}$}}
\def\p{\mbox{\boldmath$\displaystyle\boldsymbol{p}$}}
\def\e{\rm e}

\def\openone{\mathbb I}

\def\0{\bm0}
\def\vp{\bm\varphi}

\def\kb{\bm\kappa}

\def\s{\bm\sigma}
\def\p{\bm{p}}
\DeclareMathOperator{\Tr}{Tr}

\begin{document}

\begin{center}
José Antonio da Silva Neto
\par\end{center}

\vspace{3cm}

\begin{center}
{\huge{} Gravitação $f(\mathit{R})$ com torção e violação de Lorentz: cenários naturais para novos espinores singulares}
\par\end{center}{\huge \par}

\begin{center}
\vspace{3cm}

\par\end{center}

\begin{center}
\begin{minipage}[t]{0.6\columnwidth}%
\begin{center}
Tese apresentada ao Programa de Pós-Graduação em Física
da Universidade Federal do ABC (UFABC), como requisito parcial à obtenção
do título de Doutor em Física.
\par\end{center}%
\end{minipage}
\par\end{center}

\begin{center}
\vspace{2cm}
Orientador: Prof. Dr. Roldão da Rocha Junior
\par\end{center}


\begin{center}
\vspace{3cm}
Santo André - SP
\par\end{center}

\begin{center}
2017
\par\end{center}

\begin{center}
{\large{}\thispagestyle{empty}}
\par\end{center}{\large \par}

\pagebreak{}
\par

\vfill{}

\begin{center}
\fbox{\begin{minipage}[t]{0.8\columnwidth}%
Silva Neto, José Antonio
\hspace{1cm} Gravitação $f(\mathit{R})$ com torção e violação de Lorentz / José Antonio da Silva Neto - Santo André,
Universidade Federal do ABC, 2017.
  
\vspace{0.5cm}

\hspace{1cm}105 fls. XX cm\vspace{0.5cm}

\hspace{1cm}Orientador: Professor Doutor Roldão da Rocha Junior\vspace{0.5cm}

\hspace{1cm}Tese - Universidade
Federal do ABC, Programa de Pós-Graduação em Física, 2017\vspace{0.5cm}

\hspace{1cm}1. bilineares covariantes. 2. torção.  3. quebra de invariância de Lorentz.
I. Silva Neto, José Antonio. II. Programa de Pós-Graduação em Física,
2017. III. Título: subtítulo%
\end{minipage}}
\par\end{center}

\thispagestyle{empty}
\par

\pagebreak{}
\par

\begin{minipage}[t]{0.9\columnwidth}%
\begin{center}
\textbf{\large{}\rule[0.5ex]{1\columnwidth}{1pt}}
\par\end{center}{\large \par}

\begin{doublespace}
Agradecimentos \\
\indent Entrar na UFABC no ano de 2011, é algo que dividiu minha vida num antes e depois, pois recebi a oportunidade de reconstruir meu futuro. \\
\indent Devo muito ao Professor Roldão da Rocha, pela sua paciência para com as minhas limitações e também porque o Professor Roldão faz algo que não é muito comum:
convida continuamente seus orientandos para a inovação, incluindo-os em trabalhos efetivos que resultam em publicações. Isso traz visibilidade científica a todos 
os seus orientandos. Meu muito obrigado ao Professor Roldão. \\
\indent Nesses anos de UFABC, fiz muitos amigos e não teria espaço para listar todos. Citarei explicitamente apenas dois: Eliezer Batista e Rafael Budaibes, pois 
são os irmãos que a minha mãe não me deu. \\
\indent Com essa tese de Doutorado, se fecha um ciclo na minha vida, pois depois de muitos anos, estou próximo de algo que sempre acalentei, ser um profissional da ciência.\\
\indent A batalha das Termópilas, onde heróis morreram defendendo a liberdade, não foi em vão. Agora, encaro o futuro pensando em outro evento \emph{300, o nascimento de um império}. 
Que eu consiga vencê-lo também.
\end{doublespace}

\rule[0.5ex]{1\columnwidth}{1pt}%
\end{minipage}

\thispagestyle{empty}
\par

\myclearpage
\par


\rule[0.5ex]{1\columnwidth}{1pt}
\par \par

\begin{center}
\par\end{center}{\par}

\vspace{0.5cm}

RESUMO

\textbf{\Large{}\vspace{0.5cm}
}{\Large \par}

A classificação de espinores de Lounesto é uma ferramenta importante na física fundamental, pois explicita a plêiade de tipos de espinores que vão além daqueles usados na 
teoria quântica de campos (TQC).
Nesse trabalho, mostramos como a classificação surge em dois tópicos: primeiro mostramos que os bilineares covariantes surgem naturalmente 
na equação de movimento para um campo fermiônico num fundo do tipo Riemann-Cartan (com dinâmica gravitacional $f(\mathit{R})$) e identificamos um espinor singular (um \emph{flag-dipole}) como sendo solução do modelo apresentado.
Esta é a primeira solução \emph{flag-dipole} encontrada na literatura.
Investigamos o comportamento dos bilineares covariantes dentro do contexto do chamado modelo padrão estendido de Colladay e Kostelecky e apresentamos
exemplos de transformações que violam a simetria de Lorentz, provendo a extensão da classificação dos espinores, de acordo com seus bilineares covariantes, a cenários com quebra de 
simetria de Lorentz. Finalmente, provamos que espinores singulares do tipo {\it flagpole} são menos suscetíveis a efeitos de violação de Lorentz.\\

\textbf{Palavras chave: bilineares covariantes, classificação de espinores, quebra de invariância de Lorentz, torção, gravidade $f(\mathit{R})$.}

\thispagestyle{empty}
\par \par


\myclearpage
\par

\begin{center}
\textbf{\large{}\rule[0.5ex]{1\columnwidth}{1pt}}
\par\end{center}{\large \par}

\selectlanguage{english}

\begin{center}

\par\end{center}\par

\vspace{0.5cm}

ABSTRACT. \\ \\

\indent The Lounesto spinor classification is an important tool in fundamental physics, because it makes explicit the pleiade of spinors types, beyond the used in
quantum field theory (QFT).
In this work, we show how the classification emerges in two topics: first we show that the bilinear covariants arise naturally in the equation of motion for a fermionic field 
in a Riemann-Cartan background (with gravitational dynamics $f(\mathit{R})$) and we identify a singular espinor field (a \emph{flag-dipole} one) as a solution of the presented model. This is the 
first solution \emph{flag-dipole} founded in the literature. We investigated the behavior of the bilinear covariants in the context of the called standard model extension of
Colladay and Kostelecky and we present examples of transformations that violate the Lorentz symmetry, providing an extension of spinors classificaiton, according to bilinear covariants, to 
scenarios with broken Lorentz symmetry. Finally, we proved that singular spinors of type flagpole are less susceptible to effects of Lorentz violation.

\vspace{0.5cm}


\thispagestyle{empty}
\par \par

\myclearpage
\par

\tableofcontents{}

\thispagestyle{empty}
\par \par

\myclearpage
\par

\input{introducao.tex}

\input{classificacao_de_lounesto.tex}

\input{geometria_e_gravitacao_elementar.tex}
\input{gravitacao_fr_com_torcao.tex}
\input{violacao_de_lorentz.tex}

\input{conclusoes.tex}
\input{apendice_omega.tex}

\bibliographystyle{JHEP}
\bibliography{ref_Tese_Fisica_JA}

%
%
%
%
%
%
%
%
%
%
%
%
%
%
%

\end{document}

%% file: introducao.tex
\chapter*{Introdução}

$\qquad$ Com a primeira quantização na mecânica quântica, o aspecto ondulatório da natureza foi estabelecido de forma quantitativa e profunda. 
Por isso, o problema de se quantizar um campo clássico se tornou natural.
Apesar disso, como a equação de Schrödinger não era covariante por transformações de Lorentz, primeiro se buscou uma equação de onda que fosse Lorentz covariante.
Partindo-se da equação relativística para a energia de uma partícula, a equação de Klein-Gordon, foi obtida, mas há dois problemas 
ao se adotar essa equação como uma equação de onda relativística. Primeiramente a causalidade exige que a mesma seja de primeira ordem na ``coordenada temporal'' $x^0$ e segundo,
a invariância de Lorentz exige que não {}{exista} nenhuma coordenada {}{preferencial}, o que implica a necessidade da equação de onda (o operador diferencial) ser de primeira 
ordem em todas as variáveis.
O problema foi resolvido por Paul Adrien Maurice Dirac \cite{Dirac:1928hu}, que interpretou essas exigências na forma de tomar a raiz quadrada do operador d'Alembertiano $\Box$. 
O operador obtido $D$ é hoje conhecido como o operador de Dirac na teoria relativística do elétron. \\
\indent  Ainda dentro da teoria relativística do elétron (em primeira quantização), Dirac considerou os chamados bilineares covariantes que encapsulam o conteúdo físico da função de onda
$\psi$. Dado um espinor $\psi$ e as matrizes de Dirac no espaço-tempo de Minkowski, onde $\gamma_5 = -i\gamma_0\gamma_1\gamma_2\gamma_3$, os bilineares covariantes são dados por \cite{lounesto,Takahashi:1982wi,Crawford}
\begin{eqnarray}
\sigma &=& \bar{\psi}\psi,  \nonumber \\
J_{\mu} &=& \bar{\psi}\gamma_{\mu}\psi, \nonumber \\
S_{\mu\nu} &=& i\bar{\psi}\gamma_{\mu\nu}\psi,  \nonumber \\
K_{\mu} &=& \bar{\psi}\gamma_{5}\gamma_{\mu}\psi, \nonumber \\
\omega &=& \bar{\psi}\gamma_{5}\psi, \nonumber 
\end{eqnarray}
\noindent onde $\bar{\psi} = \psi^{\dagger}\gamma^0$. \\
\indent Cada uma dessas grandezas possui um significado físico que discutiremos no primeiro capítulo dessa tese. Há pouco menos de duas décadas, um avanço fundamental foi feito por 
Pertti Lounesto \cite{lounesto,Crawford} que reinterpretou os bilineares covariantes do ponto de vista da teoria das álgebras de Clifford e derivou uma classificação de campos espinoriais usando
essas classes. Por exemplo, na classificação de Lounesto, o espinor que Dirac obteve no seu trabalho é de um tipo bem específico (um espinor regular, que é autoespinor do operador de paridade
e que é solução da equação de Dirac), mas hoje existem soluções 
para equações de primeira ordem em ambientes mais intrincados, que não são espinores de Dirac \cite{fabbri_ESK}. Isso indica que uma equação de primeira ordem pode ter como solução espinores {}{de} tipos específicos, que não são os do
trabalho original de Dirac \cite{Dirac:1928hu}. \\
\indent Num primeiro nível de classifica\c c\~ao, anteriormente à
segunda quantiza\c c\~ao, isto significa que os campos espinoriais são essenciais para descrever a física. Embora espinores de Dirac sejam os objetos mais comuns que carregam
uma representa\c c\~ao espinorial do grupo de Lorentz, eles são realmente apenas a ponta do {\it iceberg} que cercam um amplo conjunto de possibilidades, descritas pela classifica\c c\~ao de
Lounesto\, \cite{lounesto,roldao_1}. Esta classifica\c c\~ao de espinores é baseada nos bilineares covariantes. Espinores de Dirac estão incluídos como casos particulares de espinores
regulares, enquanto os espinores de Majorana e Weyl são exemplos bem conhecidos nas classes de espinores {\it flagpole} e {\it dipole}, respectivamente. Além desses, a classifica\c c\~ao de Lounesto 
também descreve uma classe enorme de novas possibilidades, incluindo espinores de dimensão de massa um \,\cite{daSilva:2012wp,cyleeI,Lee:2014opa,EPJJC}, formulações exóticas com mecanismos 
dinâmicos de gera\c c\~ao de massa \cite{Bernardini:2012sc} e soluções das equações de Dirac em circunstâncias específicas, as quais não são espinores de Dirac \,\cite{daRocha:2011yr}. 
Existem ainda subconjuntos em cada classe da classifica\c c\~ao de Lounesto que permanecem inexplorados, cuja dinâmica são ainda desconhecidas\,\cite{daSilva:2012wp,cyleeI,EPJJC}.
Classificações recíprocas e equivalentes têm aberto recentes desenvolvimentos \cite{roldao_1,Cavalcanti:2014wia,Fabbri:2016msm}. \\
\indent Espinores \emph{flagpoles} foram ineditamente identificados com candidatos à matéria escura \cite{daRocha:2005ti} e foram também usados para o estudo do tunelamento de campos espinoriais singulares e sua radiação Hawking no 
contexto de cordas negras e buracos negros mais gerais \cite{Cavalcanti:2015nna,daRocha:2014dla}. Outras aplicações dos espinores singulares, ainda, foram construídas e discutidas em \cite{daSilva:2012wp,Cavalcanti:2014uta}, também no 
contexto de uma aplicação entre espinores regulares e singulares \cite{HoffdaSilva:2009is}, na derivação das ações de Einstein-Hilbert, de Holst e de Palatini através da classificação de Lounesto \cite{daRocha:2009gb,daRocha:2007sd}. 
Um protótipo de espinor singular do tipo \emph{flagpole} é o Elko, candidato à matéria escura, cujos testes e predições fenomenológicas,  experimentais 
\cite{Alves:2014kta,Ahluwalia:2015vea,Basak:2014qea,Alves:2014qua,daSilva:2014kfa} e 
observacionais \cite{Pereira:2016eez,Pereira:2016muy,Pereira:2016emd,Pereira:2014pqa,Fabbri:2014foa,S.:2014dja} 
foram amplamente propostas. 
Desenvolvimentos teóricos inesperados também foram reportados em \cite{Ahluwalia:2010zn,Ahluwalia:2009rh,Ahluwalia:2008xi}. \\
\indent Muitos esforços têm sido devotados para desenredar aquelas classes menos conhecidas de espinores. Espinores {\it flagpole} incluem por exemplo espinores Elko e Majorana e
têm sido usados em diferentes contextos, de física de partículas e fenomenologia LHC até cosmologia \cite{Agarwal:2014oaa,Pereira:2014wta,Nieto:2013kqa,Lee:2013cwa,deOliveira:2013hda,Basak:2012sn,daRocha:2011xb,deOliveira:2012wp}. 
Espinores  Elko possuem a peculiar característica de serem espinores de
dimensão de massa igual a um (eles não são os únicos, como recentemente apontado em \,\cite{EPJJC}). {}{Ainda sobre espinores {\it flagpole}, podemos citar os seguintes estudos recentes:}
possíveis assinaturas em 14 TeV no LHC e como um subproduto de uma bolha de Higgs\,\cite{Dias:2010aa}, assim como alguns aspectos 
que estudam tais espinores no contexto da gravidade conforme e com torção\,\cite{Fabbri:2011mi,Fabbri:2012qr,fabbri_fr_torsion,Fabbri:2013gza,saulo}.
Uma análise completa sobre o papel dos 
espinores {\it flagpole} e {\it flag-dipole} e suas interpretações no formalismo de Penrose foram derivados em \,\cite{lounesto,daSilva:2012wp}. Novas soluções espinoriais em supergravidade
foram também recentemente obtidas \cite{Bonora:2014dfa,Bonora:2015ppa}. \\
\indent {\it Flag-dipoles} são campos espinoriais que não foram listados em aplicações físicas até recentemente, quando mostramos que esta classe de espinores é uma solu\c c\~ao para a equa\c c\~ao de Dirac 
numa configura\c c\~ao de gravita\c c\~ao $f(\mathit{R})$ com tor\c c\~ao\,\cite{fabbri_ESK}. De fato, na teoria de gravitação de Einstein-Cartan-Sciama-Kibble (ECSK), a tor\c c\~ao é acoplada 
à densidade de spin do campo da matéria. Portanto, {}{com }todos os termos os envolvendo, as derivadas covariantes e as curvaturas podem ser 
{}{decompostas} em seus homólogos
sem tor\c c\~ao mais suas contribuições com tor\c c\~ao, os quais podem ser substituídos pelas equações de campo tor\c c\~ao-spin em termos da densidade de spin do campo da matéria espinorial 
\cite{fabbri_ESK,Fabbri:2011mi,Fabbri:2012qr,fabbri_fr_torsion}. A teoria ECSK é equivalente a uma teoria complementada por potenciais de intera\c c\~ao spin-spin, com não linearidades
nas equações de campo da matéria. O fundo gravitacional específico, por exemplo $f(\mathit{R})$ ou conforme e o tipo de espinor (Dirac ou outro espinor regular: 
{\it flag-dipoles}, {\it flagpole}s, ou espinores {\it dipole} \textendash{} os singulares \cite{daSilva:2012wp}) determinarão a estrutura exata dessas não linearidades nas equações dos 
campos da matéria. Por exemplo, na ordem mais baixa a gravidade ECSK com campos fermiônicos de Dirac, as não linearidades são realizadas por interações do quadrado 
de correntes de contato axiais, fornecidas pelos potenciais de Nambu-Jona-Lasinio (NJL). Quando o campo espinorial é um campo fermiônico {\it flag-dipole}, a intera\c c\~ao 
é modificada\,\cite{fabbri_ESK}. \\
\indent Por outro lado, a simetria de Lorentz é a simetria fundamental subjacente ao modelo padrão (MP) da física de partículas elementares, sendo experimentalmente verificada 
{}{em um} nível
requintado de precisão. Mesmo assim, o paradigma da simetria de Lorentz pode ser modificado em regimes de altas energias. De fato, sempre que efeitos gravitacionais quânticos
não podem ser descartados, a simetria de Lorentz deve ser maleável para adequar esses regimes extremos na física. Como um exemplo, a simetria de Lorentz pode ser quebrada
espontaneamente por alguns campos tensoriais adquirindo valores esperados não nulos no vácuo em regimes de energia baixa de teorias de cordas efetivas \,\cite{Kostelecky:2007kx}.
Um enfoque bastante geral no qual a viola\c c\~ao da simetria de Lorentz (VSL) é incorporada com o modelo padrão foi desenvolvida por Colladay
e Kostelecky\,\cite{Colladay:1996iz,Colladay:1998fq}. Essa configura\c c\~ao tem sido usada como uma estrutura para estudar consequências de VSL numa pletora de fenômenos físicos,
impondo limites muito rigorosos nos parâmetros VSL\,\cite{Kostelecky:2007kx,Bernardini:2007uj,Bernardini:2007ez,Bernardini:2007ex,Charneski:2012py,Maluf:2015hda}.
Alguns fenômenos VSL que têm recentemente
recebido aten\c c\~ao na literatura são efeitos em buracos negros acústicos\,\cite{Anacleto:2010cr}, vórtices BPS num fundo com quebra de Lorentz e CPT\,
\cite{Bernardini:2007ez,Cantanhede:2011nm,roldao_4,Casana:2016gek,Hernaski:2016dyk,Santos:2016uds,Bazeia:2016pra,Borges:2016ntu,Accioly:2016yht,Casana:2015pra,Casana:2015hda,Borges:2016uwl}{}{ 
, branas} espessas em VSL\,\cite{Bazeia:2010yp}, dentre muitos outros. \\
\indent De acordo com a classificação de Wigner, em termos das representações irredutíveis do grupo de Poincaré,
as partículas são classificadas em termos de suas massas e spin. As quantidades correspondentes para os campos espinoriais são dadas em termos das energias e
densidades espinoriais. {}{Para perseguirmos} o mesmo espírito que Einstein seguiu para desenvolver uma teoria para a gravidade, expressando as equações de campo via
um acoplamento da curvatura com a energia, no caso mais geral, onde a torção está presente, somos compelidos a recuperar as equações de campo acoplando a energia com a curvatura
mas acompanhadas por equações de campo similares acoplando a torção com o spin. \\
\indent Quando isto é realizado da forma mais direta, as equações de Einstein para o acoplamento curvatura-energia são generalizadas para se incluir as equações de Sciama-Kibble para
o acoplamento torção-spin. {}{O sistema} de equações de campo ECSK, pode ser obtido generalizando-se o escalar de Ricci escrito em termos da métrica $\mathit{R}(g)$ pelo
escalar de Ricci escrito em termos da métrica e da torção $\mathit{R}(g,T)$ na ação e subsequentemente variando a mesma em relação aos dois campos independentes \cite{fabbri_ESK}.
Não obstante, isto é apenas a generalização mais direta de gravidade com torção. Outras teorias mais gerais podem ser obtidas adicionando torção não apenas 
implicitamente através da curvatura, mas explicitamente também, com termos quadráticos além da curvatura $\mathit{R}(g,T)+T^{2}$ na ação \cite{sotiriou}. 
Uma vez que as equações de campo são escritas e todas as contribuições da torção são separadas e evidenciadas como interações espinoriais, os efeitos destas extensões
são {}{reduzidos} a um simples escalonamento dos termos torsionais, ou equivalentemente da interação espinorial. Isto é evidenciado pela introdução de novas constantes de acoplamento
para tais potenciais espinoriais. Um dos problemas mais importantes sobre torção e gravidade, a saber o fato de que a torção deve ser relevante somente nas escalas de Planck,
pode, assim, ser superado pois nestas teorias a torção possui sua própria constante de acoplamento, que não coincide necessariamente com a constante gravitacional 
\cite{fabbri_geomet_UF,fabbri_gravity_torsion}. 
Por outro lado, contudo, estas teorias não cercam a possibilidade de termos extensões dinâmicas, tais como as fornecidas por equações de campo de ordens 
superiores. As duas {}{extensões} mais importantes são as do caso no qual a curvatura escalar $\mathit{R}$  é trocada por uma função arbitrária $f(\mathit{R})$ e no caso que é possível 
a implementação de uma simetria conforme na ação \cite{fabbri_conformal_gravity_DM,fabbri_conformal_gravity_elko}. A seguir, trataremos de ambos. Nessa tese, abordamos cenários onde surgem campos espinoriais singulares
de forma natural. \\
\indent {}{A tese está dividida da seguinte forma.}\\
\indent No capítulo $1$ apresentamos o grupo de Lorentz e  construímos o recobrimento duplo do mesmo, o grupo spin, identificado com o grupo 
$SL(2,\mathbb{C})$. Em seguida caracterizamos as representações irredutíveis desse grupo spin de uma forma construtiva. Ao revermos a equação de onda de Dirac para a teoria
quântica do elétron, chegamos {}{ao} conceito de bilinear covariante, que foi crucial no trabalho de Dirac. Abordamos os bilineares covariantes sob o ponto de vista da 
classificação de Lounesto e discutimos o seu significado físico e finalmente, {}{vemos} como as transformações de Fierz mostram como as representações do grupo de Lorentz agem 
na representação complexa da álgebra de Clifford do espaço-tempo. \\
\indent No capítulo $2$ fazemos um resumo da geometria diferencial relevante para o nosso trabalho: variedades diferenciáveis, campos vetoriais e tensores.  
Revemos brevemente a noção de uma variedade pseudo-Riemanniana, construindo exemplos de tensores métricos e suas curvaturas associadas, como o tensor de curvatura de Riemann e 
a curvatura escalar. Também vemos que dada a métrica nesse de tipo de espaço, temos associada uma única conexão simétrica e compatível com a métrica, a chamada conexão de Levi-Civita,
que fixa a geometria do espaço. Focando agora na Física, apresentamos alguns conceitos da teoria da gravitação de Einstein e deduzimos de forma cuidadosa, a equação de campo de Einstein 
partindo da ação de Einstein-Hilbert. Em seguida apresentamos um caminho evolutivo de teorias de gravitação, abordando as teorias de gravitação de 
Einstein-Cartan-Sciama-Kibble (ECSK), a qual utiliza uma conexão com torção e depois apresentamos as teorias chamadas de $f(\mathit{R})$, que são generalizações da ação de 
Einstein-Hilbert, pois a função $f(\mathit{R})$ é tomada como sendo analítica. Enfatizamos que, apesar de uma revisão para estabelecermos as ideias, notação e pré-requisitos aos
capítulos subsequentes, a rede de apresentação é original. Os capítulos $3$,$4$ e apêndice são originais, sendo o cerne do trabalho de doutorado.\\
\indent No capítulo $3$, discutimos um pouco mais a teoria $f(\mathit{R})$ da gravitação, construindo um modelo de campo fermiônico com fundo $f(\mathit{R})$.
Derivamos as equações de movimento para as partes gravitacional e fermiônica da teoria, deixando explícito o fato de que bilineares covariantes aparecem na
equação de movimento fermiônica, restringindo desse modo o tipo de campo espinorial que pode ser solução da mesma e além disso, isso implica condições 
a serem obedecidas pela métrica ou sobre o campo de matéria. Mostramos também a existência de um campo
espinorial singular (um {\it flag-dipole}) que é solução da equação de onda de um campo fermiônico sujeito a um fundo $f(\mathit{R})$, um resultado original, totalmente inédito na literatura.\\
\indent No capitulo $4$ o foco é a questão atual muito importante da quebra de simetria de Lorentz, uma teoria proposta e desenvolvida 
principalmente por Colladay e Kostelecky. Um ponto fundamental na moderna física, é que o espaço-tempo é uma variedade diferenciável, mas usando argumentos quânticos,
tal estrutura não deve se sustentar na escala de Planck. Como não dispomos de acesso experimental em tal escala, devemos usar algum método indireto de investigação e uma quebra
de simetria de Lorentz (via dinâmica, na Lagrangiana dos modelos) é uma possibilidade importante. 
Investigamos  o comportamento dos bilineares covariantes nesse contexto, exibindo transformações concretas sobre {}{os mesmos}. Como resultado original, construímos uma teoria com violação de Lorentz para um campo eletromagnético
(e fermiônico) sujeito a um fundo $f(\mathit{R})$ com torção e, dentro 
dessa teoria, mostramos que para espinores singulares específicos, o acoplamento com a torção é menos sensível, {}{no sentido de que os bilineares 
covariantes aparecem explicitamente nos termos de acoplamento, por isso, para espinores singulares, vários termos são nulos}.\\
\indent O capítulo final da tese é sobre as conclusões e desenvolvimentos futuros. \\


%% file: classificacao_de_lounesto.tex
\chapter{Classificação de Lounesto de espinores}

\section{Introdução aos espinores: caso relativístico}

$\qquad$ As referências básicas para esta seção, são clássicos da literatura \cite{peskin,maggiore,ticciati,ryder,weinberg}. Neste capítulo, denotaremos por $\eta$ a forma bilinear de 
Minkowski
\begin{eqnarray}
\eta(x,y) = x^0y^0 - x^1y^1 - x^2y^2 - x^3y^3,
\end{eqnarray}
onde $x = (x^0,x^1,x^2,x^3)\in \mathbb{R}^4$. Uma simetria ubíqua na natureza
é  o grupo de Lorentz (posteriormente veremos possibilidades 
de violação da simetria de Lorentz), que visto de uma forma geométrica, é o conjunto das transformações lineares que preservam a forma bilinear $\eta$, ou seja
\begin{eqnarray}
 \eta(T(x),T(y)) = \eta(x,y),\quad \forall x,y\in \mathbb{R}^{1,3}. 
\end{eqnarray}
\indent Este grupo atua nos eventos, que podem ser identificados com os pontos do espaço-tempo, o espaço
de Minkowski $\mathbb{R}^{1,3}$. Na física de partículas elementares, toda a construção parte inicialmente de campos tensoriais (que tem como casos particulares, campos vetoriais e escalares) 
e espinoriais.
Neste contexto, o grupo de Lorentz age nesses campos através de representações adequadas. 
\begin{obs}
O grupo de Lorentz geral $SO(1,3)$ possui quatro componentes conexas, mas fisicamente, {}{interessa-nos} a componente conexa que contêm a unidade do grupo e é 
temporalmente orientada para o futuro, que denotaremos como  $SO(1,3)_{+}^{\uparrow}$.
\end{obs}
\indent Podemos pensar nas representações lineares de um grupo, de modo mais preciso. 
\begin{define}
Seja $G$ um grupo, e $W$ um espaço vetorial de dimensão finita sobre os reais $($ou complexos$)$.Uma representação linear de $G$  é um homomorfismo de grupos
$\pi: G \to GL(W)$.
\end{define}
\indent Podemos pensar na representação trivial do grupo de Lorentz $G = SO(1,3)$, ou seja, a função $\pi$ que associa cada matriz de $G$ a si mesma. Dispondo disso, temos imediatamente
construções canônicas, que nos fornecem inúmeras representações:
\begin{eqnarray}
\bigotimes^{k}\pi: G &\rightarrow&  GL\left (\bigotimes^{k}W \right ) \nonumber \\
g &\mapsto& \bigotimes^{k}\pi_g : \bigotimes^{k}W \rightarrow \bigotimes^{k}W  \nonumber \\
&&\hspace{0.3cm}  v_1\otimes\dots\otimes v_k \mapsto  \pi_g(v_1)\otimes\dots\otimes \pi_g(v_k),
\end{eqnarray}
\begin{eqnarray}
\bigoplus^{k}\pi: G &\rightarrow&  GL\left (\bigoplus^{k}W \right ) \nonumber \\
g &\mapsto& \bigoplus^{k}\pi_g : \bigoplus^{k}W \rightarrow \bigoplus^{k}W  \nonumber \\
&&\hspace{0.3cm}  v_1\oplus\dots\oplus v_k \mapsto  \pi_g(v_1)\oplus\dots\oplus \pi_g(v_k), 
\end{eqnarray}
\begin{eqnarray}
\bigwedge^{k}\pi: G &\rightarrow&  GL\left (\bigwedge^{k}W \right ) \nonumber \\
g &\mapsto& \bigwedge^{k}\pi_g : \bigwedge^{k}W \rightarrow \bigwedge^{k}W  \nonumber \\
&&\hspace{0.08cm}  v_1\wedge\dots\wedge v_k \mapsto  \pi_g(v_1)\wedge\dots\wedge \pi_g(v_k), 
\end{eqnarray}
\noindent onde $\pi_g = \pi(g)$. Estas são as chamadas representações tensoriais, de soma direta e de potência exterior, respectivamente. Caracterizando estas representações, teremos o ambiente a que os
campos físicos pertencem (pensando no grupo de Lorentz). No caso específico dos espinores, estes 
objetos serão elementos de um espaço vetorial no qual o grupo espinorial age e, no caso relativístico, o grupo em questão
é o $SL(2,\mathbb{C})${}{. Portanto}, os espinores clássicos \cite{roldao_1} serão elementos de $\mathbb{C}^4$. Ao contrário do que se afirma na literatura, o grupo espinorial (caso relativístico), o 
$SL(2,\mathbb{C})$ não é uma representação no  sentido usual (como na definição e construções acima). Para vermos isto, vamos calcular explicitamente os grupos espinoriais 
não-relativístico (espinores de Pauli) e relativístico (espinores de Weyl). Construiremos os recobrimentos duplos nos dois casos:
\begin{enumerate}
 \item  $SU(2)$ como o recobrimento duplo de $SO(3)$. A esfera $S^3$ pode ser vista como sendo os quatérnios ($\mathbb{H}$)  
normalizados, ou seja
\begin{eqnarray}
 S^{3} = \{q\in \mathbb{H}\,\vert\, \overline{q}q = 1\}, 
\end{eqnarray}
 sendo $\overline{q} = x_0 - x_1i - x_2j - x_3k$ o conjugado quaterniônico, onde $i,j,k$ são as unidades quaterniônicas. Agora, consideremos a ação de $S^{3}$ em $\mathbb{R}^{4}$ dada por
\begin{eqnarray}
 \varphi_q: &\mathbb{R}^{4}& \to \mathbb{R}^4 \nonumber \\
 &x& \mapsto qx\overline{q}. 
\end{eqnarray}
$\qquad$Esta aplicação possui as seguintes propriedades:
\begin{itemize}
\item [(i)]  $\mathbb{R}$-linearidade,
\item [(ii)]  Preserva norma e portanto $\varphi_q \in  SO(4)$,
\item [(iii)] $\varphi_{q_1q_2} = \varphi_{q_1}\varphi_{q_2}$,
\item [(iv)] $\varphi_q$ restrita a $\mathbb{R}^3$ (injetado em $\mathbb{R}^{4}$ via $\iota(x) = (0,x)$), é uma aplicação \linebreak  $\varphi_{q}\Big\vert_{\mathbb{R}^{3}}: 
\mathbb{R}^{3} \to \mathbb{R}^{3}$. Com isso, segue-se que $\varphi:S^{3} \to SO(3)$.
\end{itemize}
$\qquad$Finalmente utilizando a aplicação
\begin{eqnarray}
 \Xi: &S^{3}& \to SU(2) \nonumber \\
  &x& \mapsto \begin{pmatrix} x_0 + ix_1 & x_2 +ix_3 \\ -(x_2 - ix_3) & x_0 - ix_1\end{pmatrix}, 
\end{eqnarray}
temos que $S^3 \cong SU(2)$ e, portanto, segue-se que a aplicação $\varphi: S^3 \to SO(3)$ é um homomorfismo de grupos com núcleo 
$\ker \varphi = \{I,-I\} \cong \mathbb{Z}_2$
e um recobrimento duplo. Daí concluímos que a seta funcional correta é ao contrário do que se entende por
uma representação de grupo (na verdade, $SO(3)$ representa $SU(2)$!) e só temos um isomorfismo passando ao quociente.

\item $SL(2,\mathbb{C})$ como o recobrimento duplo $SO(1,3)$ (o grupo de Lorentz). Primeiro consideremos a aplicação
\begin{eqnarray}
 \varphi: &\mathbb{R}^{1,3}& \to Herm(2) \nonumber \\
 &x& \mapsto x_{\mu}\sigma^{\mu},
\end{eqnarray}
 onde $Herm(2) = \{T \in M_{2}(\mathbb{C})\,\vert\, T^{\dagger} = T\}$ e $\sigma^{\mu}$ são as matrizes de Pauli. Esta aplicação possui as seguintes propriedades:
\begin{itemize}
 \item [(A)] $\det \varphi(x) = \eta(x,x)$, ou seja, a aplicação preserva a norma de Minkowski,
 \item [(B)] $\varphi$ é uma bijeção $\mathbb{R}$-linear entre estes espaços.
\end{itemize}
$\qquad${}{Agora, por definição, um elemento $A$ de $SL(2,\mathbb{C})$ é uma transformação  linear de $\mathbb{C}^{2}$, com $\det A = 1$. Vamos} considerar uma ação $\Theta:SL(2,\mathbb{C})\times Herm(2) \to Herm(2)$ dada por 
\begin{eqnarray}
\Theta: &SL(2,\mathbb{C}) \times Herm(2)& \to Herm(2) \nonumber \\
&(A,T)& \mapsto ATA^{\dagger},
\end{eqnarray} 
como $\mathbb{R}^{1,3}\cong Herm(2)$ temos uma aplicação $\pi: SL(2,\mathbb{C}) \to SO(1,3)$. Esta aplicação está bem definida, é um homomorfismo de grupos e 
um recobrimento duplo.
\end{enumerate}
\par Então, no sentido acima, o grupo $SL(2,\mathbb{C})$ representa o grupo de Lorentz  e o espaço que carrega sua representação fundamental (a trivial), no caso $W = \mathbb{C}^2$ é um espaço
de espinores, os chamados espinores de Weyl pertencentes à representação com quiralidade negativa $D^{(\frac{1}{2},0)}$. 
Este ponto de vista, é reforçado por vários fatores, como por exemplo, o fato de que perdemos representações se não usarmos o recobrimento duplo 
do grupo de Lorentz. E agora, usando a representação fundamental, vamos apresentar de forma construtiva todas as representações irredutíveis (com quiralidade negativa)
para o grupo $SL(2,\mathbb{C})$, que encapsulam todos os campos fermiônicos de spin $j$ (inteiro ou semi-inteiro {}{em $D^{(\frac{1}{2},0)}$}).
Para isso, consideremos \cite{hall} o conjunto $\mathbb{C}^{m}[(z_1,z_2)]$ que consiste de todos os polinômios nas variáveis $z_1,z_2$ que são homogêneos de grau $m$. Este
espaço vetorial é isomorfo ao espaço vetorial da $m$-ésima potência simétrica $S^{m}(\mathbb{C}^{2})$. Caracterizamos $\mathbb{C}^{m}[(z_1,z_2)]$ usando uma base
\begin{eqnarray}
\mathbb{C}^{m}[(z_1,z_2)] = span_{Lin} \{z_1^{m - k}z_2^{k},\, \small{k} = 0,1,\dots,m\}. 
\end{eqnarray}
$\qquad$Portanto temos que $\dim_{\mathbb{C}}\mathbb{C}^{m}[(z_1,z_2)] = m + 1$. 
Denotemos por $ z = (z_1,z_2)\in \mathbb{C}^2$. Então, podemos induzir uma sequência infinita de representações de $SL(2,\mathbb{C})$ 
do seguinte modo:
\begin{eqnarray}
 \Pi_{m}: SL(2,\mathbb{C}) &\rightarrow& GL(S^{m}(\mathbb{C}^{2}))\nonumber  \\
A &\mapsto& \Pi_m(A): S^{m}(\mathbb{C}^{2}) \to S^{m}(\mathbb{C}^{2}) \nonumber \\
&& \hspace{2.0cm} p(z) \mapsto p(A^{-1}z) 
\end{eqnarray}
$\qquad$Um elemento de $p\in \mathbb{C}^{m}[z]$ tem a forma $p(z) = \sum_{k = 0}^{m} a_kz_1^{m - k}z_2^{k}$ e tomando-se um $A\in SL(2,\mathbb{C})$ 
dado explicitamente por
\begin{eqnarray}
A = \begin{pmatrix} a_{11} & a_{12} \\ a_{21} & a_{22}\end{pmatrix}, 
\end{eqnarray}
\noindent temos que $D^{(\frac{1}{2},0)}$
$A^{-1} =  \begin{pmatrix} a_{22} & -a_{12} \\ -a_{21} & a_{11}\end{pmatrix}$. Com isso, temos que a representação toma
a seguinte forma:
\begin{eqnarray}
\Pi_m (p)(z) = \sum_{k = 0}^{m} a_k(a_{22}z_1 - a_{12}z_2)^{m - k}(-a_{21}z_1 + a_{11}z_2)^{k}.
\end{eqnarray}
$\qquad$O polinômio acima também é homogêneo de grau  $m$ e, portanto, a função está bem definida (pois leva polinômios homogêneos em 
polinômios homogêneos). É imediato que a definição adotada é um homomorfismo de grupos.
Além da representação trivial, o grupo $SL(2,\mathbb{C})$ possui outra representação \cite{roldao_2}, não equivalente à fundamental, dada por
\begin{eqnarray}
\overline{\pi}: &SL(2,\mathbb{C})& \to GL(\mathbb{C}^{2}) \nonumber \\
&A& \mapsto (A^{\dagger})^{-1}. 
\end{eqnarray}
$\qquad$Neste caso, os espinores associados, são os chamados de espinores de Weyl pertencentes à representação com quiralidade positiva $D^{(0,\frac{1}{2})}$. Repetindo o procedimento acima, 
temos uma nova série infinita de representações irredutíveis.
\section{A classificação de espinores}
$\qquad$ Dentro do âmbito da matéria fermiônica, desde Dirac usamos os chamados campos espinoriais na descrição cinemática e dinâmica da mesma. Em termos de classificação, primeiro Eugene Wigner \cite{Wigner:1939cj}
classificou as partículas possíveis
de serem encontradas na natureza através da caracterização de todas as representações irredutíveis do grupo {}{de Poincaré}. Depois Dirac (em 1928) introduz os bilineares covariantes que encapsulam a física contida num campo 
{}{espinorial}.
Finalmente, Pertti Lounesto, utilizando-se de álgebras de Clifford deu uma contribuição fundamental: no livro \cite{lounesto}, 
Lounesto caracteriza os bilineares covariantes em termos dessa álgebra, mostrando que os campos espinoriais 
se dividem em seis classes. É o que apresentaremos agora. \\
\indent Seja $A$ uma álgebra associativa e $W$ um espaço vetorial, uma representação $\pi$ de $A$ é um homomorfismo de álgebras  $\pi: A \to End(W)$. \\
\indent Dirac definiu uma equação de evolução relativística (em primeira quantização) para o elétron
\begin{eqnarray}
 (D - m)\psi = 0,
\end{eqnarray}
\noindent onde $\psi$ é um espinor coluna de $\mathbb{C}^{4}$ (uma função diferenciável $\psi: \mathbb{R}^{4} \to \mathbb{C}^{4}$, identificada com 
uma matriz coluna)  e $D$ é um operador diferencial\footnote{Daqui para frente usaremos $\partial_\mu = \frac{\partial}{\partial x^\mu}$.}
\begin{eqnarray}
 D = i\gamma^{\mu}\partial_{\mu}, 
\end{eqnarray}
\noindent cujo quadrado é o Laplaciano $\Delta$ do espaço de Minkowski (o d'Alembertiano no caso), recuperando a equação de Klein-Gordon\footnote{Para espaços curvos, temos o teorema de Lichnerowicz: $D^2 = \Delta + \frac{\mathbf{R}}{4}$, 
onde $\mathbf{R}$ é a curvatura escalar do espaço \cite{lawson,friedrich}.} $D^{2} = \Delta = \nabla^2$.\\
\indent As matrizes $\gamma_{\mu},\,\mu = 0,1,2,3$, por causa {}{dessa condição}, obedecem as seguintes {}{relações}
\begin{eqnarray}
 & &  \gamma_{0}^2 = I,\,\, \gamma_1^2 = \gamma_2^2 = \gamma_3^2 = -I, \nonumber \\
 & & \gamma_{\mu}\gamma_{\nu} + \gamma_{\nu}\gamma_{\mu} = 0,\quad \mu\neq \nu. \nonumber
 \end{eqnarray}
$\qquad$As equações acima são as relações fundamentais numa álgebra de Clifford e as matrizes $\gamma_{\mu}$  são uma representação das mesmas{}{: a} chamada representação de Dirac, que é dada por:
\begin{eqnarray}
\left . \begin{matrix} \gamma_{0} = \begin{pmatrix} I & 0 \\ 0 & -I \end{pmatrix} & \gamma_{k} = \begin{pmatrix} 0 & -\sigma_k \\ \sigma_k & 0 \end{pmatrix}, \quad k = 1,2,3 \end{matrix} \right ., \nonumber 
\end{eqnarray}
 onde as $\sigma_k$ são as matrizes de Pauli.

\vskip 0.5cm

\section{Os bilineares covariantes}

$\qquad$Originalmente, os bilineares covariantes foram tomados por Dirac como observáveis exclusivamente para o elétron 
\footnote{Isso foi feito no trabalho\cite{Dirac:1928hu}.}. \\
\indent Dado um espinor $\psi$ e as matrizes de Dirac no espaço-tempo de Minkowski, onde $\gamma_5 = -i\gamma_0\gamma_1\gamma_2\gamma_3$, os bilineares covariantes são dados por \cite{lounesto,Takahashi:1982wi,Crawford} 
\begin{eqnarray}
\sigma &=& \bar{\psi}\psi,  \nonumber \\
J_{\mu} &=& \bar{\psi}\gamma_{\mu}\psi, \nonumber \\
S_{\mu\nu} &=& i\bar{\psi}\gamma_{\mu\nu}\psi,  \nonumber \\
K_{\mu} &=& \bar{\psi}\gamma_{5}\gamma_{\mu}\psi, \nonumber \\
\omega &=& \bar{\psi}\gamma_{5}\psi, \nonumber 
\end{eqnarray}
\noindent onde $\bar{\psi} = \psi^{\dagger}\gamma^0$ {}{e $\gamma_{\mu\nu} = \frac{i}{2}[\gamma_\mu,\gamma_\nu]$}. \\
\indent Estas grandezas são ditas serem covariantes por serem preservadas pelo grupo de Lorentz. Vamos discutir o significado dos bilineares covariantes (válido apenas para o caso 
da teoria do elétron de Dirac):
\begin{itemize}
 \item [(I)] ($\sigma = \bar{\psi}\psi$) \\
  A quantidade $\sigma$ é proporcional ao termo de  massa na Lagrangiana de Dirac.
 \item [(II)] ($J_{\mu} = \bar{\psi}\gamma_{\mu}\psi$)  \\
  $J_0 = \psi^{\dagger}\psi$, integrada sobre um domínio do tipo espaço, fornece a probabilidade de se encontrar o elétron naquele domínio. 
 As quantidades  $J^k = \psi^{\dagger}\gamma^{0}\gamma^{k}\psi$ \linebreak ($k = 1,2,3$) fornecem a densidade de  corrente de probabilidade {}{e $\mathbf{J} = J_\mu\gamma^\mu$.}
 \item [(III)] ($S_{\mu\nu} = i\bar{\psi}\gamma_{\mu\nu}\psi$) \\
  Os $S_{\mu\nu}$ denotam as componentes de  um tensor que descreve a densidade de  momento angular intrínseco {}{e $\mathbf{S} = S_{\mu\nu}\gamma^{\mu\nu}$}.
 \item [(IV)] ($K_{\mu} = \bar{\psi}\gamma_{5}\gamma_{\mu}\psi$) \\
  O campo vetorial $K_{\mu}$  descreve a direção do spin do elétron, {}{mas se trata de uma caracterização que depende do referencial e se 
  enfocamos partícula ou antipartícula. Temos aqui a corrente quiral.}
 \item [(V)] ($\omega = \bar{\psi}\gamma_{5}\psi$) \\
  Esta última quantidade serve para sondar  violações de paridade. 
\end{itemize}
$\qquad$ Estas quantidades classificam o espinor no seguinte sentido: temos duas classes {}{principais, }de  espinores singulares e não-singulares. 
Os espinores singulares ocorrem quando $\omega =\sigma = 0$. No caso dos não-singulares, temos que
os campos $\mathbf{S}\text{ e }\mathbf{K}$ são não nulos simultaneamente. A classificação é resumida na seguinte forma \cite{lounesto}:
\begin{itemize}
\item[1)] $\sigma\neq0,\;\;\; \omega \neq0$\qquad\qquad\qquad\qquad\qquad4) $\sigma= 0 = \omega , \;\;\;\mathbf{S}\neq 0, \;\;\;\mathbf{K}\neq0,$%
\label{Elko11}
\item[2)] $\sigma\neq0,\;\;\; \omega = 0$\label{dirac1}\qquad\qquad\qquad\qquad\qquad5) $\sigma= 0 = \omega , \;\;\;\mathbf{S}\neq0,\;\;\; \mathbf{K}=0,$%
\label{tipo41}
\item[3)] $\sigma= 0, \;\;\;\omega \neq0$\label{dirac21} \qquad\qquad\qquad\qquad\qquad\!6) $\sigma= 0 = \omega , \;\;\; \mathbf{S}=0, \;\;\; \mathbf{K} \neq 0.$%
\end{itemize}
$\qquad$ Os tipos $1,2\text{ e } 3$ são chamados de campos espinoriais regulares, os espinores do tipo-($4$) são os {\it flag-dipole}, o tipo-($5$) são os {\it flagpole} e o 
$6$ são os {\it dipole}.\\
\indent {}{Os exemplos clássicos de espinores, no âmbito da classificação, são identificados como: espinores de Majorana são do tipo-($5$), espinores de Weyl são do tipo-($6$)
e os espinores Elko são do tipo-($5$).}
A ideia básica do nosso
trabalho é o de classificar espinores em vários contextos e quando possível, ir além do nível cinemático, mostrando como o tipo espinorial (segundo Lounesto) influencia na
dinâmica. \\ 
\indent O capítulo 3 do nosso trabalho, irá usar detalhes mais específicos de duas classes de espinores na classificação de Lounesto, por isso as duas próximas seções tratam das mesmas.

\subsection{Álgebras de Clifford e campos espinoriais de tipo-(4) e tipo-(5)}

$\qquad$Seja $V$ um espaço vetorial $n$ dimensional real e $\Lambda(V) = \bigoplus_{k=0}^n\Lambda^k(V)$ o espaço dos multivetores sobre $V$, onde $\Lambda^k(V)$
denota o espaço vetorial das $k$-formas. Para definir a reversão, dado $\tau,\psi,\xi\in\Lambda(V)$, a {contração a esquerda} é definida implicitamente
por $\eta(\tau\lrcorner\psi,\xi)=\eta(\psi,\tilde\tau\w\xi)$. O produto de Clifford entre $\vv\in V$ e $\psi$ é  fornecido por $\vv\psi = \vv\w \psi + \vv\lrcorner \psi$.
Dada a métrica $\eta$, o par $(\La(V),\eta)$ munido do produto de Clifford é a álgebra de Clifford $\cl_{1,3}$ de $\RR^{1,3}$. Todos os campos espinoriais são definidos 
numa variedade que é localmente um espaço-tempo de Minkowski $(M,\eta,\mathring{D},\tau_{\eta },\uparrow )$ no que se segue, onde $M$ é uma variedade, $\mathring{D}$
denota a conexão de Levi-Civita associada a $\eta$, $M$ é orientada pelo 4-volume $\tau _{\eta }$  e orientada temporalmente por $\uparrow $. Além disso, $\{\mathrm{e}_{\mu }\}$ 
é uma seção do fibrado de referenciais $\mathbf{P}_{\mathrm{SO}_{1,3}^{e}}(M)$. O conjunto $\{\mathrm{e}^{\mu }\}$  é o referencial dual: 
$\mathrm{e}^{\mu }(\mathrm{e}_{\nu})=\delta _{\nu }^{\mu }$ {}{e} denotamos 
$\mathrm{e}_{\mu\nu} = \mathrm{e}_{\mu}\mathrm{e}_{\nu}$ e $\mathrm{e}_{\mu\nu\rho} = \mathrm{e}_{\mu}\mathrm{e}_{\nu}\mathrm{e}_{\rho}$.\\
$\qquad$Para um entendimento melhor da estrutura dos campos espinoriais do tipo-(4) e seus casos limite tipo-(5), analisaremos a forma mais geral desses 
tipos de espinores.

\subsection{Campos espinoriais do tipo-(4)}\label{field4}

$\qquad$Seja um campo espinorial $\psi: \mathbb{R}^{1.3} \rightarrow \mathbb{C}^4$, 
dado por $\psi(x) = (f(x),g(x), \zeta(x), \xi(x))$.
Nosso objetivo é caracterizar o mais geral campo espinorial $\psi$ {\it flag-dipole} do tipo-(4) (para aplicações, ver \cite{fabbri_ESK,Cavalcanti:2014wia}), 
a condição $\sigma = 0 =\omega$  resulta em $\zeta f^{*}+\xi g^{*}=0$. {}{No que segue, estamos utilizando a representação de Weyl.}
Temos que analisar as possibilidades indicadas por esta equação. Se $f=0=g$ ou $\zeta=0=\xi$, isto implica um campo espinorial do tipo-(6), com $\textbf{S}=0$ e portanto
esta possibilidade deve ser descartada aqui. Permanecem as condições: ou $\zeta=0=\xi,\; f=0=g$, ou nenhuma das componentes pode ser nula. Neste último caso, podemos isolar
uma parte delas, por exemplo $f=\frac{g\zeta \xi^*}{\Vert \zeta \Vert^2}$. Além disso, a condição $\textbf{K}\neq 0$ induz as seguintes possibilidades: 
\begin{enumerate}
\item Se $\zeta=0=g$, então $K_1 = K_2=0$, e $K_0 \neq 0 \neq K_3 \Rightarrow \Vert f \Vert^2 \neq \Vert \xi \Vert^2$;
\item Se $f=0=\xi$, isto implica a $K_1 = K_2=0$, e $K_0 \neq 0 \neq K_3 \Rightarrow \Vert g\Vert^2 \neq \Vert \zeta\Vert^2$;
\item Se todas as componentes são não-nulas, $K_1 \neq 0 \neq K_2 \Rightarrow \Vert g\Vert^2 \neq \Vert \zeta\Vert^2$.
\end{enumerate}
$\qquad$No terceiro caso, se $\Vert g\Vert^2 = \Vert \zeta\Vert^2$, portanto $\textbf{K}=0$. Além disso, ainda no terceiro caso, 
$\Vert g\Vert^2 \neq \Vert \zeta\Vert^2 \Leftrightarrow \Vert f\Vert^2 \neq \Vert \xi\Vert^2$. Portanto, os possíveis campos espinoriais {}{(independentes, inequivalentes)} do tipo-(4) são: 
\begin{eqnarray}\label{errr}\psi_{_{(4)}}&=&(f, 0, 0, \xi)^\intercal\;,\quad\qquad\quad\qquad \; \Vert f\Vert^2 \neq \Vert \xi\Vert^2\;\,,\;\;\;\;\;\;\;\;\;\text{ou}\nonumber\\
\psi_{_{(4)}}&=&(0, g, \zeta, 0)^\intercal\;,\quad\qquad\quad\qquad \;\Vert g\Vert^2 \neq \Vert \zeta\Vert^2\;\,,\;\;\;\;\;\;\;\;\;\text{ou}\nonumber\\
\psi_{_{(4)}}&=&\left(\frac{g\zeta \xi^*}{\Vert \zeta\Vert^2}\;, g, \zeta, \xi\right)^\intercal,\qquad\quad\;\;\; \Vert g\Vert^2 \neq \Vert \zeta \Vert^2 \;.\end{eqnarray}
Se alguma desigualdade associada a um destes espinores acima não vale, verifica-se imediatamente como sendo um espinor de tipo-(5), o qual será analisado no que  segue
\cite{fabbri_ESK,Cavalcanti:2014wia}.

\subsection{Campos espinoriais do  tipo-(5)}\label{field5}

$\qquad$Começamos observando como as condições dos bilineares covariantes associados a um campo espinorial do tipo-(5) implicam as seguintes condições sobre as componentes
do campo espinorial
\begin{eqnarray}\label{eq1}
\sigma&=&\overline{\psi} \psi = 0 = -\overline{\psi}\gamma_{0123} \psi = \omega \Rightarrow \zeta f^{*}+\xi g^{*}=0,\\\label{eq2}
K_1&=&\overline{\psi}i \gamma_{0123} \gamma_1 \psi = 0 = \overline{\psi}\gamma_5\gamma_2 \psi = K_2 \Rightarrow g f^{*}+\xi \zeta^{*}=0,\\
\label{eq3}
K_0&=&\overline{\psi}i \gamma_{0123} \gamma_0 \psi = 0 =\overline{\psi}\gamma_5 \gamma_3 \psi = K_3 \Rightarrow \|f\|^2=\|\xi\|^2 \mbox{ e } \|g\|^2=\|\zeta\|^2.
\end{eqnarray}
\indent A equação \eqref{eq3} pode ser obtida de \eqref{eq1} e \eqref{eq2}, as quais são essenciais para caracterizar campos espinoriais do tipo-($5$). Neste sentido, uma equação
candidata para descrever a dinâmica destes campos espinoriais gerais deve manter \eqref{eq1} e \eqref{eq2} invariantes. Essas equações implicam
\begin{equation}\label{eq4}
f=-\xi^*(\zeta + g)(\zeta^*+g^*)^{-1}=-\xi^*\left(\frac{\zeta+g}{\Vert \zeta+g \Vert}\right)^2
\end{equation}
e tomando $\tan\varphi_1=-i\frac{\zeta+g-(\zeta+g)^*}{\zeta+g+(\zeta+g)^*}$, podemos escrever 
$
f=-\xi^*e^{2i \varphi_1}$ e $g=-\zeta^*e^{2i \varphi_2},
$
onde $\varphi_1$ e $\varphi_2$ são relacionados por \footnote{Quando  $\varphi_1 \neq n \pi$, isto é, $\zeta$ + g não é real.}
\begin{equation}\label{eq7}
\tan\varphi_2=-i\frac{\xi(1+e^{-2i\varphi_1})-[\xi(1+e^{-2i\varphi_1})]^*}{\xi(1-e^{-2i\varphi_1})+[\xi(1-e^{-2i\varphi_1})]^*}=-\cot \varphi_1.\nonumber
\end{equation}
\indent Contudo, $\tan\varphi_2= -\cot \varphi_1 \Rightarrow \varphi_2= \varphi_1+(2k+1)\frac{\pi}{2}$ e então \linebreak 
$e^{2i \varphi_2}=e^{2i \varphi_1}e^{i(2k+1)\pi}=-e^{2i \varphi_1}$, para todo {}{$k\in \{0,1,2,\ldots\}$}. \\
\indent Portanto, um espinor geral do tipo-($5$) pode ser representado por
\begin{eqnarray}
&\psi_{_{(5)}}=\left(-\xi^*e^{2i \varphi_1}, \zeta^*e^{2i \varphi_1}, \,\zeta, \,\xi\right)^\intercal\,.\label{eq8}
\end{eqnarray}
\indent Escrevendo $\psi_{_{(5)}} = (\chi_2, \chi_1)^\intercal$, é imediato realizar que 
$\chi_2= -i\sigma_2\chi_1^* e^{2i\varphi_1}=\sigma_2\chi_1^* e^{i(2\varphi_1-\frac{\pi}{2})}$.
Tomando $\varphi\equiv 2\varphi_1-\frac{\pi}{2}$, uma forma mais compacta de \eqref{eq8} é
\begin{eqnarray}
\psi_{_{(5)}}=\left(e^{i\varphi}\sigma_2\chi_1^*\,,\,\chi_1\right)^\intercal\,.\label{eq10}
\end{eqnarray}
\indent Agindo agora o operador de conjugação de carga \cite{allu,allu1}, com $i\Theta = \sigma_2$, o mesmo fornece
\begin{equation}
C\psi_{_{(5)}}=\mu \psi_{_{(5)}},\qquad{\rm para}\quad C=%
\bigl(\begin{smallmatrix}
\OO & i\Theta \\
-i\Theta & \OO \nonumber
\end{smallmatrix}\bigr) {\cal K}
\qquad {\rm e}\quad \mu =-e^{i\varphi}
 .\label{conj}
\end{equation}
\indent Aqui ${\cal K}$ conjuga as componentes espinoriais. Portanto os autovalores tomam valores na esfera $S^1$. Quando estes autovalores são reais e $\chi_1, \chi_2$ são
autoestados de helicidade duais, campos espinoriais Elko são obtidos. Os campos espinoriais {\it {\it flagpole}} do tipo-($5$) têm um papel proeminente na derivação de todas
as Lagrangianas para a gravidade a partir de uma para supergravidade, são estabelecidas em \cite{osmano} e são fibrações de Hopf \cite{daRocha:2009gb,daRocha:2007sd,daRocha:2008we}.
{}{Por ser um tipo} de campo espinorial pouco conhecido, faremos uma breve revisão sobre os espinores Elko.

\subsubsection{Espinores Elko}

$\qquad$ Campos espinoriais Elko 
\cite{Ahluwalia:2015vea,Ahluwalia:2010zn,allu,allu1,Ahluwalia:2016rwl,roldao_5,Ahluwalia:2013uxa,Ahluwalia:2016jwz,Lee:2015jpa} 
$\lambda(p^\mu)$ são autoespinores do operador de conjugação de carga $C$, dado por, 

{}{$$
C\lambda(p^\mu)=\pm \lambda(p^\mu)
$$}
\noindent (aqui o espaço dos momentos é usado apenas para fixar a notação). A representação de Weyl de  $\gamma^{\mu}$ é usada de agora em diante.
Os sinais $+$($-$) indicam campos espinoriais auto-conjugados (anti auto-conjugados), denotados por $\lambda^{S}(p^\mu)$ [$\lambda^{A}(p^\mu)$]. 
Explicitamente, assim que os espinores no referencial de repouso $\lambda(k^\mu)$ são obtidos, para um $p^\mu$ arbitrário o mesmo fornece 
\begin{equation}
\lambda(p^\mu) = e^{i \kb\cdot\vp} \lambda(k^\mu),   \label{boost}
\end{equation} 
\noindent onde $k^\mu = \left(m,\lim_{p\rightarrow 0}\frac{\p}{p}\right),$ {}{para} $p = \vert\p\vert.$ O operador de {\it boost} é fornecido por 
\begin{eqnarray}
 e^{i \kb\cdot\vp} = \sqrt{\frac{E + m }{2 m}}{\rm diag}\left(\mathbb{I} + \frac{\s\cdot\p}{E +m},  \mathbb{I} - \frac{\s\cdot\p}{E +m} 
\right).  \nonumber
\end{eqnarray}
\indent  Os $\phi_{}^\pm(k^\mu)$ são definidos como sendo auto-espinores do operador de helicidade $\s\cdot\hat \p$:
\begin{equation}
\s\cdot\hat \p\, \phi_{}^\pm(k^\mu) = \pm \phi_{}^\pm(k^\mu),\nonumber
\end{equation}
\noindent onde  $\hat \p = \frac{\p}{\vert\p\vert} = (\sin\theta\cos\phi, \sin\theta\sin\phi,\cos\theta)$, e as fases utilizadas são tais que
\begin{eqnarray}
\phi^+_{}(k^\mu) &=& \sqrt{m} \left(\begin{array}{c}
\cos\left(\frac{\theta}{2}\right)e^{- i \phi/2}\\
\sin\left(\frac{\theta}{2}\right)e^{+i \phi/2}
\end{array}
\right)\,, \label{phim}\\ 
\phi^-_{}(k^\mu) &=& \sqrt{m} \left(\begin{array}{c}
-\sin\left(\frac{\theta}{2}\right)e^{- i \phi/2}\\
\cos\left(\frac{\theta}{2}\right)e^{+i \phi/2}
\end{array}
\right)\,. \label{phime}
\end{eqnarray}
\indent Campos espinoriais Elko $\lambda(k^\mu)$ são definidos por 
\begin{eqnarray}
 \lambda^S_\pm(k^\mu) & =& \left(\begin{array}{c}
i \Theta\left[\phi_{}^\pm(k^\mu)\right]^\ast\\
\phi_{}^\pm(k^\mu)
\end{array}
\right), 
\label{pppm}\\ 
\lambda^A_\pm(k^\mu) &=& \pm\left(
\begin{array}{c}
- i \Theta\left[\phi_{}^\mp(k^\mu)\right]^\ast\\
\phi_{}^\mp(k^\mu)
\end{array}
\right),
\label{ppm}
\end{eqnarray}
\noindent onde  $\Theta$ denota o operador de reversão temporal de Wigner para {\it spin} $1/2$. A notação $\phi^\pm_{}(k^\mu) = \phi^\pm$ 
será usada por uma questão de simplicidade. 
A expressão
\begin{equation}
\s\cdot\hat{\p} \,
\big[ \Theta (\phi^\pm)^\ast \big] = \mp \big[ \Theta (\phi^\pm)^\ast \big] \nonumber
\end{equation}
indica que a helicidade de $ \Theta [\phi_{}(k^\mu)]^\ast$ como sendo oposta a de
$\phi_{}(k^\mu)$ e, portanto 
\begin{eqnarray}
\lambda^S_\pm(p^\mu) = \sqrt{\frac{E+m}{2 m} }\left( 1\mp\frac{p}{E+m}\right)\lambda^S_\pm,\label{jj}\\
\lambda^A_\pm(p^\mu) = \sqrt{\frac{E+m}{2 m} }\left( 1\pm\frac{p}{E+m}\right)\lambda^A_\pm,\label{jj1}
\end{eqnarray}
\noindent são os coeficientes de expansão de um campo quântico de {}{dimensão de massa um \cite{Ahluwalia:2016rwl}}. \\
\indent {}{Para mostrarmos isso, usaremos as funções $\lambda^S_\alpha(p^\mu)$ e $\lambda^A_\alpha(p^\mu)$ como coeficientes de expansão de um campo quântico}
\begin{eqnarray}
\mathfrak{f}(x) := \int \frac{\text{d}^3p}{(2\pi)^3}  \frac{1}{2 \sqrt{m E(\p)}} \sum_\alpha \Big[ a_\alpha(\p)\lambda^S_\alpha(\p) \exp(- i p_\mu x^\mu)
+\, b^\dagger_\alpha(\p)\lambda^A_\alpha(\p) \exp(i p_\mu x^\mu){\Big]}.
\label{eq:newqf}
\end{eqnarray}
\indent {}{Os operadores de criação e aniquilação satisfazem a estatística de Fermi}
\begin{subequations}
\begin{align}
&& \left\{a_\alpha(\p),a^\dagger_{\alpha^\prime}(\p^\prime)\right\} = \left(2 \pi \right)^3 \delta^3\hspace{-2pt}\left(\p-\p^\prime\right) \delta_{\alpha\alpha^\prime} \label{eq:a-ad}\\
&& \left\{a_\alpha(\p),a_{\alpha^\prime}(\p^\prime)\right\} = 0,\quad \left\{a^\dagger_\alpha(\p),a^\dagger_{\alpha^\prime}(\p^\prime)\right\} =0\label{eq:aa-adad},
\end{align}
\end{subequations}
\noindent {}{ com anti-comutadores similares para  $b_\alpha(\p)$ e 
$b^\dagger_\alpha(\p)$. Para calcular a dimensão de massa de $\mathfrak{f}(x)$, definimos o adjunto }
 \begin{equation}
\gdualn{\mathfrak{f}}(x) :=  \int \frac{\text{d}^3p}{(2\pi)^3}   \frac{1}{2 \sqrt{m E(\p)}} \sum_\alpha \Big[ a^\dagger_\alpha(\p)\gdualn{\lambda}^S_\alpha(\p) \exp( i p_\mu x^\mu)
 + b_\alpha(\p)\gdualn{\lambda}^A_\alpha(\p) \exp(-i p_\mu x^\mu){\Big]}.\label{eq:newadjoint}
\end{equation}
\indent {}{ A dimensão de massa desse novo campo é determinada pelo propagador de correspondente. Usando as definições anteriores de
$\mathfrak{f}(x)$ e seu adjunto $\gdualn{\mathfrak{f}}(x)$, além das somas de spins, obtemos }
\begin{equation}
S_{\text{FD}}(x^\prime- x) =  i    \int\frac{\text{d}^4 p}{(2 \pi)^4} 
{\e}^{- i p^\mu \left(x^{\prime}_\mu - x_\mu\right)} \left[  \frac{ \openone_4}{p_\mu p^\mu - m^2 + i \epsilon} \right], \label{eq:FD-prop-b}
\end{equation}
\noindent {}{com $\epsilon = 0^+$.  Como consequência, temos que a dimensão de massa do campo $\mathfrak{f}(x)$ é igual a 1}
\begin{equation}
\mathfrak{D}_{\mathfrak{f}} = 1,
\end{equation}
\noindent {}{ e não $3/2$, como no caso do campo de Dirac.}

De fato, o operador de Dirac $(\gamma_\mu p^\mu \pm m \mathbb{I}_4)$ não aniquila 
$\lambda(p^\mu)$ e o seguinte resultado vale:
\begin{eqnarray}
\gamma_\mu p^\mu  \lambda^S_+(p^\mu) &=& i m  \lambda^S_-(p^\mu),    \label{1}\\ \gamma_\mu p^\mu \lambda^S_-(p^\mu) &=& - i m \lambda^S_+(p^\mu), \label{2}\\
\gamma_\mu p^\mu \lambda^A_-(p^\mu) &=& i m \lambda^A_+(p^\mu),  \label{3}\\
\gamma_\mu p^\mu \lambda^A_+(p^\mu) &=& - i m \lambda^A_-(p^\mu) . \label{4}
\end{eqnarray}
\noindent {}{Mesmo assim}, ainda implica a aniquilação do Elko pelo operador de Klein-Gordon. 
$\qquad$Existem resultados e quantidades importantes sobre espinores, que apresentaremos agora.

\subsection{Identidades de Fierz e bumerangues}
\subsubsection{Identidades de Fierz}
$\qquad$Identidades de Fierz são vínculos que os bilineares covariantes (associados a um espinor $\psi$) devem satisfazer \cite{lounesto}
\begin{eqnarray}
 & & {}{\mathbf{J}^2 = J^\mu J_\mu} = \sigma^2 + \omega^2, \qquad \mathbf{J}^2 = - \mathbf{K}^{2},   \nonumber \\
 & & {}{\mathbf{J}\cdot\mathbf{K} = J^\mu K_\mu = 0}, \qquad \quad\mathbf{J}\wedge \mathbf{K} = -(\omega + \sigma \gamma_{0123})\mathbf{S}, \nonumber 
\end{eqnarray}
onde $\gamma_5 = i\gamma_{0123} $. Estas identidades são importantes pois permitem reconstruir o espinor a partir de seus bilineares covariantes pelo algoritmo de Takahashi
\cite{Takahashi:1982wi}.\\
\indent Consideremos a grandeza multivetorial
\begin{eqnarray}
\mathbf{Z} = \sigma + \mathbf{J} + i\mathbf{S} + i\mathbf{K}\gamma_{0123} + \omega\gamma_{0123}. \nonumber 
\end{eqnarray}
$\qquad$Tome  um espinor $\eta$ tal que 
$M_{4}(\mathbb{C})\ni \eta^{\dagger}\gamma_{0}\psi \neq 0$, então $\psi$  e  $\mathbf{Z}\eta$, são proporcionais (diferindo por uma fase). A igualdade entre os 
dois espinores decorre do chamado  teorema de Takahashi \cite{Takahashi:1982wi}, via as seguintes expressões
\begin{eqnarray}
 N &=& \frac{1}{2}\sqrt{\eta^{\dagger}\gamma_0\mathbf{Z}\eta}, \nonumber \\
 e^{-i\alpha} &=& \frac{1}{N}\eta^{\dagger}\gamma_0\psi, \nonumber \\
 \psi &=& \frac{1}{4N}e^{-i\alpha}\mathbf{Z}\eta. 
\end{eqnarray}
\indent Para caracterizar um pouco mais o teorema de Takahashi, precisamos do conceito de bumerangue.

\subsubsection{Bumerangues}
\begin{define}
 Sejam os multivetores $\sigma,\mathbf{J},\mathbf{S},\mathbf{K},\omega$ que satisfazem as identidades de Fierz, então o seu agregado 
 $\mathbf{Z} = \sigma + \mathbf{J} + i\mathbf{S} + i\mathbf{K}\gamma_{0123} + \omega\gamma_{0123}$ é chamado um agregado de Fierz.
\end{define}
\begin{define}
Um multivetor $\mathbf{Z} = \sigma + \mathbf{J} + i\mathbf{S} + i\mathbf{K}\gamma_{0123} + \omega\gamma_{0123}$ que é auto-adjunto 

\begin{eqnarray}
\mathbf{Z} = \gamma^{0}\mathbf{Z}^{\dagger}\gamma^{0},
\end{eqnarray}
\noindent é chamado de bumerangue se suas componentes $\sigma,\mathbf{J},\mathbf{S},\mathbf{K},\omega$ são bilineares covariantes para algum  espinor $\psi$.
\end{define}

\subsubsection{Transformações de Fierz}
$\qquad$As transformações de Fierz \cite{ticciati} fornecem uma caracterização de como o grupo de Lorentz $SO(1,3)$ age na representação complexa da álgebra de Clifford do espaço-tempo, a 
$ A = Cl_{1,3}\otimes \mathbb{C} \cong  M(4,\mathbb{C})$. Operacionalmente, as transformações de Fierz são consideradas tomando-se traços para se calcular a decomposição de um produto de bilineares covariantes numa soma de 
bilineares covariantes, via um rearranjo da ordem dos mesmos. 
Isso envolve a escolha de uma base $\mathcal{B} = \{B^1,\dots,B^{16}\}$ e a respectiva dual $\mathcal{B}^{\prime}$, mas um ponto fundamental que em geral {}{não é abordado} na apresentação da transformação de Fierz, é que 
a base dual $\mathcal{B}^{\prime}$ são funcionais traço parametrizados por matrizes. Para ficar mais claro, consideremos a álgebra $M(n,\mathbb{C})$ com a base $\{E_{ij}\}$, as matrizes $E_{ij}$ possuem entrada igual a $1$
em $ij$ e igual a $0$ no resto. Essas matrizes obedecem à lei de multiplicação $E_{ij}E_{kl} = \delta_{jk}E_{il}$. Agora sejam os funcionais lineares 
\begin{eqnarray}
\varphi_{ij}: &M(n,\mathbb{C})& \rightarrow \mathbb{C} \nonumber \\
&X& \mapsto \Tr (XE_{ij}).  
\end{eqnarray}
\indent É imediato que $\varphi_{ji}(E_{ij}) = 1$ {}{e que} $\varphi_{ji}$ se anula nos outros elementos da base $\{E_{ij}\}$. Com isso temos que os funcionais $\varphi_{ij}$ constituem a base dual à $\{E_{ij}\}$. \\
\indent Voltando para a álgebra $A = M(4,\mathbb{C})$, temos que dada $M\in A$, a mesma pode ser decomposta como
\begin{eqnarray}
 M = \Tr(MB_i)B^i = \Tr(MB^i)B_i,
\end{eqnarray}
\noindent agora, tomemos espinores $u$ e $v$ e formemos a matriz $M = u\bar{v}$. A expansão da mesma é dada por 
\begin{eqnarray}\label{fundametalfierz}
 M =  u\bar{v} = Tr( u\bar{v} B_i)B^i = (\bar{v}B_iu)B^i = (\bar{v}B^iu)B_i.
\end{eqnarray}
$\qquad$A equação \eqref{fundametalfierz} é central na transformação de Fierz, pois reverte a ordem dos espinores $u$ e $\bar{v}$. Como aplicação, seja $u(n)$ como sendo  $u^{r_n}(p_n)$ ou $v^{r_n}(p_n)$ (os espinores
polarizados usuais). Consideremos agora a expressão $(M,N)$ e sua versão {\it flipped} (revertida) $(M,N)^{f}$ dadas por 
\begin{eqnarray}
 (M,N) &=& \bar{u}(4)Mu(2)\bar{u}(3)Nu(1), \nonumber \\
 (M,N)^{f} &=& \bar{u}(4)Mu(1)\bar{u}(3)Nu(2).
\end{eqnarray}
$\qquad$Transformando o produto $(M,N)$ usando \eqref{fundametalfierz}, obtemos:
\begin{eqnarray}
 (M,N) &=& \bar{u}(4)Mu(2)\bar{u}(3)Nu(1) \nonumber \\
 &=& = \bar{u}(4)M(\bar{u}(3)B_ku(2))B^kNu(1) \nonumber \\
 &=& (\bar{u}(4)MB^kNu(1))(\bar{u}(3)B_ku(2)) \nonumber \\
 &=& (MB^kN,B_k)^f.
\end{eqnarray}
$\qquad$Se tomarmos $M = B^i$ e $N = B^j$ e se decompormos $B_k$ e $B^iB^kB^j$ em termos da base $\mathcal{B}$, então temos que $(B^i,B^j)$ é uma combinação linear dos $(B^r,B^s)$. A matriz 
$256 \times 256$ de coeficientes resultante chamaremos de {\it matriz de Fierz}  para a base $\mathcal{B}$. \\
\indent Como aplicação da transformação de Fierz, consideremos a  base de $M(4,\mathbb{C})$ (com sua dual) formada pelas matrizes $\Gamma$:
\begin{eqnarray}
 \mathcal{B}_{\Gamma} &=& \left\{I, \gamma^\mu,\sigma^{\mu\nu},\gamma^5\gamma^\mu.\gamma^5\right \}, \nonumber \\
 \mathcal{B}_{\Gamma}^{\prime} &=& \left\{\frac{1}{4}I, \frac{1}{4}\gamma_\mu,\frac{1}{4}\sigma_{\mu\nu},\frac{1}{4}\gamma_\mu\gamma_5,\frac{1}{4}\gamma_5\right \}.
\end{eqnarray}
$\qquad$A ação de $SO(1,3)$ na álgebra $M(4,\mathbb{C})$ é dada por 
\begin{eqnarray}\label{lorentzgroupaction}
 \Lambda.M &=& \bar{D}(\Lambda)MD(\Lambda), \nonumber \\
 \bar{D}(\Lambda)\gamma^\mu D(\Lambda) &=& \Lambda^{\mu}_{\,\,\nu}\gamma^{\mu},\nonumber \\
 \bar{D}(\Lambda)\gamma^5D(\Lambda) &=& \gamma^5,
\end{eqnarray}
\noindent de \eqref{lorentzgroupaction} concluímos que $\mathcal{B}_{\Gamma}$ é uma base de tensores e a ação de $SO(1,3)$ em $M(4,\mathbb{C})$ se decompõe numa soma direta de cinco ações:
\begin{eqnarray}
 {}{\left ( \left (0,\frac{1}{2}\right ) \oplus \left (\frac{1}{2},0\right ) \right )^2 = \overbrace{(0,0)}^{I} \oplus \overbrace{\left (\frac{1}{2},\frac{1}{2}\right )}^{\gamma^\mu } 
 \oplus \overbrace{(1,0)}^{\sigma^{\mu\nu}P_L } 
 \oplus \overbrace{(0,1)}^{\sigma^{\mu\nu}P_R } \oplus \overbrace{\left (\frac{1}{2},\frac{1}{2}\right )}^{\gamma^5\gamma^\mu } \oplus \overbrace{(0,0)}^{\gamma^5},}
\end{eqnarray}
\noindent onde $P_L$ e $P_R$ são os operadores de quiralidade. Estas são as caracterizações de \linebreak $\sigma,\mathbf{S},\mathbf{J},\mathbf{K},\omega$ como representações
irredutíveis de $SO(1,3)$, {}{onde respectivamente, associamos representações irredutíveis com os bilineares covariantes.}
\begin{eqnarray}
 & & (0,0) \leftrightarrow \sigma,\omega \\
 & &  \left(\frac{1}{2},\frac{1}{2}\right ) \leftrightarrow \mathbf{J},\mathbf{K} \\
 & &  (1,0)\oplus(0,1) \leftrightarrow \mathbf{S} 
 \end{eqnarray}

%% file: geometria_e_gravitacao_elementar.tex
\chapter{Geometria e gravitação: um prelúdio} 

\section{Introdução}

$\qquad$ A teoria da gravitação de Einstein é a mais simples no sentido de usar a conexão de
Levi-Civita da variedade espaço-tempo adotada e por usar uma ação que depende linearmente da curvatura escalar. Mas não temos, em termos de primeiros princípios, nada
que nos proíba utilizar uma ação mais complexa, ou que obrigatoriamente temos que nos restringir à conexão de Levi-Civita (simétrica e compatível com a métrica). Neste
modelo, adotamos uma conexão geral, com torção, o que nos fornece mais graus de liberdade na teoria {}{(para incluir o spin da matéria por exemplo)} e uma ação munida de uma função da curvatura escalar. Em tal
cenário, adicionamos um campo fermiônico (espinorial) e provamos que a equação de movimento do mesmo, depende de bilineares covariantes, restringindo o tipo de campo
espinorial admissível. Finalmente exibimos uma solução de tal equação que é um tipo de espinor não encontrado previamente na literatura.

\section{Variedades diferenciáveis} 

$\qquad$ Precisamos de uma estrutura geométrica fundamental subjacente para as teorias físicas, {}{o conceito de variedade diferenciável \cite{Lee}}.
\begin{define}
Uma variedade diferenciável de dimensão $n$, é um espaço topológico $M$  {}{que obedece} aos seguintes axiomas: \normalfont

\begin{itemize}
 \item [(i)] $M$ é um espaço topológico de Hausdorff com base de abertos enumerável,
 \item [(ii)] Para todo ponto $p \in M$, existe um aberto $ p \in V \subseteq M$ e um homeomorfismo local $\varphi: V \to \mathbb{R}^n$. O par $(V,\varphi)$ é a chamada carta 
 local ou sistema de coordenadas local, e $n$ é a dimensão da variedade (a mesma para qualquer carta local),
 \item [(iii)] Sejam $(V,\varphi)$ e $(W,\psi)$ cartas locais,  com $p\in V\cap W \neq \varnothing$, a função mudança de coordenadas
  $\psi\circ\varphi^{-1}: \varphi(V\cap W) \to \psi (V\cap W)$ é suave (infinitamente diferenciável),
 \item [(iv)] Seja $\mathfrak{F}(M) = \{(V\,\varphi)\,\vert\, \text{ é carta local de} M \}$ um conjunto de cartas locais para $M$ (um atlas para $M$). Devemos ter:
 \begin{eqnarray}
  M = \bigcup_{V\in \mathfrak{F}(M)} V, \nonumber
 \end{eqnarray}
\noindent ou seja, os domínios dos sistemas de coordenadas devem cobrir o espaço.
\end{itemize}
\end{define}

\subsection{Exemplos:}
\begin{enumerate}
 \item $M = \mathbb{R}^{n}$ e $\varphi = id_{\mathbb{R}^{n}}: \mathbb{R}^n \to \mathbb{R}^n$. $\mathfrak{F}(M) = \{(\mathbb{R}^{n},\varphi) \}$ é um atlas para $M$.
 \item $M = S^n = \{x\in \mathbb{R}^{n + 1}\,\vert\, \sum_{j = 1}^{n + 1} x_j^2 = 1\}$, a esfera $n$-dimensional usual. Sejam os subconjuntos de $M$ dados por 
 $P_{\pm} = S^{n} \setminus \{0,\dots,0,\pm 1\}$, e as funções
 \begin{eqnarray}
  \varphi_{\pm}: &P_{\pm}& \to \mathbb{R}^{n} \nonumber \\
  &x& \mapsto \sum_{j = 1}^{n} \frac{x_j}{1 \mp x_{n + 1}} e_j, \nonumber 
 \end{eqnarray}
 \noindent O conjunto $\mathfrak{F}(M) = \{(P_{\pm},\varphi_{\pm})\}$ é um atlas para $S^n$.
\end{enumerate}

\section{Campos de vetores}
$\qquad$ Para generalizar o conceito de campo vetorial no caso do $\mathbb{R}^n$, vamos usar o modelo de derivação.
\begin{define}
Seja $M$ uma variedade suave, um campo vetorial $X$ é um função $\mathbb{R}$-linear $X: C^{\infty}(M) \to C^{\infty}(M)$ que obedece ao seguinte axioma
 (regra de Leibniz para uma derivação) :\normalfont
 $$
 X(fg) = X(f)g + fX(g).
 $$ 
\end{define}
\indent O conjunto de todos os campos vetoriais sobre uma variedade M será denotado por $\mathfrak{X}(M)$. Para mostrarmos que o conjunto $\mathfrak{X}(M)$ sempre 
possui elementos (além do campo nulo), são necessários métodos sofisticados de topologia diferencial, mas localmente sempre temos campos vetoriais.
\begin{define}
 Seja M uma variedade suave e $(V, \varphi)$ uma carta local. O $i$-ésimo campo coordenado $X_i$ associado a $i$-ésima coordenada $x_i$ é definido como:
 \begin{eqnarray}
  X_i(f) = \frac{\partial f}{\partial x_i} := \frac{\partial (f\circ \varphi^{-1})}{\partial e_i}, \nonumber
 \end{eqnarray} \normalfont
\noindent onde $e_i$ é o $i$-ésimo vetor da base canônica do $\mathbb{R}^{n}$.  Para o que segue, denotaremos $\partial_i = \frac{\partial }{\partial x_i}$. 
{}{É} simples mostrar que os $\partial_i$ formam um base local para os campos vetoriais da variedade.
\end{define}

\subsection{Exemplos}
\begin{enumerate}
 \item Para $M = \mathbb{R}^{n}$ temos $\partial_i = \frac{\partial }{\partial e_i}$.
 \item Consideremos $M = \mathbb{C} = \mathbb{R}^{2}$, ou seja, os números complexos como uma $\mathbb{R}$-álgebra. Definimos os campos globais, como sendo:
 \begin{eqnarray}
  & & X_1(z) = ze_1 = (a,b)(1,0) = (a,b) = z, \nonumber \\
  & & X_2(z) = ze_2 = (a,b)(0,1) = (-b,a) = iz. \nonumber 
 \end{eqnarray}
\noindent Uma interpretação geométrica para o campo $X_2$ é que se restringirmos o mesmo à esfera $S^1$, este é um campo tangente a mesma.
\end{enumerate}
$\qquad$ Para encerrar esta seção, precisamos do conceito de tensores do ponto de vista moderno. A definição que segue é para um campo tensorial covariante de ordem 
{\it r}  e contravariante de ordem {\it s}.
\begin{define}
 Seja $M$ uma variedade suave, um tensor $T$ {\it r}-vezes covariante e {\it s}-vezes contravariante é uma função  {\it r+s}-linear em relação ao anel 
 $A = C^{\infty}(M)$.
 \begin{eqnarray}
 T: \overbrace{\mathfrak{X}(M)}^{{\it r} \text{ vezes}} \times \underbrace{\mathfrak{X}(M)^{*}}_{{\it s} \text{ vezes}} \to C^{\infty}(M). 
 \end{eqnarray}
\end{define}
\indent Vamos denotar o espaço dos tensores do tipo $(${\it r,s}$)$ por $\otimes_{s}^{r}TM$. Vendo os tensores como
funções a valores no anel $A = C^{\infty}(M)$, podemos somar tensores  e multiplicar tensores por funções escalares (e
números reais). Uma operação fundamental no cálculo tensorial, é a chamada contração, que mapeia tensores do tipo ({\it r},{\it s}), em tensores do tipo \linebreak
({\it r} - 1, {\it s} - 1). Usando o caso de um tensor $T\in \otimes_{1}^{1} TM$, basta estender a seguinte aplicação canônica:
\begin{eqnarray}
 C: &\bigotimes_{1}^{1} TM& \to C^{\infty}(M)  \nonumber \\
 &\omega \otimes X& \mapsto \omega(X).
\end{eqnarray}
$\qquad$ Em coordenadas locais, a contração toma a seguinte forma $C(\omega\otimes X) = \omega_iX^{i}$.

\section{Geometrias relacionadas}

\subsection{Conexão de Koszul}
$\qquad$ De modo informal, uma conexão $\nabla$ numa variedade $M$ é uma estrutura que nos permite derivar campos (vetoriais, tensoriais, formas diferenciais e 
demais objetos geométricos) do mesmo modo 
que derivamos funções usuais (funções de $n$ variáveis reais). Existem vários modelos equivalentes de conexão numa variedade, mas usaremos o modelo algébrico de Koszul
\cite{spivak}.
\begin{define}
 Dada uma variedade suave $M$, uma conexão sob $M$ é uma função $\mathbb{R}$-linear  $\nabla: \mathfrak{X}(M) \times \mathfrak{X}(M) \to \mathfrak{X}(M)$ que obedece os
 seguintes axiomas:\normalfont
 
\begin{enumerate}
 \item $\nabla_{fX} Y = f\nabla_X Y, \quad \forall X,Y\in \mathfrak{X}(M),\forall f\in C^{\infty}(M)$
 \item $\nabla_X fY = X(f)Y + f\nabla_XY.\quad \forall X,Y\in \mathfrak{X}(M),\forall f\in C^{\infty}(M)$
 \item $\nabla_X(Y + Z) = \nabla_XY + \nabla_X Z,\quad \forall X,Y,Z\in \mathfrak{X}(M)$
 \item $\nabla_{X + Y}Z = \nabla_X Z + \nabla_Y Z,\quad \forall X,Y,Z\in \mathfrak{X}(M) $
\end{enumerate}
\end{define}
\begin{obs}
A definição anterior independe de coordenadas locais. Porém, muitas vezes, realizamos cálculos usando
coordenadas{}{. Para} isso, precisamos de uma caracterização local da conexão. De fato, é possível provar que $($em termos de coordenadas$)$, a conexão depende localmente apenas
do campo vetorial usado.
\end{obs}
\begin{define}
Dada uma carta local $(V, \varphi)$ de $M$, definimos os símbolos de Christoffel (as componentes da conexão) como:
 $$ 
 \nabla_{\partial_i} \partial_j = \Gamma_{ij}^{k}\partial_k\,.
 $$
\end{define}

\subsection{Geometria pseudo-Riemanniana}

$\qquad$ Uma variedade pseudo-Riemanniana \cite{petersen} é a generalização natural dos trabalhos clássicos de Gauss para a teoria das superfícies em $\mathbb{R}^3$. 
O objeto geométrico 
fundamental é o tensor métrico, no sentido de que todas as quantidades geométricas do sistema são derivadas do mesmo. Em termos mais precisos, temos:
\begin{define}
{}{Uma variedade pseudo-Riemanniana} é um par $(M,g)$, onde $M$ é uma variedade suave de dimensão $n$ e $g$ é um tensor $($o tensor métrico$)$ de 
assinatura ({\it p},{\it q}) covariante de segunda
ordem, simétrico e não degenerado $(\det \vert g(x)\vert \neq 0,\forall x\in M)$.
\end{define}
\noindent {\bf Exemplos }
\begin{enumerate}
 \item Consideremos $M = \mathbb{R}^{n}$ e $g$ dado por $g = \sum_{i = 1}^{n} dx_i^2$. O par $(M,g)$ é uma variedade Riemanniana, veremos posteriormente que tal espaço possui
 curvatura escalar nula.
 \item Tomemos $M = H^{2} = \{(x,y)\in\mathbb{R}^{2}\,\vert\, y > 0\}$ e $g = \frac{dx^2 + dy^2}{y^2}$. O par $(M,g)$ é um modelo bidimensional para a geometria de 
 Lobatchevsky (um dos modelos de Poincaré). A curvatura escalar deste espaço é constante e igual a $-1$.
\end{enumerate}
$\qquad$ Munindo uma variedade de uma métrica pseudo-Riemanniana $g$, temos canonicamente associada, uma conexão, a chamada conexão de Levi-Civita \cite{daRocha:2005vm,Rodrigues:2005yz}.
\begin{teo}
Dada uma variedade pseudo-Riemanniana $(M, g)$, existe uma única conexão $\nabla$ que é compatível com a métrica e possui torção nula (veremos mais adiante o
conceito de torção de uma conexão), ou seja:\normalfont
\begin{itemize}
 \item [(C1)] $Xg(Y,Z) = g(\nabla_XY, Z) + g(Y,\nabla_X Z),\quad \forall X,Y,Z\in \mathfrak{X}(M)$,
 \item [(C2)] $\nabla_X Y - \nabla_Y X = [X,Y],\quad X,Y\in \mathfrak{X}(M)$.
\end{itemize}
\end{teo}
\indent  Em termos teóricos e práticos, um recurso importante é que o tensor métrico fornece uma dualização entre o espaço dos campos vetoriais 
$\mathfrak{X}(M)$ e o seu dual $\mathfrak{X}(M)^{*}$ (as chamadas $1$-formas diferenciais de grau 1). Em termos mais precisos, consideremos a aplicação 
(um exemplo de correlação \cite{roldao_1,porteous})
\begin{eqnarray} 
\tau: \mathfrak{X}(M) &\rightarrow&  \mathfrak{X}(M)^{*} = \bigwedge^{1}TM \nonumber \\
X &\mapsto& \tau(X): \mathfrak{X}(M) \rightarrow C^{\infty}(M) \nonumber \\ 
&& \hspace{1.5cm} Y \hspace{0.3cm}\mapsto  g(X,Y). 
\end{eqnarray}
\indent Esta aplicação emula o subir e abaixar índices, que em coordenadas locais é {}{dado} por $a_i = g_{ij}a^j$. Realizando uma extensão imediata, obtemos uma aplicação 
$$
\tau: \bigotimes_{s}^{r} TM \to \bigotimes_{s - 1}^{r + 1}TM.
$$

\subsection{O tensor de curvatura de Riemann}

$\qquad$ De posse de uma métrica $g$, o passo natural é definirmos tensores para avaliar os mais diversos tipos de curvatura. Começamos com o tensor de curvatura de Riemann.
\begin{define}
 Seja $(M,g)$ uma variedade pseudo-Riemanniana, o tensor de curvatura de Riemann é uma função $R: \mathfrak{X}(M)^3 \to \mathfrak{X}(M)$, dada por:
 \begin{eqnarray}
  R(X,Y,Z) = \nabla_X\nabla_Y Z - \nabla_Y\nabla_X Z - \nabla_{[X,Y]} Z\,. \nonumber 
 \end{eqnarray}
\end{define}
\indent Um cálculo imediato mostra que a função acima é $3$-linear em relação a álgebra \linebreak  $A = C^{\infty}(M)$, sendo portanto, um tensor (três vezes covariante e uma 
vez contravariante).
\begin{obs}
 Em coordenadas locais, o tensor de curvatura tem a seguinte forma:\normalfont
 \begin{eqnarray}
  R_{\,\,\,ijk}^{l}\partial_l &=& R(\partial_i,\partial_j,\partial_k) = \nabla_{\partial_i}\nabla_{\partial_j}\partial_k - 
  \nabla_{\partial_j}\nabla_{\partial_i}\partial_k - \overbrace{\nabla_{[\partial_i,\partial_j]}\partial_k}^{\,=0} \nonumber \\
 &=& \nabla_{\partial_i}\Gamma_{jk}^{r}\partial_r - \nabla_{\partial_j}\Gamma_{ik}^{r}\partial_r \nonumber \\
 &=& \partial_i(\Gamma_{jk}^{r})\partial_r + \Gamma_{jk}^{r}\nabla_{\partial_i}\partial_r - \partial_j(\Gamma_{ik}^{r})\partial_r - 
 \Gamma_{ik}^{r}\nabla_{\partial_j}\partial_r \nonumber \\
 &=& \partial_i(\Gamma_{jk}^{r})\partial_r + \Gamma_{jk}^{r}\Gamma_{ir}^{t}\partial_t - \partial_j(\Gamma_{ik}^{r})\partial_r - \Gamma_{ik}^{r}\Gamma_{jr}^{t}\partial_t \nonumber \\
 &=& \left \{\partial_i(\Gamma_{jk}^{r})  -\partial_j(\Gamma_{ik}^{l}) +  \Gamma_{jk}^{r}\Gamma_{ir}^{l} - \Gamma_{ik}^{r}\Gamma_{jr}^{l} 
 \right \}\partial_l \nonumber \\
 &\therefore& \boxed{R_{\,\,\,ijk}^{l} = \partial_i(\Gamma_{jk}^{l}) - \partial_j(\Gamma_{ik}^{l}) + \Gamma_{jk}^{r}\Gamma_{ir}^{l} - \Gamma_{ik}^{r}\Gamma_{jr}^{l} }\nonumber 
 \end{eqnarray}
\end{obs}
\indent O tensor de curvatura de Riemann é muito intrincado e por isso foram definidos vários tensores de curvatura a partir do mesmo, tais como 
o tensor de curvatura seccional, o tensor de Ricci e a curvatura escalar.
\begin{define}
 Seja $(M,g)$ uma variedade pseudo-Riemanniana, e $\nabla$ a conexão de Levi-Civita associada. Define-se o tensor de curvatura de Ricci como sendo o tensor:
 \begin{eqnarray}
  Ric(\partial_i,\partial_j) = R_{ij} = R_{\,\,\,ikj}^{k}.
 \end{eqnarray}
\end{define}
\indent Ou seja, o tensor de Ricci é  a contração do segundo índice covariante, com o índice contravariante.\\
\indent Passemos agora à curvatura escalar $R$. Como  o tensor de Ricci $Ric\in \bigotimes_{2}^{0}TM$, dualizamos o mesmo e depois contraímos, ou seja
\begin{eqnarray}
 R = R_{\,i}^{i}. \nonumber
\end{eqnarray}
\indent Até chegarmos ao resultado do nosso artigo \cite{fabbri_ESK}, se faz necessária a apresentação de uma sequência evolutiva de teorias de gravitação.
\noindent Em todas as teorias de gravitação que apresentaremos, problemas importantes devem ser abordados \cite{sotiriou}: na cosmologia, a dinâmica da mesma deve estar correta,
o comportamento sob perturbações gravitacionais deve ser o correto e as teorias devem gerar perturbações cosmológicas compatíveis com os vínculos do fundo de micro-ondas cósmico, 
com estruturas de larga escala, a núcleo-síntese do Big Bang e ondas gravitacionais \cite{sotiriou,wheeler}.

\section{Teorias de Gravitação}

\subsection{Gravitação de Einstein}\label{gravieinst}

$\qquad$ Como aplicação imediata da maquinaria algébrico-geométrica apresentada, passemos a uma análise da equação de Einstein. 
Podemos conceber o lado esquerdo da equação do campo gravitacional,
como sendo o tensor de segunda ordem mais simples que satisfaz {}{os} seguintes requisitos:
\begin{enumerate}
\item O limite de campo fraco da teoria deve resultar na gravitação Newtoniana (uma equação do tipo Poisson),
\item A lei da conservação da energia deve ser respeitada,
\item O tensor procurado precisa ser simétrico.
\end{enumerate}
$\qquad$ Após várias tentativas, Einstein chegou a versão final da sua equação de campo \footnote{Uma abordagem rigorosa sobre os vínculos corretos que devem ser satisfeitos por qualquer teoria (numa variedade de dimensão $4$), 
que pretenda fornecer uma 
apresentação dual das equações do campo gravitacional para uma teoria de Riemann-Cartan geral foi desenvolvida  na Ref. \cite{daRocha:2009sq}. }
\begin{eqnarray}
G_{\mu\nu} = R_{\mu\nu} - \frac{1}{2}g_{\mu\nu}R = \frac{8\pi G}{c^4}T_{\mu\nu},
\end{eqnarray}
\noindent onde $G$ é a constante gravitacional e $c$ é a velocidade da luz no vácuo. Esta equação iguala um objeto puramente geométrico (o tensor $G_{\mu\nu}$) 
a outro objeto de conteúdo físico, o tensor de energia-momento $T_{\mu\nu}$. 
Esta equação está dizendo que a presença da matéria-energia numa região do espaço-tempo (representada pelo tensor $T_{\mu\nu}$), 
gera uma curvatura no mesmo. Esta curvatura, 
não é uma deformação morfológica do espaço-tempo, mas sim uma associação de curvaturas que são sentidas de forma universal via a equação das geodésicas (como o campo gravitacional 
afeta a matéria-energia).
Em termos mais estruturais, esta equação de movimento pode ser derivada da ação de Hilbert-Einstein (com fonte)
\begin{eqnarray}
 S = \int \left ( \frac{1}{2\kappa}R + \mathcal{L}_{m}(g_{\mu\nu},\psi) \right ) \sqrt{-g} \,\mathrm{d}^4x,  \nonumber 
\end{eqnarray}
\noindent onde $\mathcal{L}_{m}(g_{\mu\nu},\psi)$ é a Lagrangiana da matéria $\psi$ denota de forma genérica os campos da matéria definidos sob esse fundo e
$\kappa = \frac{8\pi G}{c^4}$.
Devemos ressaltar a estrutura do ambiente geométrico: a conexão adotada é a de Levi-Civita e, portanto, todas as grandezas geométricas envolvidas são funções do tensor métrico $g_{\mu\nu}$, que é a variável dinâmica do
sistema. {}{A equação de Einstein que consideramos, não possui um termo correspondente à constante cosmológica, originalmente adicionada por Einstein, 
por acreditar num Universo estático.} Passemos agora à construção detalhada da equação de movimento \cite{carroll}. \\
\indent A variação (a derivada funcional) em relação ao tensor métrico $g^{\mu\nu}$ deve ser igual a zero.
\begin{eqnarray}
\delta S &=& \int 
         \left[ 
            {1 \over 2\kappa} \frac{\delta (\sqrt{-g}R)}{\delta g^{\mu\nu}} + 
            \frac{\delta (\sqrt{-g} \mathcal{L}_{m})}{\delta g^{\mu\nu}}
         \right] \delta g^{\mu\nu}\mathrm{d}^4x \nonumber \\
       &=& \int 
        \left[ 
           {1 \over 2\kappa} \left( \frac{\delta R}{\delta g^{\mu\nu}} +
             \frac{R}{\sqrt{-g}} \frac{\delta \sqrt{-g}}{\delta g^{\mu\nu} } 
            \right) +
           \frac{1}{\sqrt{-g}} \frac{\delta (\sqrt{-g} \mathcal{L}_{m})}{\delta g^{\mu\nu}} 
        \right] \delta g^{\mu\nu} \sqrt{-g}\, \mathrm{d}^4x = 0, \nonumber 
\end{eqnarray}
\noindent e como a variação $\delta g^{\mu\nu}$ é arbitrária, segue a equação de movimento para o tensor métrico $g_{\mu\nu}$:
\begin{eqnarray}
  \frac{\delta R}{\delta g^{\mu\nu}} + \frac{R}{\sqrt{-g}} \frac{\delta \sqrt{-g}}{\delta g^{\mu\nu}} 
= - 2 \kappa \frac{1}{\sqrt{-g}}\frac{\delta (\sqrt{-g} \mathcal{L}_{m})}{\delta g^{\mu\nu}}, \nonumber 
\end{eqnarray}
\noindent onde 
$$
T_{\mu\nu} = - \frac{1}{\sqrt{-g}}\frac{\delta (\sqrt{-g} \mathcal{L}_{m})}{\delta g^{\mu\nu}} = 
-2 \frac{\delta \mathcal{L}_{m}}{\delta g^{\mu\nu}} - 2\frac{1}{\sqrt{-g}}\frac{\delta \sqrt{-g}}{\delta g^{\mu\nu}}\mathcal{L}_{m},
$$
\noindent é o tensor de energia-momento da fonte. Para continuarmos com a construção da equação de movimento, 
precisamos determinar a variação da curvatura escalar $\delta\mathit{R}$.
\\
\indent Primeiro, vamos calcular  a variação do tensor de curvatura de Riemann, $\delta {R^\rho}_{\sigma\mu\nu}$:
\begin{eqnarray}
\delta{R^\rho}_{\sigma\mu\nu} = \partial_\mu\delta\Gamma^\rho_{\nu\sigma} - \partial_\nu\delta\Gamma^\rho_{\mu\sigma} + \delta\Gamma^\rho_{\mu\lambda} \Gamma^\lambda_{\nu\sigma} + 
\Gamma^\rho_{\mu\lambda} \delta\Gamma^\lambda_{\nu\sigma}
- \delta\Gamma^\rho_{\nu\lambda} \Gamma^\lambda_{\mu\sigma} - \Gamma^\rho_{\nu\lambda} \delta\Gamma^\lambda_{\mu\sigma}. 
\end{eqnarray}
$\qquad$ Como $\delta\Gamma_{\nu\mu}^{\rho}$ é a diferença entre duas conexões, podemos calcular a derivada covariante do mesmo:
\begin{eqnarray}
& & \nabla_\mu (\delta \Gamma^\rho_{\nu\sigma} ) = \partial_\mu (\delta \Gamma^\rho_{\nu\sigma} ) + \Gamma^\rho_{\alpha\mu} \delta\Gamma^\alpha_{\nu\sigma} - 
\Gamma^\alpha_{\nu\mu} \delta \Gamma^\rho_{\alpha\sigma} - \Gamma^\alpha_{\sigma\mu} \delta \Gamma^\rho_{\nu\alpha}.\nonumber \\
& & \nabla_\nu (\delta \Gamma^\rho_{\mu\sigma} ) = \partial_\nu (\delta \Gamma^\rho_{\mu\sigma} ) + \Gamma^\rho_{\alpha\nu} \delta\Gamma^\alpha_{\mu\sigma} - 
\Gamma^\alpha_{\nu\mu} \delta \Gamma^\rho_{\alpha\sigma} - \Gamma^\alpha_{\sigma\nu} \delta \Gamma^\rho_{\mu\alpha}.
\end{eqnarray}
\par Com isso temos que:
\begin{eqnarray}
 \delta R^\rho{}_{\sigma\mu\nu} = \nabla_\mu (\delta \Gamma^\rho_{\nu\sigma}) - \nabla_\nu (\delta \Gamma^\rho_{\mu\sigma}).
\end{eqnarray}
\par O próximo passo, é calcular a variação do tensor de curvatura de Ricci:
\begin{eqnarray}\label{riccivariation}
\delta R_{\mu\nu} = \delta R^\rho{}_{\mu\rho\nu} = \nabla_\rho (\delta \Gamma^\rho_{\nu\mu}) - \nabla_\nu (\delta \Gamma^\rho_{\rho\mu}).
\end{eqnarray}
\par A curvatura escalar é dada por $\mathit{R} = R_{\mu\nu}g^{\mu\nu}$, isso implica 
\begin{eqnarray}
 \delta R &=& R_{\mu\nu} \delta g^{\mu\nu} + g^{\mu\nu} \delta R_{\mu\nu}\nonumber \\
         &=& R_{\mu\nu} \delta g^{\mu\nu} + \underbrace{\nabla_\sigma \left( g^{\mu\nu} \delta\Gamma^\sigma_{\nu\mu} - g^{\mu\sigma}\delta\Gamma^\rho_{\rho\mu} \right)}_{\nabla_\sigma g^{\mu\nu} = 0}. \nonumber
\end{eqnarray}
\par Agora $\nabla_{\sigma} V^{\sigma}$ toma a seguinte forma:
\begin{eqnarray}
 \nabla_\sigma \overbrace{\left( g^{\mu\nu} \delta\Gamma^\sigma_{\nu\mu} - g^{\mu\sigma}\delta\Gamma^\rho_{\rho\mu} \right)}^{V^\sigma}\sqrt{-g} = 
 \partial_\sigma \left (\sqrt{-g}V^\sigma \right ), \nonumber 
\end{eqnarray}
um termo que levado na ação, resulta num termo de fronteira. Portanto:
$$
\frac{\delta\mathit{R}}{\delta g^{\mu\nu}} = R_{\mu\nu}. \nonumber 
$$
\indent Ainda precisamos calcular $\frac{1}{\sqrt{-g}}\frac{\delta \sqrt{-g}}{\delta g^{\mu\nu}}$. Temos que
$$ 
\delta g = \delta \det(g_{\mu\nu}) = g \, g^{\mu\nu} \delta g_{\mu\nu}.
$$
\indent Esta igualdade segue da fórmula de Jacobi para a derivada do determinante de uma matriz $d \det (T) = \mathrm{tr} (\mathrm{adj}(T) \, dT)$ \cite{magnus}. 
Utilizando este resultado, temos que:
\begin{eqnarray}
\delta \sqrt{-g} = -\frac{1}{2\sqrt{-g}}\delta g  = \frac{1}{2} \sqrt{-g} (g^{\mu\nu} \delta g_{\mu\nu}) = -\frac{1}{2} \sqrt{-g} (g_{\mu\nu} \delta g^{\mu\nu}) \,, \nonumber 
\end{eqnarray}
\noindent a última passagem decorre do seguinte:
\begin{eqnarray}
 g^{\mu\beta}g_{\beta\nu} &=& \delta_{\nu}^{\mu},  \nonumber \\
 g_{\beta\nu}\delta g^{\mu\beta} + g^{\mu\beta}\delta g_{\beta\nu} &=& 0, \nonumber \\
 g_{\beta\nu}\delta g^{\mu\beta} &=& - g^{\mu\beta}\delta g_{\beta\nu}, \nonumber \\
 g^{\nu\alpha}g_{\beta\nu}\delta g^{\mu\beta} &=& - g^{\nu\alpha}g^{\mu\beta}\delta g_{\beta\nu}, \nonumber \\
 \delta^{\alpha}_{\beta}\delta g^{\mu\beta} &=& - g^{\nu\alpha}g^{\mu\beta}\delta g_{\beta\nu}, \nonumber \\
 \delta g^{\mu\alpha} &=& - g^{\nu\alpha}g^{\mu\beta}\delta g_{\beta\nu}, \nonumber \\
 \delta g^{\mu\nu} &=& -g^{\mu\alpha}\delta g_{\alpha\beta}g^{\beta\nu}. \nonumber  
\end{eqnarray}
\indent Finalmente concluímos que:
\begin{eqnarray}
\frac{1}{\sqrt{-g}}\frac{\delta \sqrt{-g}}{\delta g^{\mu\nu} } = -\frac{1}{2} g_{\mu\nu}. \nonumber
\end{eqnarray}
\indent Antes de apresentarmos a próxima teoria de gravitação, vamos resumir as ideias da aplicação das teorias de gravitação no estudo do Universo em larga escala 
(cosmologia).

\subsection{Cosmologia FLRW}
$\qquad$ Para criarmos um modelo do Universo em larga escala \cite{ellis}, abstraímos a imensa quantidade de objetos cósmicos existentes, modelamos a matéria como sendo um fluido 
caracterizado por uma pressão $p$ e uma quadrivelocidade $u$. O acrônimo FLRW denota o modelo de Friedmann-Lemaitre-Robertson-Walker, que consiste num
Universo isotrópico e em expansão, munido de uma 
métrica do tipo Robertson-Walker. Nesta seção apresentamos os rudimentos de cosmologia \cite{straumann,wheeler,hawking}.

\subsubsection{Introdução}
$\qquad$  A teoria da gravitação de Einstein (e modelos subsequentes) tem um papel fundamental na descrição do Universo em larga escala. 
Muito do sistema conceitual da cosmologia atual foi criado ao 
longo do século XX depois do surgimento da teoria da gravitação, tendo hoje um forte inter-relacionamento com a realidade observacional, resultando numa
descrição bem estabelecida do Universo. 
Além disso, as descobertas na cosmologia têm impactado {}{a} física fundamental: existe uma forte evidência da existência de partículas além 
das previstas pelo modelo padrão, a assim chamada matéria 
escura e uma energia exótica e quase homogênea chamada de energia escura \cite{sotiriou}. O candidato mais simples para esta energia desconhecida é um termo cosmológico nas equações de campo de 
Einstein. Independentemente
do que seja esta densidade de energia exótica, a densidade de energia pertencente a constante cosmológica não é maior que a densidade crítica cosmológica, 
sendo {}{portanto} extremamente pequena para os padrões
da física de partículas. {}{Trata-se} de algo pouco compreendido, pois espera-se que todos os tipos de energias do vácuo contribuam para uma constante cosmológica efetiva.
Por exemplo, de 
flutuações quânticas em campos conhecidos até a escala eletrofraca, {}{contribuições são esperadas como sendo 50 ordens de magnitude maiores que a densidade observada da 
energia escura \cite{straumann}}.
Os modelos cosmológicos de Friedmann-Lemaitre são notavelmente simples matematicamente, pois são altamente simétricos e as métricas adotadas são produtos não fatorizáveis. 

\subsubsection{Espaços-tempo FLRW}
$\qquad$ Após esforços observacionais que demandaram décadas, hoje existe uma boa evidência de que em largas escalas o Universo é homogêneo e isotrópico. A contribuição mais fundamental que 
corrobora tal afirmação são os dados sobre isotropia coletados pelo CMB (radiação cósmica de fundo) \cite{Ade:2015xua,Gawiser:2000az,Liddle_2003d} e amostras sobre o desvio para o vermelho de galáxias. A distribuição de galáxias até
uma distância de 4 bilhões de anos-luz mostra que existem aglomerados gigantes e longos filamentos, mas o mapeamento também mostra que não há estruturas maiores.\\
\indent Obtemos um Universo do tipo FLRW postulando que para cada observador, se movendo ao longo de uma curva integral de um quadrivetor $u$, o Universo se parece espacialmente isotrópico.
De forma mais precisa, fazemos:
\begin{define}
Um espaço-tempo de Friedmann $(M,g)$ é um produto não fatorizável da forma $M = I \times \Sigma$, onde $I\subseteq \mathbb{R}$ e a métrica $g$ é da seguinte forma: \normalfont
\begin{eqnarray}\label{fridwalkermetric}
g  = -dt^2 +a^2(t)h,
\end{eqnarray}
\noindent onde $(\Sigma,h)$ é uma variedade Riemanniana (com dimensão $3$) de curvatura escalar constante $k = 0,\pm 1$. O tempo \emph{t} é o tempo cósmico, e $a(t)$ é \emph{fator de escala}.
\end{define}
\indent Ao invés de $t$, frequentemente usaremos o \emph{tempo conforme} $\eta$, definido por $d\eta = \frac{dt}{a(t)}$. O campo de velocidades é perpendicular 
{}{às} fatias de tempo cósmico constante 
$u = \frac{\partial}{\partial t}$.

\subsubsection{Espaços de curvatura constante}
$\qquad$ Para o espaço $(\Sigma,h)$ de curvatura constante, a curvatura é dada por:
\begin{eqnarray}
 R^{(3)}(X,Y,Z) = k\{h(Z,Y)X - h(Z,X)Y\},
\end{eqnarray}
\noindent que em coordenadas locais {}{fica}
\begin{eqnarray}\label{riccicomp}
 R^{(3)}_{ijkl} = k(h_{ik}h_{jl} - h_{il}h_{jk}).
\end{eqnarray}
\indent Portanto, o tensor de Ricci e a curvatura escalar são
\begin{eqnarray}
 R^{(3)} = 2kh_{jl},\quad R^{(3)} = 6k.
\end{eqnarray}
\indent As 2-formas de curvatura são obtidas de \eqref{riccicomp} relativas a um co-referencial $\{\theta^i\}$:
\begin{eqnarray}
 \Omega^{3}_{ij} = \frac{1}{2}R^{(3)}_{ijkl} \theta^k\wedge \theta^l = k\theta_i\wedge \theta_j,
\end{eqnarray}
\noindent com $\theta_i = h_{ik}\theta^k$. Como realizações destes espaços, temos $S^3$($k = 1$), $\mathbb{R}^{3}$($k = 0$) e $\mathbb{H}^3$($k = -1$).

\subsubsection{Curvatura de espaços-tempo FLRW}
$\qquad$ Seja $\{\bar{\theta}^i\}$ co-referencial ortonormal em $(\Sigma, h)$. Nesta variedade Riemanniana, as grandezas geométricas são indicadas por barras e a primeira equação de estrutura de 
Cartan tem a seguinte forma:
\begin{eqnarray}\label{firstspacestructeq}
 d\bar{\theta}^i + \bar{\omega}_{j}^{i}\wedge \bar{\theta}^{j} = 0.
\end{eqnarray}
\indent Sobre $(M,g)$ definimos o seguinte co-referencial ortonormal
\begin{eqnarray}\label{corefonM}
 \theta^{0} = dt, \quad \theta^{i} = a(t)\bar{\theta}^{i}.
\end{eqnarray}
\indent Da primeira equação de estrutura \eqref{firstspacestructeq} obtemos
\begin{eqnarray}
 d\theta^0 = 0,\quad d\theta^{i} = da(t) \wedge \bar{\theta}^{i}  + a(t)d\bar{\theta}^{i} = \frac{\dot{a}}{a}\theta^0\wedge \theta^i - a\bar{\omega}_{j}^{i}\wedge \bar{\theta}^{j}.
\end{eqnarray}
\indent Comparando estas últimas expressões com a primeira equação de estrutura do espaço-tempo FLRW, segue-se
\begin{eqnarray}
 \omega^i_0\wedge \theta^i = 0,\quad \omega^i_0\wedge \theta^0 + \omega^i_j\wedge \theta^j = \frac{\dot{a}}{a}\theta^i\wedge \theta^0 + a\bar{\omega}_{j}^{i}\wedge \bar{\theta}^{j},
\end{eqnarray}
\noindent isso implica 
\begin{eqnarray}\label{linkinforms}
 \omega_i^0 = \frac{\dot{a}}{a}\theta^i, \quad \bar{\omega}_{j}^{i} = \omega_{j}^{i}.
\end{eqnarray}
\indent As linhas-mundo de observadores co-móveis são curvas integrais do campo de quadri-velocidade $u = \partial_t$, os quais são geodésicas. De fato,  seja o referencial
móvel $\{e_{\mu}\}$ dual a \eqref{corefonM}. Como $e_0 = u$ e usando as equações \eqref{linkinforms}, segue-se
\begin{eqnarray}\label{formconex}
 \nabla_u u = \nabla_{e_0} e_0 = \omega^{\lambda}_0(e_0)e_{\lambda} = \omega^{i}_0(e_0)e_{i} = 0.
\end{eqnarray}
\subsubsection{Equações de Einstein para espaços-tempo FLRW}
$\qquad$ Levando as formas de conexão \eqref{linkinforms} na segunda equação de estrutura, obtemos as \linebreak 2-formas de curvatura $\Omega^{\mu}_{\nu}$ de  $(M,g)$
\begin{eqnarray}
 \Omega^{0}_{i} = \frac{\ddot{a}}{a}\theta^0\wedge \theta^i, \quad \Omega^{i}_{j} = \frac{k + \dot{a}^2}{a^2}\theta^i\wedge \theta^j,
\end{eqnarray}
\noindent que nos fornece as seguintes componentes do tensor de  Einstein (em relação ao co-referencial \eqref{corefonM}):
\begin{eqnarray}\label{fieldequa1}
 & & G_{00} = 3 \left ( \frac{\dot{a}^2}{a^2} + \frac{k}{a^2}\right ), \\
 & & G_{11} = G_{22} = G_{33} = -2 \frac{\ddot{a}}{a} - \frac{\dot{a}^2}{a^2} - \frac{k}{a^2}, \\
 & & G_{\mu\nu} = 0\,(\mu\neq \nu).
\end{eqnarray}
\indent Para satisfazer as equações de campo \eqref{fieldequa1}, o tensor de energia momento deve ter a forma de um fluido perfeito:
\begin{eqnarray}\label{tensorenermom}
 T^{\mu\nu} = (\rho + p)u^{\mu}u^{\nu} + pg^{\mu\nu},
\end{eqnarray}
\noindent onde $u$ é o campo de velocidades. \\
\indent Isso implica  as equações de campo (com constante cosmológica) {}{serem} dadas por:
\begin{eqnarray}\label{friedmanneq1}
 & & 3 \left ( \frac{\dot{a}^2}{a^2} + \frac{k}{a^2}\right ) = 8\pi G\rho + \Lambda, \\
 & & -2 \frac{\ddot{a}}{a} - \frac{\dot{a}^2}{a^2} - \frac{k}{a^2} = 8\pi Gp - \Lambda.
\end{eqnarray}
\indent {}{Devemos frisar que estamos no contexto de um espaço-tempo de Einstein-de Sitter.}
Agora devemos verificar a conservação do tensor de energia-momento. Para a conservação da energia, usamos a seguinte relação
\begin{eqnarray}\label{conservenergy}
 \nabla_{u}\rho = -(p+\rho)\nabla \cdot u\,.
\end{eqnarray}
\indent Sendo $\nabla \cdot u$ a taxa de expansão dada por
\begin{eqnarray}\label{taxaexpan}
\nabla . u = \omega^{\lambda}_{\,0}(e_{\lambda})u^{0} = \omega^{i}_{\,0}(e_i),
\end{eqnarray}
\noindent portanto, usando as equações \eqref{formconex}, temos
\begin{eqnarray}
\nabla . u = 3\frac{\dot{a}}{a}.
\end{eqnarray}
\indent A equação \eqref{conservenergy} toma a seguinte forma
\begin{eqnarray}\label{conservenergy2}
 \dot{\rho} + 3\frac{\dot{a}}{a}(\rho + p) = 0.
\end{eqnarray}
\indent Esta última equação não deve ser considerada uma lei de conservação de energia, porque pelo princípio da equivalência da relatividade geral, temos que 
não há uma lei de conservação local, {}{mas a mesma é útil em análises sobre entropia \cite{straumann}.} \\
\indent Agora supondo uma dada equação de estado $p = p(\rho)$, podemos transformar  a equação \eqref{conservenergy2} do seguinte modo
\begin{eqnarray}
 \frac{d}{da}(\rho a^3) = -3pa^2,
\end{eqnarray}
\noindent para determinar $\rho$ como uma função de $a$. Como exemplos, temos:
\begin{enumerate}
 \item Para radiação (partículas sem massa), temos $p = \frac{\rho}{3}$ e portanto temos $\rho \propto  a^{-4}$.
 \item Para poeira ($p = 0$) obtemos $\rho \propto a^{-3}$.
\end{enumerate}
$\qquad$ Com isso temos que a equação de Friedmann \eqref{friedmanneq1} determina a evolução temporal de $a(t)$. Para obtermos a equação da aceleração da expansão, 
multiplicamos por $3$ a segunda equação em \eqref{friedmanneq1} e somamos com a primeira, obtendo
\begin{eqnarray}\label{acelerexpansion}
 \ddot{a} = -\frac{4\pi G}{3}(\rho + 3p)a + \frac{\Lambda a}{3}.
\end{eqnarray}
\indent Esta equação mostra que se $\rho + 3p > 0$, o primeiro termo em \eqref{acelerexpansion} é de desaceleração, enquanto que uma constante cosmológica positiva é 
um termo repulsivo. Para {}{apreciarmos} melhor o que acabamos de dizer, vamos escrever a equação de campo na seguinte forma:
\begin{eqnarray}
 G_{\mu\nu} = 8\pi G (T_{\mu\nu} + T_{\mu\nu}^{\Lambda}), 
\end{eqnarray}
\noindent onde $T_{\mu\nu}^{\Lambda} = -\frac{\Lambda}{8\pi G}g_{\mu\nu}$. Este termo corresponde a contribuição de vácuo do tensor de energia-momento de
um fluido ideal, com densidade de energia $\rho_{\Lambda} = \frac{\Lambda}{8\pi G}$ e pressão $p_{\Lambda} = -\rho_{\Lambda}$. Portanto, temos que 
$\rho_{\Lambda} + 3p_{\Lambda} = -2\rho_{\Lambda} \,<\,0$.
\subsubsection{Desvio para o vermelho}
$\qquad$ Como um resultado da expansão do Universo, a luz de fontes distantes parece deslocada para o vermelho. A quantidade do deslocamento para o vermelho pode 
ser expressa em termos do fator de escala $a(t)$. Consideremos duas curvas integrais do campo de velocidades médio $u$, com uma descrevendo a linha de mundo de uma fonte 
co-móvel e a outra de um observador num telescópio (ver figura em \cite{straumann}, capítulo $10$, página $554$). Como a luz está se propagando ao longo de geodésicas nulas, 
concluímos de \eqref{fridwalkermetric}
que ao longo do raio de luz $dt = a(t)d\sigma $, onde $d\sigma$  é o elemento de arco no espaço $3-$dimensional $(\Sigma, h)$ de curvatura constante $k=0,\pm 1$.
Portanto, a integral do lado esquerdo da equação abaixo
\begin{eqnarray}
 \int_{t_e}^{t_o} \frac{dt}{a(t)} = \int_{\text{fonte}}^{\text{observador}} d\sigma,
\end{eqnarray}
\noindent entre o tempo de emissão $t_e$ e o tempo de chegada no observador $t_o$, é independente de $t_e$ e $t_o$. Portanto, se considerarmos um segundo raio de luz
que é emitido no tempo $t_e + \Delta t_e$ e é recebido no tempo $t_o + \Delta t_o$, obtemos da equação acima que 
\begin{eqnarray}
 \int_{t_e + \Delta t_e}^{t_o + \Delta t_o} \frac{dt}{a(t)} = \int_{t_e}^{t_o} \frac{dt}{a(t)}.
\end{eqnarray}
\indent Para $\Delta t_e$ pequeno obtemos
\begin{eqnarray}
 \frac{\Delta t_o}{a(t_o)} = \frac{\Delta t_e}{a(t_e)}.
\end{eqnarray}
\indent As frequências observadas e emitidas $\nu_o$ e $\nu_e$ respectivamente, são portanto relacionadas de acordo com 
\begin{eqnarray}
 \frac{\nu_o}{\nu_e} = \frac{\Delta t_e}{\Delta t_o} = \frac{a(t_e)}{a(t_o)}.
\end{eqnarray}
\indent O parâmetro de desvio para o vermelho $z$ é definido por 
\begin{eqnarray}
 z = \frac{\nu_e - \nu_o}{\nu_o},
\end{eqnarray}
\noindent e dado pela equação
\begin{eqnarray}
 1 + z = \frac{a(t_o)}{a(t_e)}.
\end{eqnarray}
\indent Podemos expressar esta última relação como $\nu a = \text{constante}$ ao longo de uma geodésica nula.

\subsection{Teoria de gravitação ECSK (Einstein-Cartan-Sciama-Kibble)}
$\qquad$ Na teoria da gravitação de Einstein, a fonte de curvatura é a matéria-energia encapsulada no tensor de energia-momento. Mas a matéria também possui um grau de
liberdade intrínseco que é o spin e isso foi levado em conta por Cartan \cite{trautman}. {}{Esse grau de liberdade }foi implementado
considerando-se os seguintes ingredientes no modelo: o espaço-tempo $M$ 
possui um tensor métrico $g$, a conexão adotada $\nabla$ é compatível com a métrica $Xg(Y,Z) = g(\nabla_X Y, Z) + g(X,\nabla_X Z)$, a conexão possui torção não-nula
$T(X,Y) = \nabla_X Y - \nabla_Y X - [X,Y]$, as variáveis dinâmicas são o tensor métrico $g$ e a torção $T$, a ação adotada é formalmente a mesma da teoria de Einstein. \\
\indent Manter o tensor métrico $g$ é necessário para podermos construir uma curvatura escalar $\mathit{R}$, mas tomando uma conexão com torção não-nula, 
ampliamos o número de
graus de liberdade da teoria, e no caso, relacionamos a torção com a densidade de momento angular intrínseco da matéria (o spin da matéria é a fonte da torção), o 
que não é possível na teoria da gravitação de Einstein. Como temos agora duas variáveis dinâmicas, temos duas variações na ação:
\begin{itemize}
 \item [(1)]Variação em relação ao tensor métrico $g_{\mu\nu}$
 $$
 R_{\mu\nu} - \frac{1}{2}g_{\mu\nu}R = \kappa T_{\mu\nu}.
 $$
\indent Apesar desta equação ser formalmente idêntica à equação clássica de Einstein, o tensor de Ricci $R_{\mu\nu}$ não é mais simétrico, pois a conexão possui torção
não-nula e portanto o tensor de energia-momento no caso não é simétrico também (além do tensor energia-momento clássico, há um termo adicional que depende do tensor 
densidade de spin \cite{poplawski}).
\item [(2)] A variação em relação à torção, fornece a outra equação de movimento
$$
{T_{\mu\nu}}^{\sigma} + {g_{\mu}}^{\sigma}{T_{\nu\rho}}^{\rho} - {g_{\nu}}^{\sigma} {T_{\mu\rho}}^{\rho} = \kappa {s_{\mu\nu}}^{\sigma},
$$
\noindent onde ${s_{\mu\nu}}^{\sigma}$ é o tensor densidade de spin \cite{trautman}. Note que temos uma equação puramente algébrica para a torção 
(a torção não {}{se  propaga}), {}{mas isso depende do modelo considerado, há situações onde a torção é dinâmica \cite{Kostelecky:2007kx}}. Além disso,
se considerarmos uma Lagrangiana da matéria composta de um campo fermiônico, chegamos numa teoria de gravitação que elimina a singularidade inicial ({\it big-bang}) \cite{poplawski}.
\end{itemize}

\subsection{Teoria de gravitação {\it f}($\mathit{R}$)}


$\qquad$ Desde o início da relatividade geral, surgiram iniciativas para ampliar a mesma \cite{Weyl:1919fi,eddington}, a motivação no caso foi o de obter um melhor entendimento da teoria,
 sem nenhuma exigência experimental para isso, que surgiram décadas depois. Por volta do início da década de 60, a comunidade de gravitação foi percebendo que 
 uma ação gravitacional mais intrincada era útil. Por exemplo, a relatividade geral não é renormalizável o que impede uma quantização convencional, mas em 1962, Utiyama e
 De Witt mostraram que renormalização em $1$-{\it loop} exige que na ação de Hilbert-Einstein sejam adicionados termos de curvatura de ordem superior \cite{Utiyama:1962sn}.
 Resultados bem mais contemporâneos, por exemplo, via quantização semi-clássica da gravitação, mostram que a ação gravitacional efetiva em baixas energias exibe naturalmente invariantes de 
 curvatura de
 ordem superior \cite{quant1, quant2, quant3}. Mas a importância de tais termos na ação foi considerada como sendo restrita a regimes de gravidade forte, sendo suprimidos
 por acoplamentos pequenos, como esperado numa teoria efetiva. Deste modo, correções a gravitação de Einstein foram consideradas importantes apenas em escalas próximas às de Planck.
 Ou seja, no Universo inicial ou próximo a uma singularidade de um buraco negro \cite{Starobinsky:1980te}. \\
 \indent Na última década, surgiram novas evidências astrofísicas e cosmológicas, revelando um quadro inesperado do Universo. O CMB e o exame de supernovas sugerem que 
 a energia do Universo se compõe do seguinte modo: 4\% de matéria bariônica, 20\% de {\em matéria escura } e 76\% de {\em energia escura }\cite{Spergel:2006hy,Riess:2004nr,Astier:2005qq,Eisenstein:2005su}. 
 Por matéria escura se entende uma forma desconhecida de matéria, que possui a propriedade de se aglomerar como a matéria usual ({}{mas só interage gravitacionalmente})
, energia escura é uma forma desconhecida de energia que não foi detectada experimentalmente de forma direta e que não se aglomera como a matéria usual.
 Usando as várias condições (desigualdades) de energia \cite{wald}, podemos distinguir a matéria escura da energia escura: a matéria usual e a matéria escura satisfazem a
 condição de energia forte, enquanto que a energia escura não a satisfaz. Além disso, a energia escura se assemelha muito a uma constante cosmológica. Como atualmente a energia
 escura domina sobre a matéria dos dois tipos, a expansão do Universo parece ser de uma forma acelerada, ao contrário do que se acreditava anteriormente. \\
\indent Observações adicionais mostram que a matéria escura surge não apenas em dados cosmológicos, mas também em observações astrofísicas: o problema da \emph{massa faltante} 
medindo-se a curva de rotação de galáxias \cite{Rubin:1980zd,Rubin:1970zza,Bosma:1978,Persic:1995ru,Ellis:2002qd,Moore:2001fc}. 
Tudo isso nos leva a admitir que a nossa visão atual {}{observacional} da evolução e do conteúdo de 
matéria/energia do Universo é surpreendente e obviamente deve ser explicada. O modelo mais simples que se adapta aos dados experimentais é chamado modelo 
$\Lambda$CDM  ($\Lambda$-Matéria Escura Fria), suplementado por um cenário inflacionário, onde um campo escalar (o inflaton) é usado. Por ser um modelo {\it ad hoc}, não
explica a natureza da matéria escura e nem a origem do {\it inflaton}. Outro problema  também não explicado por
este modelo é o da magnitude da constante cosmológica: o valor da mesma é extremamente pequeno para ser atribuído à energia de vácuo dos campos da matéria  
\cite{Weinberg:1988cp,Carroll:2000fy}. \\
\indent Passando para a questão  da evolução cosmológica: a mesma motivou o interesse recente na gravidade $f(\mathit{R})$ para explicar a aceleração cósmica atual sem utilizarmos
a hipótese da energia escura. Nesse sentido, uma teoria $f(\mathit{R})$ precisa obedecer aos seguintes critérios: possuir a dinâmica cosmológica correta, exibir o comportamento 
correto de perturbações gravitacionais e gerar perturbações cosmológicas compatíveis com vínculos cosmológicos da radiação cósmica de fundo (e estrutura de
larga escala, nucleossíntese e ondas gravitacionais) \cite{sotiriou}.\\
\indent Na cosmologia, a identificação do nosso Universo com um espaço-tempo do tipo FLRW é amplamente baseada no alto grau de isotropia medida no fundo de micro-ondas cósmico.
Esta identificação depende de um resultado matemático conhecido como teorema de Ehlers-Geren-Sachs (EGS) \cite{Ehlers:1966ad}, uma caracterização cinemática de espaços-tempo 
FLRW. Este teorema afirma que  se uma congruência de observadores em queda livre do tipo tempo veem um campo de radiação isotrópico, então o espaço-tempo é espacialmente 
homogêneo e isotrópico e portanto, um espaço-tempo do tipo FLRW. Tal resultado se aplica a um  Universo preenchido com qualquer fluido perfeito que é geodésico e barotrópico 
\cite{Clarkson:1999yj}. Modificando a teoria via uma ação $f(\mathit{R})$, gostaríamos que o teorema EGS continuasse válido, de fato, isso foi
demonstrado em \cite{Rippl:1995bg}. \\
\indent Outro aspecto é nas chamadas eras cosmológicas. O interesse recente em teorias $f(\mathit{R})$ se deve à necessidade de explicar a aceleração atual do Universo descoberta
com supernovas do tipo Ia \cite{Riess:1998cb, Riess:1999th}. Numa gravidade do tipo $f(\mathit{R})$ pode haver aceleração cósmica e uma equação de estado efetiva.
Por outro lado, para $f(\mathit{R}) = \mathit{R}^2$, temos cenários do Universo inicial deste tipo, em paralelo com o uso de campos escalares para 
conduzir a inflação inicial ou aceleração posterior dos modelos de quintessência, além de tentativas de unificar a inflação inicial e aceleração posterior em gravidade modificada
\cite{Nojiri:2007cq,Nojiri:2007uq,Nojiri:2007as}. Mas em qualquer tentativa de modelar 
a aceleração do Universo em tempos tardios, o modelo criado deve seguir os modelos cosmológicos padrão que exigem uma sequência de eras bem definidas: inflação inicial,
uma era de radiação durante a qual ocorre a nucleossíntese, uma era da matéria, a época da aceleração presente e uma era futura.
\subsubsection{A ação {\it f}($\mathit{R}$)}
$\qquad$ Um outro modo de se generalizar a teoria da gravitação de Einstein {}{ (além da teoria ECSK)}, é mantendo-se a conexão de Levi-Civita (deste modo, a variável dinâmica é 
o tensor métrico $g_{\mu\nu}$), mas adotando-se uma ação mais geral
\begin{eqnarray}\label{fr_gravity}
 S(g) = \int \left ( f(\mathit{R}) + \mathcal{L}_{m}(g_{\mu\nu},\psi)\right )\, \sqrt{-g}\,\mathrm{d}^4x, 
\end{eqnarray}
\noindent onde $f(\mathit{R})$ é uma função analítica da curvatura escalar $\mathit{R}$. A questão imediata é do porque fazermos esta generalização. Como motivações, temos:
explicar a velocidade de expansão do Universo, descrever  o processo de formação de estruturas no Universo, explicar a massa faltante no Universo, 
sem recorrer às hipóteses de existência de matéria e energia escuras \cite{fabbri_fr_cosmology,fabbri_fr_torsion}. \\
\indent A equação de movimento, variando a ação em relação ao tensor métrico $g^{\mu\nu}$ é dada por:
\begin{eqnarray}
 f^{\prime}(R)R_{\mu\nu}-\frac{1}{2}f(\mathit{R})g_{\mu\nu}+\left[g_{\mu\nu} \Box-\nabla_\mu\nabla_\nu \right]f^{\prime}(R) = \kappa T_{\mu\nu}.
\end{eqnarray}
\indent É imediato que tomando $f(\mathit{R}) = R$, reobtemos a equação de campo da gravitação de Einstein.
Vamos agora apresentar uma dedução mais detalhada da equação de movimento (ver \eqref{gravieinst} para detalhes adicionais). \\
\noindent Como já vimos, a variação do determinante é dada por:
\begin{eqnarray}
 \delta \sqrt{-g}= -\frac{1}{2} \sqrt{-g} g_{\mu\nu} \delta g^{\mu\nu}. \nonumber
\end{eqnarray}
\indent Como a curvatura escalar é dada por
\begin{eqnarray}
\mathit{R} = g^{\mu\nu} R_{\mu\nu}, \nonumber
\end{eqnarray}
\noindent temos que a variação da mesma em relação ao tensor métrico $g^{\mu\nu}$ é dada por:
\begin{eqnarray}\label{scalarvariation}
\delta \mathit{R} &=& R_{\mu\nu} \delta g^{\mu\nu} + g^{\mu\nu} \delta R_{\mu\nu} \nonumber \\
         &=& R_{\mu\nu} \delta g^{\mu\nu} + g^{\mu\nu}(\nabla_\rho \delta \Gamma^\rho_{\nu\mu} - \nabla_\nu \delta \Gamma^\rho_{\rho\mu}), 
\end{eqnarray}
\noindent na segunda igualdade, usamos o resultado \eqref{riccivariation}:
\begin{eqnarray}
\delta R_{\mu\nu} = \delta R^\rho{}_{\mu\rho\nu} = \nabla_\rho (\delta \Gamma^\rho_{\nu\mu}) - \nabla_\nu (\delta \Gamma^\rho_{\rho\mu}).
\end{eqnarray}
\indent Agora, como $\delta\Gamma^{\lambda}_{\,\,\mu\nu}$ é um tensor (por ser a diferença entre duas conexões), temos que a mesma é dada por:
\begin{eqnarray}\label{gammatensorvariation}
 \delta \Gamma^\lambda_{\mu\nu}=\frac{1}{2}g^{\lambda a}\left(\nabla_\mu\delta g_{a\nu}+\nabla_\nu\delta g_{a\mu}-\nabla_a\delta g_{\mu\nu} \right).
\end{eqnarray}
\indent Substituindo \eqref{gammatensorvariation} em \eqref{scalarvariation}, isso implica 
\begin{eqnarray}
\delta R= R_{\mu\nu} \delta g^{\mu\nu}+g_{\mu\nu}\Box \delta g^{\mu\nu}-\nabla_\mu \nabla_\nu \delta g^{\mu\nu}, \nonumber 
\end{eqnarray}
\noindent onde $\Box = g^{\mu\nu}\nabla_{\mu}\nabla_{\nu}$ é o Laplaciano da conexão. \\
\indent Deste modo, a variação da ação $\delta S(g)$ toma a forma:
\begin{eqnarray}
\delta S(g) &=& \int {1 \over 2\kappa} \left(\delta f(\mathit{R}) \sqrt{-g}+f(\mathit{R}) \delta \sqrt{-g} + \delta(\sqrt{-g}\mathcal{L}_{m}(g_{\mu\nu},\psi))\right)\, \mathrm{d}^4x \nonumber \\
&=& \int {1 \over 2\kappa} \left(f^{\prime}(R) \delta R \sqrt{-g}-\frac{1}{2} \sqrt{-g} g_{\mu\nu} \delta g^{\mu\nu} f(\mathit{R})
+ \delta(\sqrt{-g}\mathcal{L}_{m}(g_{\mu\nu},\psi))\right) \, \mathrm{d}^4x \nonumber \\
&=& \int {1 \over 2\kappa} \sqrt{-g} f^{\prime}(R)(R_{\mu\nu} \delta g^{\mu\nu}+g_{\mu\nu}\Box \delta g^{\mu\nu}-\nabla_\mu \nabla_\nu \delta g^{\mu\nu}) -
\frac{1}{2} g_{\mu\nu} \delta g^{\mu\nu} f(\mathit{R}) + \nonumber \\
&+& \delta(\sqrt{-g}\mathcal{L}_{m}(g_{\mu\nu},\psi)) )\, \mathrm{d}^4x. 
\end{eqnarray}
\indent Finalmente, fazendo $\delta S(g) = 0$ (para obtermos o extremo da ação):
\begin{eqnarray}
f^{\prime}(R)R_{\mu\nu}-\frac{1}{2}f(\mathit{R})g_{\mu\nu}+\left[g_{\mu\nu} \Box-\nabla_\mu \nabla_\nu \right]f^{\prime}(R) = \kappa T_{\mu\nu},
\end{eqnarray}
\noindent onde $T_{\mu\nu}$ é o tensor de energia-momento dado por:
\begin{eqnarray}
T_{\mu\nu}=-\frac{2}{\sqrt{-g}}\frac{\delta(\sqrt{-g} \mathcal{L}_{m}(g_{\mu\nu},\psi))}{\delta g^{\mu\nu}}. \nonumber 
\end{eqnarray}

%% file: gravitacao_fr_com_torcao.tex
\chapter{Gravitação {\it f}($\mathit{R}$) com torção}

$\qquad$ Neste capítulo, mostramos a existência de um campo espinorial singular (um {\it flag-dipolo}) como solução de uma equação de evolução de um campo fermiônico sujeita a 
um fundo $f(\mathit{R})$ com torção. Não há na literatura, nenhum resultado desse tipo.

\section{Teoria de gravitação {\it f}($\mathit{R}$)-ECSK}

$\qquad$ Dentro da sequência de teorias de gravitação que estamos apresentando, esta é a mais completa, pois modificamos a parte gravitacional da ação, a conexão possui torção
 \cite{Arcos:2005ec} e na Lagrangiana
da matéria usamos campos fermiônicos que são afetados pela conexão com torção. Descrevemos nossos resultados obtidos em \cite{fabbri_ESK}, que remetem a 
\cite{fabbri_fr_torsion,fabbri_fr_cosmology}.
\subsection{Gravitação  {\it f}($\mathit{R}$) com torção}
$\qquad$ A extensão da ação de Hilbert-Einstein em relação a uma função arbitrária $f(\mathit{R})$ \eqref{fr_gravity} é atraente, pois é a mais geral sempre que restringimos o escalar de
Ricci como a única fonte de informação dinâmica. No caso em que ambos métrica e a torção são levadas em conta, a variação em relação a uma métrica arbitrária
$g$ e uma conexão $\Gamma$ compatível com $g$ (ou equivalentemente um campo de tétradas $\{e_\mu\}$ e uma conexão de spin $\omega$) fornecem o enfoque métrico afim (ou tétrada afim)
\cite{capozziello_fr_metric_affine,capozziello_fr_j_bundle,capozziello_fr_gravity_overview,rubilar_ECSK_matter_fields}. 
{}{A ação adotada e as equações de campo correspondentes são dadas por:
\begin{eqnarray}
S(g) = \int \left ( f(\mathit{R}) + \mathcal{L}_{m}(g_{\mu\nu},\Gamma,\psi)\right )\, \sqrt{-g}\,\mathrm{d}^4x  
\end{eqnarray}
\noindent onde 
\begin{eqnarray}
     {\cal L}_m=
\left[\frac{i}{2}\left(\bar\psi\gamma^{\sigma}D_{\sigma}\psi-D_{\sigma}\bar\psi\gamma^{\sigma}\psi\right)-m\bar\psi\psi\right]
\end{eqnarray}}
\noindent e
\begin{subequations}
\label{2.1}
{}{\begin{equation}
\label{2.1b }
\qquad\qquad\qquad T_{ij}^{\;\;\;h}
=\frac{1}{f'(R)}
\left[\frac{1}{2}\left(\frac{\partial f'(R)}{\partial x^{p}} + S_{pq}^{\;\;\;q}\right)
\epsilon_{r}^{\;\,ph}\epsilon_{i\;\,j}^{\,r}
+S_{ij}^{\;\;\;h}\right],
\end{equation}}
\begin{equation}
\label{2.1a}
\hspace*{-1.3cm}\Sigma_{ij}=f'\/(R)R_{ij} -\frac{1}{2}f\/(R)g_{ij},
\end{equation}
\end{subequations}
\noindent onde $R_{ij}$, $\epsilon_{ijk}$, e $T_{ij}^{\;\;\;h}$ são as componentes dos tensores de Ricci, Levi-Civita e de torção respectivamente.
Os $\Sigma_{ij}$ e $S_{ij}^{\;\;\;h}$ denotam as componentes dos tensores de energia-momento e de densidade espinorial associados aos campos da matéria{}{. As leis} de conservação
ffixando \begin{subequations}
\label{2.2}
\begin{equation}
\label{2.2a}
\nabla_{i}\Sigma^{ij}+T_{i}\Sigma^{ij}-\Sigma_{pi}T^{jpi}-\frac{1}{2}S_{sti}R^{stij}=0,
\end{equation}
\begin{equation}
\label{2.2b}
\qquad\qquad\nabla_{h}S^{ijh}+T_{h}S^{ijh}+\Sigma^{ij}-\Sigma^{ji}=0,
\end{equation}
\end{subequations}
\noindent seguem das identidades de Bianchi \cite{fabbri_fr_torsion,fabbri_ESK}. Nas equações \eqref{2.2} os símbolos $\nabla_i$ e $R^{ijkl}$ denotam respectivamente a derivada covariante e o tensor de
curvatura de Riemann, em relação a conexão dinâmica $\Gamma$. Denotando $\Gamma^i = e^i_\mu\gamma^\mu$, onde $e^\mu_i$ é uma tétrada associada à métrica e introduzindo
S$_{\mu\nu}:= \frac{1}{8}[\gamma_\mu,\gamma_\nu]$, as derivadas do campo da matéria $\psi$ e seus adjuntos de Dirac são denotados por 
$D_i\psi = \frac{\partial \psi}{\partial x^i} + \omega_i^{\;\;\mu\nu}{\rm S}_{\mu\nu}\psi\/$ e
$D_i\bar\psi = \frac{\partial \bar\psi}{\partial x^i} - \bar\psi\omega_i^{\;\;\mu\nu}{\rm S}_{\mu\nu}\/$, onde $\omega_i^{\;\;\mu\nu}$  é a conexão de spin. 
Podemos além disso escrever 
$D_i\psi = \frac{\partial \psi}{\partial x^i} - \Omega_i\psi$ e $D_i\bar\psi = \frac{\partial \bar\psi}{\partial x^i} + \bar{\psi}\Omega_i$ onde
\begin{equation}\label{2.3}
\Omega_i := - \frac{1}{4}g_{jh}\left(\Gamma_{ik}^{\;\;\;j} - e^j_\mu\partial_i\/e^\mu_k \right)\Gamma^h\Gamma^k,
\end{equation}
$\Gamma_{ik}^{\;\;\;j}$ denotam os coeficientes da conexão linear  $\Gamma$, pois a relação entre a conexão linear e a conexão de spin
é fornecida por 
$\Gamma_{ij}^{\;\;\;h} = \omega_{i\;\;\;\nu}^{\;\;\mu}e_\mu^h\/e^\nu_j + e^{h}_{\mu}\partial_{i}e^{\mu}_{j}$, como pode ser imediatamente calculado.
No caso dos campos da matéria, o tensor de densidade de spin é dado por 
S$_{ij}^{\;\;\;h}=\frac{i}{2}\bar\psi\left\{\Gamma^{h}, {\rm S}_{ij}\right\}\psi 
\equiv-\frac{1}{4}\eta^{\mu\sigma}\epsilon_{\sigma\nu\lambda\tau} K^\tau e^{h}_{\mu}e^{\nu}_{i}e^{\lambda}_{j}$. Lembre que $K^\tau$ é a componente do bilinear 
covariante pseudo-vetorial definido em (\ref{fierz}). As componentes do tensor de energia-momento dos campos da matéria são portanto descritas como 
\begin{equation}\label{2.5}
\Sigma^D_{ij} := \frac{i}{4}\/\left( \bar\psi\Gamma_{i}{D}_{j}\psi - {D}_{j}\bar{\psi}\Gamma_{i}\psi \right)\quad\quad\text{ e }\quad\;\;\;
\Sigma^F_{ij}:= (\rho +p)\/U_iU_j -pg_{ij}.
\end{equation}
\indent Nas equações \eqref{2.5} $\rho$, $p$ e $U_i$ denotam respectivamente a densidade da matéria-energia, a pressão, e a quadrivelocidade do fluido.
O traço das equações \eqref{2.1a}, dado por 
\begin{equation}\label{2.6}
f'(R)R -2f(\mathit{R})=\Sigma,
\end{equation} 
{}{que relaciona} a curvatura escalar de Ricci $\mathit{R}$ e o traço $\Sigma$ do tensor de energia-momento, como em 
\cite{fabbri_fr_torsion,capozziello_fr_metric_affine,capozziello_fr_j_bundle,capozziello_fr_gravity_overview}.
Além disso, é assumido que $f(\mathit{R})\not = kR^2$ --- pois o caso $f(\mathit{R})=kR^2$ é compatível somente com a condição $\Sigma=0$. Agora, da equação 
\eqref{2.6} é possível expressar $\mathit{R} = F(\Sigma)$, onde $F$ é uma função arbitrária. Além disso, introduzindo o campo escalar
$\varphi := f'\/(F\/(\Sigma))$ assim como o potencial efetivo  $V(\varphi):= \frac{1}{4}\left[ \varphi F^{-1}\/((f')^{-1}\/(\varphi))+ \varphi^2\/(f')^{-1}\/(\varphi)\right]$, 
as equações de campo \eqref{2.1a} são escritas numa forma do tipo Einstein
\begin{equation}\label{2.9}
\begin{split}
\mathring{R}_{ij} -\frac{1}{2}\mathring{R}g_{ij}= \frac{1}{\varphi}\Sigma^F_{ij} + \frac{1}{\varphi}\Sigma^D_{ij} + \frac{1}{\varphi^2}\left( - \frac{3}{2}\varphi_i\varphi_j + \varphi\mathring{\nabla}_{j}\varphi_i + \frac{3}{4}\varphi_h\varphi_k g^{hk}g_{ij} \right. \\
\left. - \varphi\mathring{\nabla}^h\varphi_hg_{ij} - V\/(\varphi)g_{ij} \right) + \mathring{\nabla}_h\hat{{\rm S}}_{ji}^{\;\;\;h} + \hat{{\rm S}}_{hi}^{\;\;\;p}\hat{{\rm S}}_{jp}^{\;\;\;h} - \frac{1}{2}\hat{{\rm S}}_{hq}^{\;\;\;p}\hat{{\rm S}}_{\;\;p}^{q\;\;\;h}g_{ij},
\end{split}
\end{equation}
\noindent onde $\mathring{R}_{ij}$, $\mathring R$ e $\mathring{\nabla}_i$ denotam respectivamente o tensor de Ricci, a curvatura escalar e a derivada covariante da conexão 
de Levi-Civita.
Aqui $\hat{{\rm S}}_{ij}^{\;\;\;h}:=-\frac{1}{2\varphi}{\rm S}_{ij}^{\;\;\;h}\/$ e $\varphi_i:=\frac{\partial\varphi}{\partial x^i}$.
Além disso, as equações de Dirac generalizadas para o campo espinorial neste contexto {}{são}
\begin{equation}\label{2.10}
i\Gamma^{h}D_{h}\psi + \frac{i}{2}T_h\Gamma^h\psi- m\psi=0,
\end{equation}
\noindent onde $T_h :=T_{hj}^{\;\;\;j}$ é a torção axial\footnote{É interessante notar que neste ponto não é formalmente explícito por \eqref{2.10} se estamos lidando 
com uma equação de Dirac com torção num espaço simplesmente conexo ou com uma equação de Dirac sem torção num espaço-tempo multiplamente conexo \cite{daRocha:2011yr}. 
Como ambas descrições são matematicamente equivalentes, pode-se passar de um formalismo para o outro, para contornar tal questão.}.
A parte simetrizada das equações do tipo Einstein \eqref{2.9} assim como as equações de Dirac \eqref{2.10} são escritas como em \cite{fabbri_fr_torsion}
\begin{equation}\label{2.15}
\begin{split}
\mathring{R}_{ij} -\frac{1}{2}\mathring{R}g_{ij}= \frac{1}{\varphi}\Sigma^F_{ij} + \frac{1}{\varphi}\mathring{\Sigma}^D_{ij}
+ \frac{1}{\varphi^2}\left( - \frac{3}{2}\varphi_i\varphi_j + \varphi\mathring{\nabla}_{j}\varphi_i +
\frac{3}{4}\varphi_h\varphi_kg^{hk}g_{ij} \right. \\
\left. - \varphi\mathring{\nabla}^h\varphi_hg_{ij} - V\/(\varphi)g_{ij} \right) + \frac{3}{64\varphi^2}K^\tau K_\tau g_{ij}\text{ e }
\end{split}
\end{equation}
\begin{equation}\label{2.16}
i\Gamma^{h}\mathring{D}_{h}\psi
-\frac{3}{16\varphi}\left[\sigma
+i\omega\gamma_5\right]\psi-m\psi=0,
\end{equation}
\noindent onde
$\mathring{\Sigma}^D_{ij} := \frac{i}{4}\/\left[ \bar\psi\Gamma_{(i}\mathring{D}_{j)}\psi - \left(\mathring{D}_{(j}\bar\psi\right)\Gamma_{i)}\psi \right]$ e $\mathring{D}_i$ 
é a derivada covariante da conexão de Levi-Civita induzida no campo fermiônico. Devemos notar a ocorrência explícita de dois bilineares covariantes na equação \ref{2.16}, 
fixando deste modo o tipo de espinor 
que será solução das equações de movimento {}{(no sentido de ao escolhermos se $\sigma\text{ e }\omega$ são ou não nulos, reduzimos o tipo de espinor admissível)}.\\ 
\indent Como campos espinoriais satisfazendo a equação de Dirac neste cenário são incompatíveis com uma simetria esférica estacionária \cite{fabbri_dirac_field_no_go}, 
a escolha mais simples para este fundo deve ser pelo menos uma métrica axialmente simétrica do tipo Bianchi-I, dada por 
 $ds^2 = dt^2 - a^2(t)\,dx^2 - b^2(t)\,dy^2 - c^2(t)\,dz^2$, 
onde  $\Gamma^i = e^i_\mu\gamma^\mu$ são dadas por
\begin{equation}\label{3.5}
\Gamma^0 = \gamma^0,\qquad \Gamma^1 = \frac{1}{a(t)}\gamma^1, \qquad \Gamma^2 = \frac{1}{b(t)}\gamma^2, \qquad \Gamma^3 = \frac{1}{c(t)}\gamma^3,
\end{equation} 
\noindent e o campo de tétrada é dado por $e^\mu_0=\delta^\mu_0$, $e^\mu_1 = a(t)\/\delta^\mu_1$, $e^\mu_2 = b(t)\/\delta^\mu_2,$ e
$e^\mu_3 = c(t)\/\delta^\mu_3$, para $ \mu =0,1,2,3$. 
O operador de spin-Dirac age sobre campos espinoriais e seus conjugados respectivamente como 
$\mathring{D}_i\psi = \partial_i\psi - \mathring{\Omega}_i\psi$ 
e $ \mathring{D}_i\bar\psi = \partial_i\bar\psi + \bar{\psi}\mathring{\Omega}_i$,
onde os coeficientes da conexão de spin $\mathring{\Omega}_i$ são dados por 
(introduzindo a notação  $a_1 = a$, $a_2 = b$, e $a_3 = c$) 
\begin{eqnarray}
\mathring{\Omega}_0=0,\qquad\qquad\quad \mathring{\Omega}_i=\frac{1}{2}{\dot{a}_i}\gamma^i\gamma^0.
\end{eqnarray}
\indent Portanto, a equação do tipo Einstein \eqref{2.15} toma a forma
\begin{subequations}\label{3.10}
\begin{equation}\label{3.10a}
\begin{split}
\frac{\dot a}{a}\frac{\dot b}{b} + \frac{\dot b}{b}\frac{\dot c}{c} + \frac{\dot a}{a}\frac{\dot c}{c} = \frac{\rho}{\varphi} - \frac{3}{64\varphi^2}K^\sigma\,K_\sigma\,
+\frac{1}{\varphi^2}\left[- \frac{3}{4}{\dot\varphi}^2 - \varphi\dot\varphi\frac{\dot\tau}{\tau} - V(\varphi)\right],
\end{split}
\end{equation}
\begin{equation}\label{3.10b}
\begin{split}
\frac{\ddot a_r}{a_r} + \frac{\ddot a_s}{a_s} + \frac{\dot a_r}{a_r}\frac{\dot a_s}{a_s} = - \frac{p}{\varphi} +
\frac{1}{\varphi^2}\left[\varphi\dot\varphi\frac{\dot a_t}{a_t}+\frac{3}{4}{\dot\varphi}^2 -\varphi\left( \ddot\varphi + \frac{\dot\tau}{\tau}\dot\varphi \right) - V(\varphi)\right] +\frac{3}{64\varphi^2}K^\sigma\,K_\sigma\, ,
\end{split}
\end{equation}
\end{subequations}
onde  $r,s,t$ denotam índices $1,2,3$ diferentes um dos outros. A equação de campo de Dirac \eqref{2.16} assume a forma
\begin{equation}\label{3.9}
\dot\psi + \frac{\dot\tau}{2\tau}\psi + im\gamma^0\psi - \frac{3i}{16\varphi}(\sigma\gamma^0 + i\omega\gamma^0\gamma^5) \psi = 0,
\end{equation}
onde $\tau := abc$ \cite{saha_nonlinear_Spinor_bianchi_I,saha_bianchi_I_cosmology}. Junto com as condições 
\begin{equation}\label{3.11}
\mathring{\Sigma}^D_{rs}=0\quad \Rightarrow \quad a_r\/\dot{a}_s - a_s\/\dot{a}_r=0 \quad \cup \quad K^\intercal =0,
\end{equation}
as equações $\mathring{\Sigma}^D_{0i}=0$ são automaticamente satisfeitas. 
Finalmente, as leis de conservação junto com uma equação de estado do tipo $p=\lambda\rho$ (aqui $\lambda$  é um escalar entre $0$ e $1$)
fornecem $\dot\rho + \frac{\dot\tau}{\tau}(1+\lambda)\rho =0$, {}{a qual} completa o conjunto total das equações de campo, tendo a solução geral dada por
\begin{equation}\label{3.12bis}
\rho = \rho_0\tau^{-(1+\lambda)}\,, \qquad \rho_0 = {\rm constante}.
\end{equation}
\indent O campo da matéria em tal fundo gravitacional é tal que as condições \eqref{3.11} são vínculos impostos sobre a métrica ou sobre o campo da matéria.
Eles existem se e somente se uma das seguintes condições valem:
\begin{enumerate}
\item Impondo vínculos de origem puramente geométrica, como $a\dot{b}-b\dot{a}=0$, $a\dot{c}-c\dot{a}=0$, $c\dot{b}-b\dot{c}=0$. 
Neste cenário existem campos da matéria fermiônicos num Universo isotrópico, o que pode \emph{a priori} causar alguma patologia, pois campos de Dirac 
são bem conhecidos por não se submeterem ao Princípio Cosmológico \cite{tsamparlis}. Mas pelo resultado de Tsamparlis \cite{tsamparlis}, embora válidos para os campos espinoriais de Dirac, 
não são válidos para outras classes de campos espinoriais, segundo a classificação de Lounesto.
\item Outra condição é impor vínculos de origem puramente material, requerendo que as componentes espaciais da direção do spin satisfaçam $K_i = 0$.
Isto representa um Universo anisotrópico desprovido de termos acoplando a matéria com a torção axial. Neste caso não há nenhuma interação fermiônica torsional.
De fato, o spin da partícula interage com a componente axial do tensor de torção e quando as componentes espaciais da direção do spin se igualam a zero temos que 
tais partículas são descritas por um campo $\psi$ que não interage com a torção. Além disso, se os campos de Dirac estão ausentes então não é claro o que possa 
justificar anisotropias \cite{fabbri_fr_cosmology}.
\item A última situação seria originada pela geometria e pela matéria também, insistindo que por exemplo $a\dot{b}-b\dot{a}=0$ e $K_1 = 0 = K_2$.
Isso fornece uma isotropia parcial para apenas dois eixos, com as respectivas componentes do vetor de spin se anulando. Isso descreve um Universo como um 
elipsóide de rotação em relação ao eixo ao longo do qual o vetor de spin não se anula. Insistindo na proporcionalidade entre dois pares de eixos, inevitavelmente obtemos
uma isotropia total do espaço $3$-dimensional. Portanto, a situação na qual temos $a=b$, com $K_1 = 0 = K_2$, é a única inteiramente satisfatória. 
Daqui em diante esta situação será considerada, onde a única componente espacial da direção do spin é $K_3\neq 0$ \cite{fabbri_ESK}.
\end{enumerate} 
$\quad$ Aqui, as equações de Dirac e do tipo Einstein \eqref{3.10} e \eqref{3.9} podem ser trabalhadas como em \cite{saha_nonlinear_Spinor_bianchi_I,saha_bianchi_I_cosmology}:
por exemplo, através de combinações adequadas de \eqref{3.10} obtemos as equações
\begin{subequations}\label{3.13}
\begin{equation}\label{3.13a}
\frac{d}{dt}(J_0\tau)=0 =
\frac{d}{dt}(\sigma\tau)
+\frac{3\omega K_0\tau}{8\varphi},
\end{equation}
\begin{equation}\label{3.13c}
-\frac{d}{dt}(\omega\tau)
+\left[2m+\frac{3\sigma}{8\varphi}\right]K_0\tau=0=\frac{d}{dt}(K_0\tau)+2m\omega\tau.
\end{equation}
\end{subequations}
\noindent {}{Das} equações \eqref{3.13} é imediato deduzir que 
{}{\begin{equation}\label{3.13abis}
(K_3)^2=\sigma^2 + \omega^2 + (K_0)^2 = \frac{C^2}{\tau^2} \Rightarrow C = K_3\tau,\quad\quad
(J_0)^2=\frac{D^2}{\tau^2}\Rightarrow D = J_0\tau,
\end{equation}}
com $C$ e $D$ constantes. É fundamental enfatizar neste caso especial que a teoria possui uma simetria discreta adicional fornecida pela transformação
$\psi \mapsto \gamma^5\gamma^0\gamma^1\psi$, fazendo com que todas as equações de campo sejam invariantes. 
Na equação de Dirac, as quatro componentes complexas neste caso se reduzem a duas componentes complexas. Tal afirmação é equivalente a tomar campos espinoriais do tipo
{\it flagpole}, que possuem quatro parâmetros reais. Portanto \eqref{3.13} são as equações de campo a serem resolvidas. Devemos frisar {}{a} originalidade deste resultado,
sendo a primeira vez na literatura que esse tipo de espinor surge como solução de uma equação de movimento fermiônica.
A compatibilidade com todos os vínculos permite apenas três classes de campos espinoriais, cada uma das quais possui um membro geral escrito numa das 
{}{três seguintes formas independentes}
\begin{eqnarray}
\label{generalspinor}
&\psi_1=\frac{1}{\sqrt{2\tau}}\left(\begin{tabular}{c}
$\sqrt{K-C}\cos{\zeta_{1}}e^{i\theta_{1}}$\\
$\sqrt{K+C}\cos{\zeta_{2}}e^{i\vartheta_{1}}$\\
$\sqrt{K-C}\sin{\zeta_{1}}e^{i\vartheta_{2}}$\\
$\sqrt{K+C}\sin{\zeta_{2}}e^{i\theta_{2}}$
\end{tabular}\right),
\end{eqnarray}
com os vínculos $\tan{\zeta_{1}}\tan{\zeta_{2}}=(-1)^{n+1}$ e $\theta_{1}+\theta_{2}-\vartheta_{1}-\vartheta_{2}=\pi n$ para qualquer $n$ inteiro e 
também
\begin{eqnarray}
\label{restrictedspinor1}
&\psi_2=\frac{1}{\sqrt{2\tau}}\left(\begin{tabular}{c}
$\sqrt{K-C}\cos{\zeta_{1}}e^{i\theta_{1}}$\\
$0$\\
$0$\\
$\sqrt{K+C}\sin{\zeta_{2}}e^{i\theta_{2}}$
\end{tabular}\right)\quad
\text{ e }\;\;\;
\psi_3=\frac{1}{\sqrt{2\tau}}\left(\begin{tabular}{c}
$0$\\
$\sqrt{K+C}\cos{\zeta_{1}}e^{i\vartheta_{1}}$\\
$\sqrt{K-C}\sin{\zeta_{2}}e^{i\vartheta_{2}}$\\
$0$
\end{tabular}\right),
\end{eqnarray}
onde $\zeta_{1}$, $\zeta_{2}$, $\theta_{1}$, $\theta_{2}$, $\vartheta_{1}$, $\vartheta_{2}$ dependem do tempo. 
O caso mais interessante é fornecido por \eqref{restrictedspinor1}. Por exemplo, o segundo campo espinorial em \eqref{restrictedspinor1} é 
\begin{eqnarray}
\label{spinorsolution}
&\psi_3=\frac{1}{\sqrt{2\tau}}\left(\begin{tabular}{c}
$0$\\
$\sqrt{K+C}e^{i\beta(t)}$\\
$\sqrt{K-C}e^{-i\beta(t)}$\\
$0$
\end{tabular}\right),\quad\text{}
\end{eqnarray} 
para $\beta(t)=-mt-\frac{3C}{16}\int{\frac{dt}{\tau}}$. Existem vínculos adicionais $\sigma=\frac{C}{\tau}$, $\psi^{\dagger}\psi=\frac{K}{\tau}$ e $\omega=0 = K_0$.
Tal campo espinorial {\it  é do tipo-(4) de acordo com a classificação de Lounesto} \cite{daSilva:2012wp}. 
Este é um fato notável: assim que é assumido um campo espinorial $\psi$ numa cosmologia do tipo Riemann-Cartan, alguns campos espinoriais do tipo-(4) são obtidos
como os campos espinoriais (\ref{restrictedspinor1}). De fato, não há nenhuma suposição na equação (\ref{2.10}) que faça $\psi$ um campo espinorial de Dirac legítimo, visto que diz
respeito \emph{a priori} apenas um campo espinorial $\psi$ que satisfaça a equação de Dirac. Até onde sabemos, este é até agora o único sistema físico cuja solução aceitável
é dada em termos de tais campos espinoriais.
Por outro lado, quando impomos $K_3 = 0$ como um vínculo de origem puramente material, as equações (\ref{3.13abis}) implicam  $K_0 = 0$.
Portanto $K^\mu = 0$ e obtemos um campo espinorial do tipo-(5) de acordo com a classificação de Lounesto para campos espinoriais, que englobam campos espinoriais
{}{de Majorana.}
Deve ser enfatizado que a condição $K_3=0$ não implica necessariamente neste caso a inexistência de campos fermiônicos satisfazendo a equação de Dirac (\ref{2.10}).
De fato, campos Elko não satisfazem a equação de Dirac\footnote{De fato, campos espinoriais Elko \cite{allu,allu1} são auto-espinores do operador de 
conjugação de carga e não satisfazem a equação de Dirac \cite{allu,allu1}. Algumas aplicações importantes são fornecidas, por exemplo, em \cite{Basak:2012sn}.
Ainda {}{existe} o conjunto complementar de campos espinoriais Elko e Majorana, em relação ao spinor de tipo-(5), cuja dinâmica é ainda desconhecida.}.
Em resumo, pelas soluções acima, o chamado campo de Dirac $\psi$ em (\ref{restrictedspinor1}, \ref{spinorsolution}) não é um campo de Dirac de acordo com a
classificação de Lounesto, mas um campo espinorial do tipo-(4), um {\it flag-dipole}. Além disso, como $K_i = 0$ e em particular $K_3=0$,
por \eqref{3.13abis} agora estamos tratando de um espinor do tipo-(5), o qual é um {\it flagpole}. Mas neste caso, é bem conhecido que os do  tipo-(5)
englobam campos espinoriais Elko, Majorana, e complementares, presentes na equação \eqref{eq10}. {}{O espinor Elko, contudo, é bem conhecido por não satisfazer as equações de
Dirac{}{. Como} partimos de \eqref{3.13abis}, Elko é excluído como solução de tal sistema}. O ponto a ser enfatizado aqui é que de acordo com a classificação de campos 
espinoriais de Lounesto, $\psi$ pode ser localizado em qualquer das seis classes disjuntas e não há \emph{a priori} nenhuma relação entre o tipo de campo espinorial
e a dinâmica associada. Como mencionado, por exemplo, os tipos-(1), (2) e (3)  são campos espinoriais regulares na classificação de Lounesto, com alguns subconjuntos
satisfazendo a equação de Dirac. Pelo mesmo motivo, campos espinoriais do tipo-(6) englobam campos espinoriais de Weyl, que de fato satisfazem equações de Dirac. 
Mesmo assim, é um problema em aberto se campos espinoriais do tipo-(4) satisfazem ou não as equações de Dirac, mas as equações de Dirac são mostradas como sendo 
dinamicamente proibidas para soluções encontradas em \cite{fabbri_fr_cosmology}.

{}{\subsubsection{Exemplo concreto de gravitação f($\mathit{R}$)}}

{}{$\qquad$ Como exemplo concreto (não-trivial) de um modelo f($\mathit{R}$) \cite{fabbri_fr_cosmology}, vamos considerar o caso $f(\mathit{R}) = \mathit{R} + \delta\alpha \mathit{R}^2$, 
onde $\delta\alpha$ é um termo de perturbação em relação a gravitação de Einstein. Tomaremos como soluções das equações de campo, soluções que são correções em $\delta\alpha$ das equações de campo de Einstein no caso sem fontes.
Escolhemos a seguinte forma como solução para a equação de campo de Dirac}

{}{
\begin{eqnarray}
\label{spinorsolution}
&\psi=\frac{1}{\sqrt{2\tau}}\left(\begin{tabular}{c}
$0$\\
$\sqrt{K+C}e^{i\left(-mt-\frac{3C}{16}\int{\frac{dt}{\tau}}\right)}$\\
$\sqrt{K-C}e^{-i\left(-mt-\frac{3C}{16}\int{\frac{dt}{\tau}}\right)}$\\
$0$
\end{tabular}\right)
\end{eqnarray}}

\noindent {}{sujeita aos vínculos $J_0 = \bar\psi\psi=\frac{C}{\tau}$, $\psi^{\dagger}\psi=\frac{K}{\tau}$ e $\psi^{\dagger}\gamma^5\psi=0$, 
$i\bar\psi\gamma^5\psi=0$. Esse tipo de solução leva a uma forma simples das equações de campo gravitacionais para a forma e volume do Universo, como}

\begin{eqnarray}\label{cosmic1}
\frac{a}{c}=De^{\left(X\int{\frac{1}{\tau}dt}\right)}
\end{eqnarray}
e também
\begin{eqnarray}\label{cosmic2}
\ddot\tau-\frac{3mC}{4}=0,
\end{eqnarray}
\noindent {}{onde $m$, $C$, $X$, $D$ são constantes. Integrando a equação \eqref{cosmic2}, obtemos a evolução do volume do Universo }
\begin{eqnarray}\label{cosmic3}
\tau=\frac{3mC}{8}(b+2\beta t +t^{2}),
\end{eqnarray}
\noindent {}{$b$ é uma constante de integração que encapsula a informação a respeito do do volume inicial do Universo, fornecendo um estado inicial sem singularidade e
$\beta$ é a constante de integração que contem a informação a respeito da velocidade de expansão do Universo. Então, como $\tau=a^{2}c$, da equação \eqref{cosmic1}
obtemos a evolução da forma do Universo}
\begin{eqnarray}\label{cosmic4}
\frac{a}{c}=D\left(\frac{t+\beta+\sqrt{\beta^{2}-b}}{t+\beta-\sqrt{\beta^{2}-b}}\right)^{-\frac{8X}{6mC\sqrt{\beta^{2}-b}}},
\end{eqnarray}
\noindent {}{na qual, quando $t$ tende ao infinito ambos os fatores se tornam proporcionais, fornecendo uma isotropia. Fazendo $D= 1$ e 
$8X=-3mC\sqrt{\beta^{2}-b}$ temos a evolução dos fatores $a$ e $c$}
\begin{subequations}\label{cosmic5}
\begin{eqnarray}
a=\sqrt[3]{\frac{3mC}{8}}\left(t+\beta-\sqrt{\beta^{2}-b}\right)^{-\frac{1}{3}}
\left(b+2\beta t +t^{2}\right)^{\frac{1}{2}},
\end{eqnarray}
\begin{eqnarray}\label{cosmic6}
c=\sqrt[3]{\frac{3mC}{8}}\left(t+\beta-\sqrt{\beta^{2}-b}\right)^{\frac{2}{3}},
\end{eqnarray}
\end{subequations} 
{}{\noindent com vínculo dado por $3m^{2}(\beta^{2}-b)=1$. Finalmente o campo espinorial evolui como}
\begin{eqnarray}
&\psi=\frac{2}{\sqrt{3m(b+2\beta t +t^{2})}}\left(\begin{tabular}{c}
$0$\\
$\sqrt{\left(\frac{K}{C}+1\right)}e^{i\left(-mt
+\sqrt{\frac{3}{16}}\ln{\left(\frac{m\sqrt{3}(t+\beta)+1}{m\sqrt{3}(t+\beta)-1}\right)}\right)}$\\
$\sqrt{\left(\frac{K}{C}-1\right)}e^{-i\left(-mt
+\sqrt{\frac{3}{16}}\ln{\left(\frac{m\sqrt{3}(t+\beta)+1}{m\sqrt{3}(t+\beta)-1}\right)}\right)}$\\
$0$
\end{tabular}\right).
\end{eqnarray}
\indent {}{ De posse das soluções exatas do caso  $f(R)\equiv R$, vamos usá-las como base para construirmos 
a correção em  $\delta\alpha$ para as soluções do caso mais geral  $f(R)\equiv R+\delta\alpha R^{2}$.}
\\
{}{ A forma do campo espinorial não muda, a mesma é dada em termos da função $\varphi$}
\begin{eqnarray}
\varphi\equiv 1-\delta\alpha\frac{ m C}{\tau},
\end{eqnarray}
com potencial
\begin{eqnarray}
V(\varphi)\equiv \delta\alpha \frac{m^{2} C^{2}}{8\tau^{2}}
\end{eqnarray}
\noindent {}{ e a equação de movimento gravitacional para o volume do Universo é }
\begin{eqnarray}
\left(\ddot\tau-\frac{3mC}{4}\right)
-\delta\alpha\frac{ mC}{2}\left(\frac{\dot\tau^{2}}{\tau^{2}}-\frac{\ddot\tau}{\tau}
-\frac{3mC}{4\tau}\right)=0\,.
\end{eqnarray}
\noindent {}{Integrando a mesma, temos a evolução do volume do Universo}
\begin{eqnarray}
&\tau=\frac{3mC}{8}(b\!+\!2\beta t\!+\!t^{2})
+\delta\alpha \frac{mC}{2}\left(\varsigma\!\left(\xi\!+\!t\right)\!
+\!m\sqrt{3}\left(\beta\!+\!t\right)
\ln{\left(\frac{m\sqrt{3}(t+\beta)+1}{m\sqrt{3}(t+\beta)-1}\right)}\right),
\end{eqnarray}
\noindent {}{onde $\varsigma$ e $\xi$ são duas novas constantes de integração para as quais o volume inicial do Universo é agora diferente, mas ainda sem singularidade.}

\subsection{Gravidade Conforme Torsional}
$\qquad$ Vale a pena apontar alguns progressos recentes no estudo de campos espinoriais em gravidade generalizada, assim como em alguns problemas em aberto 
os quais estão sob investigação em curso. 
Embora seja um pouco à parte do tema principal deste capítulo da tese, é certamente enriquecedor em termos de completude. Nesta linha, outra teoria de gravidade 
de ordem  superior é uma com duas curvaturas, porque é o único caso no qual invariância conforme pode ser obtida \cite{fabbri_conformal_gravity_DM}. Como se constata, 
existem duas maneiras de se implementar transformações conformes para a torção: a primeira é {}{exigir a} transformação mais geral (razoável) para a torção (por ``razoável''
queremos dizer de acordo com o que é discutido por exemplo em \cite{Shapiro:2001rz}). A outra é insistir no fato de que nenhuma transformação conforme é dada para a 
torção (porque transformações conformes são de origem métrica enquanto que a torção é independente de métrica). No caso anterior, como a transformação conforme liga a métrica a torção,
devemos modificar a curvatura de Riemann com termos de torção de traço quadrático para obtermos uma curvatura cuja parte irredutível é conformemente invariante \cite{fabbri_conformal_gravity_DM}.
No último caso, torção e curvatura são separados e essencialmente independentes. Consequentemente, no caso anterior \cite{fabbri_conformal_gravity_DM} as equações de campo estão
intimamente entrelaçadas, enquanto que no último caso as equações de campo são independentes, portanto mantendo o acoplamento curvatura-energia e torção-spin no 
espírito das equações de campo ECSK. Outro enfoque para o tratamento de torção na relatividade geral, são os trabalhos de F. Hehl 
\cite{Hehl:2007bn,Hehl:1976kj,Hehl:1973qn,Hehl2011}.

\subsubsection{Torção com transformações conformes gerais}

$\qquad$ No primeiro caso, o acoplamento com o campo de Dirac foi estudado em \cite{fabbri_conformal_gravity_DM}. Contudo, como neste caso as equações de campo que acoplam torção com spin
não são invertíveis em geral, a torção não pode ser substituída pela densidade de spin nas equações de campo de Dirac, as quais permanecem na forma geral
\begin{eqnarray}
&i\gamma^{\mu}\mathring{D}_{\mu}\psi+\frac{3}{4}W_{\sigma}\gamma^{5}\gamma^{\sigma}\psi=0,
\end{eqnarray}
onde $W_{\sigma}$ é o vetor axial dual da parte completamente anti-simétrica do tensor de torção. \\
\indent Portanto o argumento usado em \cite{fabbri_conformal_gravity_elko} não pode ser 
recuperado {}{e} simetrias esfericamente simétricas são possíveis. Contudo, neste caso, a completa antissimetria do campo de Dirac não implica {}{a} anti-simetria da torção.
Ao invés disso, leva a vínculos para os campos gravitacionais que não podem ser satisfeitos em situações gerais. Neste caso de transformações conformes gerais, o campo de
Dirac parece estar mal definido. Uma situação alternativa é portanto estudar campos Elko, {}{o que} foi realizado em \cite{Fabbri:2011mi}. Contudo, sua dinâmica em termos
de soluções cosmológicas ainda não foram estudadas.

\subsubsection{Torção sem transformações conformes }

$\qquad$ O acoplamento para o campo de Dirac foi estudado em \cite{fabbri_conformal_gravity_DM}, mostrando que a anti-simetria completa da densidade de spin resulta na completa anti-simetria do 
tensor de torção, cujo dual é um vetor axial dado por
\begin{eqnarray}
&W_{\rho}=\left(\frac{4a}{\hbar}\,K^{\mu}K_{\mu}\right)^{-1/3}J_{\rho},
\label{torsionequations}
\end{eqnarray}
de modo que a torção pode ser trocada pela densidade de spin do campo espinorial e a equação de campo de Dirac torna-se
\begin{eqnarray}
&i\gamma^{\mu}\mathring{D}_{\mu}\psi
-\left(\frac{256a}{27}K^{\rho}K_{\rho}\right)^{-\frac{1}{3}} \overline{\psi}\gamma_{\nu}\psi\gamma^{\nu}\psi=0,
\label{matterequations}
\end{eqnarray}
com uma auto-interação não-linear que não obstante é renormalizável \cite{fabbri_ESK}. Após um rearranjo de Fierz imediato, as mesmas podem ser escritas como
\begin{eqnarray}
&i\gamma^{\mu}\mathring{D}_{\mu}\psi
-\left(\frac{27}{256a}\right)^{{1}/{3}}
{}{}\left(\sigma^{2}
{}{}+{}i\omega^2\right)^{-{1}/{3}}
{}\left(\sigma I{}
{}-\omega\gamma_5\right)\psi=0,
\label{matteqarranged}
\end{eqnarray}
claramente mostrando que campos espinoriais do tipo-(4) deveriam obedecer a uma equação de Dirac da forma $i\gamma^{\mu}\mathring{D}_{\mu}\psi=0$, como se a torção
não estivesse presente, precisamente como na teoria ECSK. Neste caso, novamente parece que o raciocínio realizado em \cite{fabbri_conformal_gravity_elko} não se aplica e soluções
esfericamente simétricas são possíveis e as equações do campo gravitacional reduziriam-se a equações de campo de Weyl esfericamente simétricas sem torção num espaço-tempo de
Schwarzschild. 

\subsubsection{Discussão}

$\qquad$ Investigamos campos espinoriais regulares e singulares, estabelecendo um fundo gravitacional com torção geral no qual campos espinoriais {}{são definidos.}
Provamos que alguns campos espinoriais {\it flag-dipole} são soluções físicas da equação de Dirac em teorias ECSK: em particular isso foi obtido numa gravidade $f(\mathit{R})$ mas 
não pode ser recuperado numa gravidade conforme.\\
\indent No caso da cosmologia, quando consideramos campos do tipo Dirac na gravidade $f(\mathit{R})$, a presença de torção impõe o uso de um fundo anisotrópico no qual 
o lado geométrico é diagonal, enquanto que o tensor de energia momento não, devido a características intrínsecas do campo espinorial. Nesta circunstância, a parte
não-diagonal das equações do campo gravitacional resultam nos vínculos \eqref{3.11} caracterizando a estrutura do espaço-tempo, ou a helicidade do campo espinorial, ou ambos.
No nosso entendimento, a única situação fisicamente significativa é uma na qual dois eixos são iguais e a componente espacial do vetor de torção axial não se anula.
A mesma fornece um Universo que é espacialmente um elipsóide de rotação em torno do eixo cuja densidade de spin não é nula \cite{fabbri_ESK}. \\
\indent No caso da gravidade, exceto para o caso de transformações conformes torsionais, para as quais o campo de Dirac parece não estar bem definido, o caso da torção
sem transformações conformes parece ser bem posto. Neste caso, o fundo gravitacional é muito mais do tipo sem torção e embora não tenhamos provado matematicamente,
há razões para se acreditar que um campo espinorial do tipo-(4) deve ainda emergir. \\
\indent Em suma, a presença de torção induz interações não-lineares, cujos detalhes dependem de qual fundo conforme de gravidade $f(\mathit{R})$ é usado, mas em geral tais 
auto-interações induzidas torsionalmente para os espinores afetam a dinâmica do campo espinorial: especificamente, é possível encontrar soluções físicas da equação de 
Dirac as quais não são campos de Dirac, mas {\it flag-dipole}s e portanto singulares. Também encontramos que, além disso, novas soluções envolvem campos espinoriais Elko e 
Majorana, quando a direção de spin se anula, fornecendo um Universo anisotrópico sem interações torsionais fermiônicas. \\
\indent Contudo, acreditamos que a mensagem principal a ser tomada é que um campo espinorial satisfazendo a equação de campo de Dirac não é necessariamente
não-singular: com uma analogia metafórica, podemos dizer que a equação de Dirac não necessariamente cuida de si mesma ao proibir soluções singulares.

%% file: violacao_de_lorentz.tex
\chapter{Viola\c c\~ao da simetria de Lorentz}
$\qquad$ Neste capítulo, construímos uma teoria com violação de Lorentz para um campo eletromagnético (e fermiônico) sujeito a um fundo $f(\mathit{R})$ com torção. 
{}{Dentro} dessa teoria, mostramos que para espinores singulares específicos o acoplamento com a torção é menos sensível 
{}{(no sentido de que os bilineares covariantes aparecem explicitamente nos termos de acoplamento, por isso, para espinores singulares, vários termos são nulos), um resultado original.
Mostraremos que espinores singulares são menos propensos aos efeitos de violação de Lorentz em fundos de Riemann-Cartan. 
De fato, os termos de seus acoplamentos são implementados ao se acoplar os seus bilineares com termos de torção.
Como já vimos que espinores singulares, na classificação de Lounesto, possuem pelo menos três dos seus bilineares nulos, diversos termos 
de violação de Lorentz nas Lagrangianas associadas são identicamente nulos.}
\section{Introdu\c c\~ao}

$\qquad$ Nosso objetivo principal neste capítulo é o estudo do papel da violação da simetria de Lorentz (VSL) em teorias contendo campos espinoriais singulares em espaços-tempo de Riemann-Cartan,
{}{ que já são ambientes naturais para violação de Lorentz}. 
Uma vez que  as densidades de Lagrangiana de Palatini e Einstein-Hilbert são equivalentes, também no contexto da classifica\c c\~ao de espinores de Lounesto\,\cite{daRocha:2007sd},
curvatura e tor\c c\~ao são descrições equivalentes  do campo gravitacional. O tensor de energia-momento da matéria é a fonte da curvatura, 
no caso da relatividade geral e da tor\c c\~ao
no caso da gravidade teleparalela \cite{NotteCuello:2006ek}. Além disso, o acoplamento tor\c c\~ao-spin pode ser considerado, além do acoplamento curvatura-energia 
{}{e o fundo} com tor\c c\~ao viola a  invariância de Lorentz local efetiva\,\cite{Kostelecky:2007kx}. A geometria de Riemann-Cartan é a configura\c c\~ao para a teoria de
Einstein-Cartan, contudo, teorias de gravita\c c\~ao mais gerais em espaços-tempo de Riemann-Cartan podem incorporar \emph{vierbein} propagantes e conexões de spin, 
descrevendo tor\c c\~ao e curvatura dinâmicas\,\cite{waldyr,daRocha:2007sd}. Já vimos uma parte desse aspecto no capítulo 3, mas aqui gostaríamos de
considerar o acoplamento da tor\c c\~ao com os campos da matéria, {}{sendo que} usualmente os campos fermiônicos de Dirac são utilizados na literatura. Nesta configura\c c\~ao,
os efeitos da tor\c c\~ao manifestam-se como auto-interações, capazes de fornecer uma explica\c c\~ao dinâmica para o princípio da exclusão\,\cite{Fabbri:2012qr}. 
Efeitos da tor\c c\~ao influenciam a dinâmica próxima ou na escala de Planck, onde efeitos de quebra de invariância de Lorentz poderiam ser relevantes. Além disso,
tor\c c\~ao induz uma orienta\c c\~ao preferencial para um observador em queda livre, realizada como uma manifesta\c c\~ao de viola\c c\~ao de Lorentz local. Portanto, vínculos na simetria de
Lorentz levam a vínculos na tor\c c\~ao\,\cite{Kostelecky:2007kx}. Queremos analisar o acoplamento da tor\c c\~ao com campos da matéria além de campos fermiônicos de Dirac, 
abrangendo na análise o conjunto completo de campos espinoriais singulares, em particular campos espinoriais {\it flagpole} e {\it flag-dipole}. Uma das nossas motivações é que 
recentemente a geometria de Riemann-Cartan lançou uma nova luz nos papéis proeminentes de espinores singulares: {}{mostramos no capítulo anterior} que a teoria de 
Einstein-Cartan-Sciama-Kibble acoplada a espinores admite soluções que não são campos espinoriais de Dirac, mas sim {\it flag-dipole}s\,\cite{fabbri_ESK}. 
Também temos em \cite{roldao_3}, espinores do tipo {\it flag-dipole} são soluções num fundo Kerr, induzindo uma estrutura de fluxo de fluido.\\
\indent Independentemente da sua origem em alguma teoria de alta energia, a viola\c c\~ao de Lorentz é usualmente realizada numa estrutura de teoria de campo efetiva. 
A assim chamada Extensão do Modelo Padrão (EMP) \cite{Mariz:2016ooa} contem a relatividade geral (GR) e o Modelo Padrão (MP). Termos dominantes da a\c c\~ao EMP contem a gravidade pura e ações MP
minimamente acopladas, junto com todos os termos de ordem {}{dominantes} dos campos gravitacionais e MP, acoplados com tensores de fundo constantes. 
A geometria de Riemann-Cartan permite quantidades com valores esperados no vácuo não nulos que violam a invariância de Lorentz,  embora preservando invariância de
coordenadas gerais, abrangendo acoplamentos gravitacionais mínimos de espinores. Mostraremos que, neste contexto, campos espinoriais {\it flagpole} são exemplos de campos espinoriais 
singulares que não acoplam minimamente com a tor\c c\~ao. \\
\indent É bem conhecido que nem todos os parâmetros VSL no EMP possuem significado físico\,\cite{Colladay:1996iz}, no sentido de que alguns deles podem ser absorvidos por
uma redefini\c c\~ao dos campos. Além disso, uma corrente conservada devidamente definida satisfaz a álgebra de Poincaré usual, pelo menos tanto quanto estes coeficientes 
são concernidos. Este ponto foi extensivamente trabalhado em\,\cite{Colladay:2002eh}, onde um procedimento sistemático para eliminar coeficientes espúrios VSL e definir
correntes conservadas foi desenvolvido para o caso da eletrodinâmica quântica (EDQ). As implicações deste procedimento para as teorias envolvendo campos espinoriais mais gerais ainda está faltando e
esta é outra questão que abordamos neste trabalho. 
\section{\label{sec:Coupling-singular-spinors} Acoplando espinores singulares com a tor\c c\~ao numa estrutura com viola\c c\~ao de Lorentz}
$\qquad$ Até as equações \eqref{dirac21}, precisamos escrever as equações \eqref{Elko11} pelo seu repetido uso e sua importância na apresentação deste capítulo.
Optamos escolher o didatismo em tal apresentação, em detrimento da repetição da fórmula \eqref{dirac21}.
Por conveniência de leitura, repetimos aqui o resumo sobre a classificação de Lounesto de campos espinoriais. \\
\indent Consideremos o conjunto dos campos espinoriais no espaço-tempo de Minkowski \linebreak $M\simeq\mathbb{R}^{1,3}$.
Dadas seções do fibrado de referenciais ${P}_{\mathrm{SO}_{1,3}^{e}}(M)$, com base dual $\{\mathbf{{\rm e}^{\mu}}\}$, campos espinoriais clássicos carregando uma 
representa\c c\~ao 
{}{$\pi = $\, ${(1/2,0)}\oplus{(0,1/2)}$} da componente conexa do grupo de Lorentz na identidade $\mathrm{SL}(2,\mathbb{C)}\simeq\mathrm{Spin}_{1,3}^{e}$ são seções 
do fibrado vetorial associado ${P}_{\mathrm{Spin}_{1,3}^{e}}(M)\times_{\pi}\mathbb{C}^{4}.$ Denotando por  $\{\gamma^{\mu}\}$ as matrizes gama, os bilineares covariantes
são dados por
\begin{align}
\sigma & =\bar{\psi}\psi,\nonumber \\
\mathbf{J} & =J_{\mu}{\rm e}^{\mu}=\bar{\psi}\gamma_{\mu}\psi{\rm e}^{\mu},\nonumber \\
\mathbf{S} & =S_{\mu\nu}\;{\rm e}^{\mu}\wedge{\rm e}^{\nu}=\frac{1}{2}\bar{\psi}i\gamma_{\mu\nu}\psi\;{\rm e}^{\mu}\wedge{\rm e}^{\nu},\nonumber \\
\mathbf{K} & =K_{\mu}\;{\rm e}^{\mu}=\bar{\psi}\gamma_{5}\gamma_{\mu}\psi\;{\rm e}^{\mu},\nonumber \\
\omega & =\bar{\psi}\gamma_{5}\psi\thinspace,\label{fierz}
\end{align}
onde $i\gamma^{5}=\gamma^{0}\gamma^{1}\gamma^{2}\gamma^{3}$. Exclusivamente na teoria de Dirac do elétron, $\mathbf{J}$ é interpretada como sendo a densidade de corrente,
$\mathbf{K}$  fornece a dire\c c\~ao do spin do elétron e $\mathbf{S}$ relacionada com a distribui\c c\~ao do momento angular intrínseco. Esta interpreta\c c\~ao física está ausente 
nos casos mais gerais. Sempre que $\omega=0=\sigma$, o campo espinorial é dito ser singular, caso contrário é um espinor regular. Os bilineares covariantes para espinores
regulares satisfazem as identidades de Fierz,
\begin{eqnarray}
\mathbf{K}\cdot\mathbf{J}=0\,,\qquad\mathbf{S}=(\omega+\sigma i\gamma_{5})^{-1}\mathbf{K}\wedge\mathbf{J}\,,\qquad\mathbf{J}^{2}=\omega^{2}+\sigma^{2}=-\mathbf{K}^{2}\,.
\end{eqnarray}
\indent A classifica\c c\~ao de campos espinoriais de Lounesto é baseada em seis classes disjuntas de campos espinoriais~\cite{lounesto},
 \begin{eqnarray}
1)\;\;\sigma\neq0,\;\;\;\omega\neq0,\;\;\;\mathbf{J}\neq0\qquad{}\qquad{}\qquad{}4)\;\;\sigma=0=\omega,\;\;\;\mathbf{S}\neq0,\;\;\mathbf{K}\neq0,\;\;\mathbf{J}\neq0\label{Elko11}\\
2)\;\;\sigma\neq0,\;\;\;\omega=0,\;\;\;\mathbf{J}\neq0\qquad{}\qquad{}\qquad{}5)\;\;\sigma=0=\omega,\;\;\;\mathbf{S}\neq0,\;\;\mathbf{K}=0,\;\;\mathbf{J}\neq0\label{tipo41}\\
\!\!\!3)\;\;\sigma=0,\;\;\;\omega\neq0,\;\;\;\mathbf{J}\neq0\qquad{}\qquad{}\qquad{}\!6)\;\;\sigma=0=\omega,\;\;\;\mathbf{S}=0,\;\;\mathbf{K}\neq0,\;\;\mathbf{J}\neq0\label{dirac21}
\end{eqnarray}
\indent Os três primeiros tipos de campos espinoriais, caracterizados por $\mathbf{J},\mathbf{K},\mathbf{S}\neq0$, são regulares. Os três últimos são conhecidos, respectivamente, 
como campos espinoriais {\it flag-dipole}, {\it flagpole} e dipole. Vale a pena mencionar que o paradigma da classifica\c c\~ao de Lounesto é em rela\c c\~ao a ${\bf J}\neq0$. O mecanismo
que gera três classes adicionais tem sido proposto em\,\cite{EPJJC}, incluindo o caso ${\bf J}=0$, implicando os espinores subsequentes não {}{possuirem} de forma canônica 
dimensão de massa um. Isto permite a constru\c c\~ao de novos espinores de dimensão de massa um, por exemplo. {}{Espinores tipo-$5$ que caracteriza os espinores 
{\it flagpoles} possuem subconjuntos que englobam espinores neutros \footnote{Por espinores neutros, denotamos os que carregam carga eletromagnética.}, como Elko e Majorana e 
espinores que carregam carga eletromagnética como por exemplo, soluções da equação de Dirac \cite{fabbri_ESK,daRocha:2016bil}} \\
\indent Neste ponto, revemos como VSL pode ser introduzida de um modo geral numa teoria tal como a EDQ e como algo de VSL então introduzida é realmente espúria,
sendo eliminada por uma redefini\c c\~ao adequada do campo. Estaremos interessados na parte fermiônica livre da Lagrangiana da EDQ, a qual no caso 
com violação de Lorentz (VL) é escrita como 
\begin{align}
\mathcal{L}^{\text{LV-EDQ}} & ={\frac{i}{2}\bar{\psi}\Gamma^{\nu}\overset{\leftrightarrow}{\nabla_{\nu}}\psi}-\bar{\psi}M\psi\thinspace,\label{eleclagran}
\end{align}
onde $\nabla_{\mu}=\partial_{\mu}+iqA_{\mu}$ é a derivada covariante e
\begin{eqnarray}
\Gamma_{\nu} & := & \gamma_{\nu}+c_{\mu\nu}\gamma^{\mu}+d_{\mu\nu}\gamma_{5}\gamma^{\mu} + \mathfrak{e}_{\nu}+if_{\nu}\gamma_{5}+\frac{1}{2}g_{\rho\mu\nu}\sigma^{\rho\mu}\,,\label{gamaLV}\\
M & := & m+im_{5}\gamma_{5}+a_{\mu}\gamma^{\mu}+b_{\mu}\gamma_{5}\gamma^{\mu}+\frac{1}{2}H_{\mu\nu}\sigma^{\mu\nu}\thinspace,\label{mmm}
\end{eqnarray}
$m$ sendo a massa do elétron, $\sigma^{\mu\nu}=\frac{i}{2}[\gamma^{\mu},\gamma^{\nu}]$ e $a,b,c,d,\mathfrak{e},f,g,m_{5},H$ são tensores constantes reais os quais parametrizam a VSL.
Por hipótese, a a\c c\~ao\,\eqref{eleclagran} é Hermitiana, portanto restringindo os coeficientes da viola\c c\~ao de Lorentz  serem reais. Alguns destes parâmetros possuem um 
forte limite experimental/fenomenológico, como discutido em\,\cite{Kostelecky:2007kx}. Contudo, alguns parâmetros nas equações \,\eqref{gamaLV} e\,\eqref{mmm} podem ser eliminados
por uma redifini\c c\~ao adequada do campo. \\
\indent De fato, um espinor $\chi$ que satisfaz uma Lagrangiana da EDQ estendida VSL, pode ser obtido a partir de um espinor $\psi$, o qual é solu\c c\~ao de uma Lagrangiana da EDQ 
padrão que é Lorentz invariante (LI), através da transforma\c c\~ao
\begin{eqnarray}
\psi &=& (\mathbf{1}+f(x^{\mu},\partial_{\nu}))\chi \nonumber \\
&=&\chi+(v\cdot\Gamma+i{\rm \theta}+i\tilde{C}_{\mu}x^{\mu}+C_{\mu\nu}x^{\mu}\partial^{\nu}+B_{\mu}\partial^{\mu}+\gamma_{5}\tilde{B}_{\mu}\partial^{\mu})\chi\thinspace,\label{spintrans}
\end{eqnarray}
onde $f(x,\partial)$ representa uma fun\c c\~ao matricial geral  $4\times4$ das coordenadas e derivadas\,\cite{Colladay:2002eh}. Aqui, $v\Gamma=v_{I}\Gamma_{I}$, $\Gamma_{I}$ é uma base
para ${\cal {M}}(4,\mathbb{C})$ para um índice composto $I\in\{{\scriptstyle \emptyset,\mu,\mu\nu,\mu\nu\rho,{5}}\}$, onde $\Gamma_{\emptyset}=\boldsymbol{1}$. Além disso, 
${\rm \theta},\thinspace \tilde{C}_{\mu},\thinspace B_{\mu},\thinspace\tilde{B}_{\mu},\thinspace C_{\mu\nu}$ são coeficientes escalares. Os parâmetros $\Re({\rm \theta})$, $B_{\mu}$
e $C_{[\mu\nu]}$ em particular correspondem as simetrias $U(1)$ e de Poincaré da Lagrangiana padrão. Apenas os termos de ordem mais baixa na redefini\c c\~ao do campo são mantidos,
desde que os parâmetros VL sejam assumidos como sendo pequenos. Estas redefinições podem ser tomadas como {\it  mixings} dependentes do ponto das componentes no espaço espinorial.\\
\indent Por simplicidade, {}{citamos o} um resultado da redefini\c c\~ao parametrizada pelo parâmetro $v$, a qual foi descrita em\,\cite{Colladay:1996iz}. Começamos com uma Lagrangiana
explicitamente LI
\begin{equation}
\mathcal{L}=\frac{i}{2}\bar{\psi}\gamma^{\mu}\stackrel{\leftrightarrow}{\partial_{\mu}}\psi-m\bar{\psi}\psi\thinspace,\label{eq:LLI}
\end{equation}
a qual é reescrita via a redefini\c c\~ao\,\eqref{spintrans} como 
\begin{align}
\mathcal{L} & =\frac{i}{2}\bar{\chi}\gamma^{\mu}\stackrel{\leftrightarrow}{\partial_{\mu}}\chi-m\bar{\chi}\chi+\frac{i}{2}\bar{\chi}[\{\gamma^{\mu},\Gamma\cdot{\rm \Re}v\}+i[\gamma^{\mu},\Gamma\cdot{\rm \Im}v]]\stackrel{\leftrightarrow}{\partial_{\mu}}\chi-2m{\rm \Re}v\cdot\bar{\chi}\Gamma\chi\thinspace.\label{eq:LLV}
\end{align}
\indent O ponto essencial é que ambas as Lagrangianas descrevem a mesma física, portanto mesmo {}{que} a equa\c c\~ao\,\eqref{eq:LLV} inclua termos que possam parecer violar a invariância de Lorentz,
esta teoria é realmente invariante por Lorentz. De fato, podemos definir geradores de Poincaré devidamente modificados em termos de $\chi$ que satisfazem a álgebra de Poincaré
usual. Este exemplo mostra que a redefini\c c\~ao de campo geral descrita pela equa\c c\~ao \,\eqref{spintrans} pode eliminar alguns dos coeficientes numa teoria efetiva com VL geral
como a EMP, desde que estes coeficientes são realmente inobserváveis em primeira ordem numa fenomenologia relacionada a EMP. \\
\indent Para simplificar ainda mais, assumimos que $v\cdot\Gamma=v_{\mu}\gamma^{\mu}$. Denotando a Lagrangiana livre  convencional  para $\chi$ por 
\begin{equation}
\mathcal{L}_{0}=\frac{i}{2}\bar{\chi}\Gamma^{\nu}\stackrel{\leftrightarrow}{\partial_{\nu}}\chi-m\bar{\chi}\chi\thinspace,
\end{equation}
a equa\c c\~ao\,\eqref{eleclagran} pode ser escrita como
\begin{eqnarray}
{\mathcal{L}}={\mathcal{L}}_{0}+{\Re}v_{\mu}[i\bar{\chi}\stackrel{\leftrightarrow}{\partial^{\mu}}\chi-2m\bar{\chi}\gamma^{\mu}\chi]-i{\Im}v_{\mu}[\bar{\chi}\sigma^{\mu\nu}\stackrel{\leftrightarrow}{\partial_{\nu}}\chi].\label{aeredef}
\end{eqnarray}
\indent Comparando com \eqref{gamaLV}, a escolha simultânea de parâmetros VSL $\mathfrak{e}^{\mu}=2{\Re}v^{\mu}$  e  \linebreak 
$a^{\mu}=2m{\Re}v^{\mu}$ podem ser inteiramente {}{atribuída} à redefini\c c\~ao 
do campo e portanto não {}{introduz} uma VSL real. {}{Numa} perspectiva diferente, pode-se dizer que na teoria VL definida por\,\eqref{eleclagran}, ou  $\mathfrak{e}^{\mu}$ ou $a^{\mu}$
podem ser eliminados via uma redefini\c c\~ao do campo. O termo ${\Im}v_{\mu}$ indica que a escolha $2{\Im}\,v_{[\mu}g_{\rho]\nu}=g_{\rho\mu\nu}$ elimina os termos proporcionais a
$g_{\rho\mu\nu}$ em\,\eqref{eleclagran}. Os parâmetros  $m_{5}$, $a_{\mu}$, $\mathfrak{e}_{\mu}$, $f_{\mu}$ e  $c_{[{\mu\nu}]}$ em\,\eqref{eleclagran} também podem ser removidos
\,\cite{Colladay:2002eh}. \\
\indent Agora revisamos como esta discussão pode ser estendida para um contexto Riemann-Cartan com tor\c c\~ao \cite{Arcos:2005ec}. Utilizando índices latinos para rotular coordenadas de Lorentz e 
índices gregos para coordenadas do espaço-tempo, a métrica de Minkowski é relacionada à métrica $g_{\mu\nu}$  do  espaço-tempo curvo via o \emph{vierbein}
$e_{\mu}^{\,\, a}$, pela rela\c c\~ao $g_{\mu\nu}=\vb\mu a\vb\nu b\et_{ab}$.
O determinante do \emph{vierbein} é denotado por $e$ e a carga do elétron é denotada por $-q$. Para a derivada covariante do espaço-tempo, a conexão é assumida como sendo 
compatível com a métrica. Além disso, índices do espaço-tempo curvo são corrigidos pela conexão de Cartan  $\Ga_{\pt\la\mu\nu}^{\la}$, isto é, 
\begin{equation}
\nabla_{\mu}\vb\nu a=\prt_{\mu}\vb\nu a+\lulsc\mu ab\vb\nu b-\Ga_{\pt\al\mu\nu}^{\al}\vb\al a\thinspace.
\end{equation}
\indent A conexão de Cartan pode ser escrita como $\Ga_{\pt\la\mu\nu}^{\la}=\dot{\Ga}_{\pt\la\mu\nu}^{\la}+\half T_{\pt\la\mu\nu}^{\la}$, na qual $\dot{\Ga}_{\pt\la\mu\nu}^{\la}$
é o símbolo de Christoffel e $T_{\pt\la\mu\nu}^{\la}=-T_{\pt\la\nu\mu}^{\la}$ é o tensor de torção. O tensor de contorção é definido como 
\begin{equation}
K_{\mu\nu}^{\lambda} = \frac{1}{2}(T_{\mu\nu}^{\,\,\,\lambda} - T_{\nu\,\,\mu}^{\,\lambda} - T_{\,\,\,\mu\nu}^{\lambda})\thinspace,
\end{equation}
e a curvatura como $R_{\pt\ka\la\mu\nu}^{\ka}=
\mathring{R}_{\pt\ka\la\mu\nu}^{\ka}+\nabla_{[\mu}K_{\pt{\ka}\nu]\la}^{\ka}+K_{\pt\al[\mu\nu]}^{\al}K_{\pt{\ka}\al\la}^{\ka}+K_{\pt\al[\mu\la}^{\al}K_{\pt{\ka}\nu]\al}^{\ka}$,
onde $\mathring{R}_{\pt\ka\la\mu\nu}^{\ka}$ denota o tensor de curvatura de Riemann usual na ausência de torção. A fonte da contorção pode ser considerada como um campo de
Kalb-Ramond $B_{\alpha\beta}$, através de $K_{\;\,\alpha\beta}^{\rho}=-\frac{1}{\kappa^{3/2}}H_{\;\,\alpha\beta}^{\rho}$, onde $H_{\rho\alpha\beta}=\partial_{[\rho}B_{\alpha\beta]}$
e $\kappa$ denota a constante de acoplamento\,\cite{HoffdaSilva:2008hc}. Contudo, nossa discussão não dependerá desta identificação, sendo uma contorção geométrica genérica
considerada de agora em diante. \\
\indent Assinaturas da torção, no contexto de acoplamentos minimais e não minimais com férmions, são fenomenologicamente e experimentalmente abundantes \,\cite{Kostelecky:2007kx}.
O acoplamento minimal entre torção e campos MP é realizado através de derivadas covariantes. Não obstante, acoplamentos não minimais são também uma opção. De fato, provaremos que,
quando espinores {\it flagpole} são considerados, acoplamentos não minimais são a única possibilidade. \\
\indent As variáveis essenciais aqui são o \emph{vierbein} e a conexão de spin, uma vez que outras variáveis tais como curvatura e torção podem ser expressas em termos dessas.
Por exemplo, a conexão de Cartan é dada por $\Ga_{\pt\la\mu\nu}^{\la}=\uvb\la a(\prt_{\mu}\lvb\nu a-\lulsc\mu ba\lvb\nu b),$ enquanto que a torção é dada por 
$T_{\la\mu\nu}=\vb\la a(\prt_{[\mu}e_{\nu]a}+\omega_{[\mu|ab|}e_{\nu]}^{b})$. Além disso, a conexão de spin é relacionada com o \emph{vierbein} por 
\begin{eqnarray}
\nsc\mu ab & = & \half e^{\nu[a}\prt_{[\mu}e_{\nu]}^{\;b]}-\half\uvb\al a\uvb\be b\vb\mu c\prt_{[\al}e_{\be]c}+K_{\nu\mu\la}\uvb\nu a\uvb\la b.
\end{eqnarray}
\indent De agora em diante, consideramos campos gravitacionais fracos, $g_{\mu\nu}=\et_{\mu\nu}+h_{\mu\nu},$ onde $h_{\mu\nu}$ é uma flutuação. Em termos de primeira ordem, 
o \emph{vierbein} e a conexão de spin podem ser expressos em termos de quantidades pequenas, 
\begin{eqnarray}
\lvb\mu a & = & \et_{\mu a}+\ep_{\mu a}\approx\et_{\mu a}+\half h_{\mu a}+\ch_{\mu a},\qquad\quad e\approx1+\half h,\\
\lsc\mu ab & \approx & -\half\prt_{a}h_{\mu b}+\half\prt_{b}h_{\mu a}+\prt_{\mu}\ch_{ab}+K_{a\mu b}.\label{123321}
\end{eqnarray}
\indent Os campos básicos não gravitacionais para as violações de Lorentz e CPT da extensão da EDQ num espaço-tempo de Riemann-Cartan são um campo fermiônico $\ps$ e o campo
do fóton $A_{\mu}$. A ação para a teoria pode ser expressa como uma soma de ações parciais para o férmion, para o fóton e para a gravidade. A parte do férmion da ação contém 
termos que são dominantes em baixas energias, envolvendo férmions e seus acoplamentos mínimos com fótons e gravidade. Em geral, também devemos considerar termos de ordens 
superiores envolvendo férmions e fótons que são não renormalizáveis, não minimais e de ordem superior nos acoplamentos gravitacionais, assim como operadores de campo de
dimensão maior do que quatro que acoplam curvatura e torção com os campos da matéria e do fóton. \\
\indent A parte fermiônica da ação \eqref{eleclagran} para a extensão da EDQ pode ser escrita como   
\begin{equation}
S = \int d^ 4x\,e\left ( \half i\ivb\mu a\bar{\ps}\Ga^{a}\lrDmu\ps-\bar{\ps}M\ps\right ),\label{qedxps}
\end{equation}
onde a derivada covariante usual $U(1)$ é dada por $\nabla_{\mu}\ps\equiv\prt_{\mu}\ps+\frac{1}{4}i\nsc\mu ab\si_{ab}\ps-iqA_{\mu}\ps$\,\cite{fabbri_fr_torsion}.
Além disso, as equações \,\eqref{gamaLV} e \,\eqref{mmm} se expressam, respectivamente, em termos da estrutura multivetorial do \emph{vierbein}, 
\begin{eqnarray}
\Ga^{a} & = & \ga^{a} - c_{\mu\nu}\uvb\nu a\ivb\mu b\ga^{b} + d_{\mu\nu}\ivb\mu b\uvb\nu a\ga^{b}\ga_{5} - \mathfrak{e}_{\rho}\uvb\rho a \nonumber \\
&& - if_{\mu}\ga_{5}\uvb\mu a-\half g_{\la\mu\nu}\ivb\la b\uvb\nu a\mathfrak{e}^\mu c\si^{bc}\thinspace,\label{gamdef}\\
M & = & m+a_{\mu}\ivb\mu a\ga^{a}+\half H_{\mu\nu}\ivb\mu a\ivb\nu b\si^{ab}+b_{\mu}\ivb\mu a\ga_{5}\ga^{a}+im_{5}\ga_{5}.\label{mdef}
\end{eqnarray}
Portanto a equação de Dirac em espaços-tempo de Riemann-Cartan é dada por
\begin{eqnarray}
 i\ivb\mu b\Ga^{b}\nabla_{\mu}\ps+\half\ivb\mu a\nsc\mu bc(i\et_{\pt{a}b}^{a}\Ga_{c}-\frac{1}{4}[\si_{bc},\Ga^{a}])\ps-\half i\tor\rho\rho\mu\ivb\mu a\Ga^{a}\ps-M\ps=0.\label{direq}
\end{eqnarray}
\indent Os termos VL envolvendo $M$ contribuem para a equação de Dirac de um modo minimal, como assumido nos acoplamentos não derivados. Não obstante, estes termos envolvendo 
$\Ga^{a}$ emergem como minimais e através da comutação com os geradores de Lorentz na derivada covariante. Em particular, as partes Lorentz invariantes dos termos na equação
\eqref{direq} cancelam-se.\\
\indent Correções à violação de Lorentz (VL) podem ser exploradas no contexto de campos espinoriais singulares. De fato, as extensões da EDQ nos espaços-tempo de Minkowski e
Riemann-Cartan diferem por acoplamentos gravitacionais fracos. Neste regime, dado pela equação\,\eqref{123321}, os termos lineares da Lagrangiana são dados por
\begin{equation}\label{lagapprox}
\cl_{\ps}\sim-i(c_{{\rm efetivo}})_{\mu\nu}\bar{\ps}\ga^{\mu}\prt^{\nu}\ps-(b_{{\rm efetivo}})_{\mu}\overbrace{\bar{\ps}\ga_{5}\ga^{\mu}\ps}^{=\,K^\mu},
\end{equation}
onde  
\begin{eqnarray}
(c_{{\rm efetivo}})_{\mu\nu} \equiv c_{\mu\nu}+\ch_{\mu\nu}-\half h_{\mu\nu}\,\,\,,(b_{{\rm efetivo}})_{\mu}\equiv b_{\mu}+\frac{1}{8}\ep_{\mu\nu\rho\sigma}T^{\nu\rho\sigma}-\frac{1}{4}\prt^{\nu}\ch^{\rho\sigma}\ep_{\mu\nu\rho\sigma}.\label{effcoeff}
\end{eqnarray}
\indent Nesta expressão, os termos de ordem principal decorrentes do escalonamento do determinante $e$ do \emph{vierbein} são negligenciados, por serem LI.
As equações \,\eqref{effcoeff} mostram que, em primeira ordem, uma métrica de fundo fraca é governada por $c_{\mu\nu}$, enquanto que  a torção é efetivamente governada
por um termo $b_{\mu}$\,\cite{Shapiro:2001rz}. O termo  $b_{\mu}$ é um termo de violação CPT, portanto a presença de uma torção de fundo pode imitar uma violação CPT. 
Se a Lagrangiana\,\eqref{lagapprox} modela um campo espinorial {\it flagpole}, que satisfaz $K^\mu = \bar{\ps}\ga_{5}\ga^{\mu}\ps=0$, o termo 
$b_{{\rm efetivo}}$ é irrelevante em tal modelo.
Portanto, os férmions {\it flagpole} não {}{são} sensíveis a esse tipo de VSL. \\
\indent Considerando a aplicação entre campos espinoriais LI e VSL dada pela equação\,\eqref{spintrans}, eles podem ser usados para mostrar que, em primeira ordem nos 
coeficientes para a violação de Lorentz, não há efeitos físicos dos coeficientes $\mathfrak{e}_\mu$, \f, ou da parte anti-simétrica de $c_{\mu\nu}$. \\
\indent Quando acoplamentos gravitacionais não minimais são levados em conta, operadores de dimensão de massa quatro ou menos podem ser analisados. Na extensão da EDQ, 
tais operadores não minimais não são acoplamentos principais e os únicos acoplamentos de calibre são produtos da torção com bilineares covariantes de férmions. 
Os invariantes de Lorentz possíveis são
\begin{eqnarray}
\cl_{{\rm LI}}\! & = & \!(aT_{\pt{\la}\la\mu}^{\la} + a_5\ep_{\mu\nu\rho}\,T^{\mu\nu\rho})\overbrace{\bar{\ps}\ga^{\mu}\ps}^{=\, J^\mu}+
(bT_{\pt{\la}\la\sigma}^{\la} + b_{5}T^{\mu\nu\rho}\ep_{\mu\nu\rho\sigma})\overbrace{\bar{\ps}\ga_{5}\ga^{\mu}\ps}^{=\,K^\sigma}.\label{LLI}
\end{eqnarray}
\indent O acoplamento $b_{5}$ é minimal, enquanto os outros são não minimais. As possibilidades com VL são
\begin{eqnarray}
\cl_{{\rm VL}} & \!=\! & k_{\mu\nu\rho}T^{\mu\nu\rho}\overbrace{\bar{\ps}\ps}^{=\,\sigma}\! + 
\!k_{\mu\nu\rho\sigma}T^{\mu\nu\rho}\overbrace{\bar{\ps}\ga^{\sigma}\ps}^{=\,J^ \sigma} + 
k_{\mu\nu\rho\sigma\tau}T^{\mu\nu\rho}\overbrace{\bar{\ps}\si^{\sigma\tau}\ps}^{-iS^{\sigma\tau}}\! + 
\!k_{5\mu\nu\rho\sigma}T^{\mu\nu\rho}\overbrace{\bar{\ps}\ga_{5}\ga^{\sigma}\ps}^{K^\sigma}\!\nonumber \\
 &  & \qquad+k_{5\mu\nu\rho}T^{\mu\nu\rho}\overbrace{\bar{\ps}\ga_{5}\ps}^{\omega}.\label{LLV}
\end{eqnarray}
\indent As $k_{\mu \gamma \alpha \beta }$ são funções dependendo do ponto, adicionalmente as mesmas devem possuir as simetrias do tensor
de curvatura de  Riemann  \cite{Colladay:1998fq}. No nosso caso, usamos a equação (33) de \cite{kostelecky4}, com a seguinte forma 
$k_{\mu \gamma \alpha \beta }=\frac{1}{2}(\eta _{\mu \alpha
}h_{\gamma \beta }-\eta _{\gamma \alpha }h_{\mu \beta }+\eta
_{\gamma \beta }h_{\mu \alpha }-\eta_{\mu \beta }h_{\gamma
\alpha })$, onde $h_{\mu\nu}$ é a métrica de fundo de campo fraco. \\
\indent Se a violação de Lorentz é suprimida e a torção é pequena também, então todos os termos na equação\,\eqref{LLV} são subdominantes. Não obstante, todos os outros operadores
acima podem ser de interesse em cenários mais exóticos. Por exemplo, a presença de um dubleto de Higgs na EMP permite outros tipos de acoplamentos gravitacionais
não minimais de dimensão quatro ou mais, incluindo aqueles envolvendo tanto curvatura quanto torção. Operadores de dimensão maior que quatro genericamente 
{}{vêm} com a 
supressão da escala de Planck\,\cite{Kostelecky:2007kx}. Portanto, efeitos de operadores invariantes de Lorentz de dimensão quatro suprimidos pelo inverso da massa de
Planck $m_{P}$ são comparáveis, em magnitude, com aqueles de um operador de dimensão quatro envolvendo um coeficiente para VL suprimido por $m_{P}$ por exemplo. \\
\indent Vale a pena enfatizar que, quando campos espinoriais {\it flagpole} são levados em conta, {}{pois $K^\mu = 0$ em \eqref{LLI} e \eqref{LLV}. Portanto,}
férmions {\it flagpole} possuem um alcance restrito de acoplamento, quando comparados
com férmions de Dirac.\\
\indent Levando-se em conta a classificação de Lounesto, já sabemos que numa geometria de Riemann-Cartan numa configuração conforme do tipo $f(R)$, campos espinoriais
{\it flag-dipole} são soluções da equação de Dirac\,\cite{fabbri_ESK}. Neste contexto, {\it flagpoles} também {}{desempenham} um papel proeminente quando a torção é levada em conta, 
{}{pois de forma análoga ao parágrafo anterior, $K^\mu = 0$ em \eqref{LLI} e \eqref{LLV}. Portanto,}
campos espinoriais {\it flagpole} são exemplos de campos espinoriais singulares os quais provamos como sendo os menos sensíveis a violação de Lorentz.\\
\indent Campos da matéria fermiônicos em espaços-tempo de Riemann-Cartan podem ser governados por uma Lagrangiana com acoplamentos de torção arbitrários. Numa configuração
com aproximação de torção constante, os acoplamentos de torção podem ser substituídos por soluções de fundo para as equações da torção. A densidade Lagrangiana efetiva 
correspondente é dada por\,\cite{Kostelecky:2007kx} 
\begin{eqnarray}
\cl & \sim & \half i\bar{\ps}\ga^{\mu}\stackrel{\leftrightarrow}{\partial_{\mu}}\ps-m\bar{\ps}\ps+\cl_{{\rm LI(4)}}+\cl_{{\rm LI(5)}}\thinspace,
\end{eqnarray}
onde todos os acoplamentos independentes constantes na torção de dimensão de massa quatro ou cinco são respectivamente dadas por
\begin{eqnarray}
\cl_{{\rm LI(4)}} &=&(a_{1}T_{\;\rho\mu}^{\rho}+a_{3}\mathfrak{A}_{\mu})\overbrace{\bar{\ps}\ga^{\mu}\ps}^{=\,J^\mu}+
(a_{2}T_{\;\rho\mu}^{\rho}+a_{4}\mathfrak{A}_{\mu})\overbrace{\bar{\ps}\ga_{5}\ga^{\mu}\ps}^{=\,K^\mu}+\half i\mathring{a}_{1}T^{\mu}\bar{\ps}\stackrel{\leftrightarrow}{\partial_{\mu}}\ps\label{lag1} \\
\cl_{{\rm LI(5)}} &=&\half\left(\mathring{a}_{2}T^{\mu}+\mathring{a}_{4}\mathfrak{A}^{\mu}\right)\bar{\ps}\ga_{5}\stackrel{\leftrightarrow}{\partial_{\mu}}\ps 
\half i\mathring{a}_{3}\mathfrak{A}^{\mu}\bar{\ps}\stackrel{\leftrightarrow}{\partial_{\mu}}\ps \nonumber \\
&+&\half i\left(\mathring{a}_{5}{M^{\nu}}_{\mu\lambda}+\mathring{a}_{6}T_{\;\rho\mu}^{\rho}+\mathring{a}_{7}\mathfrak{A}_{\mu}\right)\bar{\ps}\stackrel{\leftrightarrow}{\partial_{\nu}}\si^{\mn}\ps \nonumber \\
&+&\half i\left(\mathring{a}_{8}\cep^{\la\ka\mn}T_{\rho\lambda}^{\rho}
+\mathring{a}_{9}\cep^{\la\ka\mn}\mathfrak{A}_{\la}\right)\bar{\ps}\stackrel{\leftrightarrow}{\partial_{\ka}}\si_{\mn}\ps,\label{lag}
\end{eqnarray}
onde os $a^{A}$ {[}$\mathring{a}^{A}${]} denotam constantes de acoplamento com dimensão de massa quatro{[}cinco{]},
$M_{\alpha\mu\nu} = \frac{1}{3} \left ( T_{\alpha\mu\nu} + T_{\mu\alpha\nu} + T_{\mu}g_{\alpha\mu}\right ) - 
\frac{1}{3}(\mu \leftrightarrow \nu)$, e $ \mathfrak{A}^\mu = \frac{1}{6} \epsilon^{\alpha\beta\gamma\mu}T_{\alpha\beta\gamma}$ 
\cite{Kostelecky:2007kx,kostelecky4,Pereira:2001xf}.
O acoplamento minimal é obtido no caso particular onde 
$a_{4}=3/4$ e todos os outros acoplamentos se anulam. Considerando as equações \,\eqref{Elko11} para\,\eqref{dirac21}, podemos ver que um campo espinorial tipo-(5), {\it flagpole},
são descritos por
\begin{eqnarray}
\cl_{(4)}^{{\rm type-(5)}} & \sim & (a_{1}T_{\;\rho\mu}^{\rho}+
a_{3}\mathfrak{A}_{\mu})\overbrace{\bar{\ps}\ga^{\mu}\ps}^{=\,J^\mu}\thinspace,\label{lag}
\end{eqnarray}
e pelo menos metade dos acoplamentos entre férmions flapole, associados com constantes de acoplamento de dimensão quatro e torção, anulam-se. Mostramos que, em
espaços-tempo de Riemann-Cartan, espinores {\it flagpole} são menos sensíveis a violação de Lorentz. Buscas experimentais recentes para violação de Lorentz são exploradas 
para extrair novos vínculos envolvendo componentes independentes da torção para níveis abaixo de $10^{-31}$ GeV\,\cite{Kostelecky:2007kx}. Embora sensibilidade excepcional 
para a densidade da torção pode ser alcançada buscando por seus acoplamentos com férmions de Dirac, espinores {\it flagpole} são  menos sensíveis à torção.

\section{\label{sec:A-bridge-between}Uma ponte entre a simetria de Lorentz e sua violação}
$\qquad$ Nesta {}{seção apresentamos} exemplos de mapeamento entre espinores e na próxima seção, construiremos o propagador de uma das nossas escolhas. \\
\indent Uma vez que provamos que espinores singulares são menos sensíveis a acoplamentos com a torção em cenários de VL, agora discutimos a relação entre bilineares covariantes
numa estrutura aparentemente com VL e os bilineares covariantes padrão. Como os bilineares covariantes {}{podem realizar} os observáveis em teorias envolvendo campos fermiônicos, esta
é uma questão de máxima importância física. Discutiremos como um campo espinorial geral com VL pode ser transliterado num espinor LI e {}{mostraremos} que esta relação mistura
as classes espinoriais como definidas na classificação de Lounesto.\\
\indent Uma consideração formal quando consideramos VSL é que parte da simetria de Lorentz é quebrada e que parte permanece. Portanto, se a VSL é obtida permitindo-se 
uma classe mais ampla de transformações que são, em geral, não-lineares e dependentes das coordenadas e derivadas, este formalismo deve mergulhar o grupo de Lorentz num grupo 
maior. Isto torna necessário descrever como o grupo de Lorentz se situa dentro deste grupo de simetria maior. Contudo, concentrando-nos no estudo de espinores
via bilineares covariantes, não precisamos visar estas questões. De fato, bilineares covariantes de qualquer espinor {}{não dependem da} representação com respeito a simetria residual
e {}{nem ao grupo associado a simetria residual}. Isto é devido ao fato que qualquer grupo de simetria pode ser mergulhado em algum grupo Spin que, 
até dimensão cinco, pode ser definido pelos elementos invertíveis $R$ do grupo contorcido de Clifford-Lipschitz que satisfaz $R^{\dagger}R=I$\,\cite{lounesto,roldao_1}.
Portanto, qualquer simetria residual não será aparente quando bilineares covariantes são levados em conta. Quando espinores são levados em conta, então obviamente 
o conteúdo da simetria residual é importante. Não obstante, o nosso objetivo de termos em conta os {}{possíveis observáveis}, isto é os bilineares covariantes, torna o conteúdo do grupo 
ser reduzido. Vale a pena mencionar que tomando o espinor clássico $\xi$ que satisfaz $\xi^{\dagger}\gamma_{0}\psi\neq0$, o espinor original $\psi$  pode ser recuperado
do seu agregado de Fierz $\mathbf{Z}$, que é dado por
\begin{eqnarray}
\mathbf{Z}=\sigma+\mathbf{J}+i\mathbf{S}+i\mathbf{K}\gamma_{0123}+\omega\gamma_{0123}\,,\label{Z}
\end{eqnarray}
utilizando-se o algoritmo de Takahashi\,\cite{roldao_1}. \\
\indent Além de removermos da Lagrangiana termos de violação de Lorentz espúrios, queremos investigar o efeito destas redefinições de campo na classificação de Lounesto para os
espinores transformados. Realizamos isto relacionando os observáveis na estrutura de violação de Lorentz com os bilineares covariantes na teoria padrão LI. Portanto os 
espinores regulares (Dirac em particular) e os espinores singulares (abrangendo espinores de Weyl, Majorana e Elko, dentre outros) podem ter uma descrição dual em termos da 
estrutura de violação de Lorentz, {}{pois espinores de um tipo (como soluções) numa teoria com LI, podem ser mapeados em espinores de outro tipo (como soluções) em 
outra teoria com VL.}\\
\indent Começamos listando os bilineares na estrutura com violação de Lorentz, {}{obtidos através da transformação \eqref{spintrans}}
\begin{align}
\sigma_{\chi} & =\bar{\chi}\chi,\nonumber \\
\mathbf{J}_{\chi} & =\bar{\chi}\gamma_{\mu}\chi{\rm e}^{\mu},\nonumber \\
\mathbf{S}_{\chi} & =\frac{1}{2}\bar{\chi}i\gamma_{\mu\nu}\chi{\rm e}^{\mu}\wedge{\rm e}^{\nu},\nonumber \\
\mathbf{K}_{\chi} & =\bar{\chi}\gamma_{5}\gamma_{\mu}\chi{\rm e}^{\mu},\nonumber \\
\omega_{\chi} & = \bar{\chi}\gamma_{5}\chi.
\end{align}
\indent A seguir, relacionamos estes invariantes com os correspondentes espinores {}{transformados (LI) \eqref{spintrans}}. Após alguns cálculos, encontramos
\begin{align}
\sigma_{\psi} & =\Delta\sigma_{\chi}+\Omega_{\sigma}\thinspace,\label{eq:upsigma}
\end{align}
onde $\Delta=1+2\Im{\rm \theta}+\vert{\rm \theta}\vert^{2}+2\tilde{C}_{\mu}x^{\mu}\Re{\rm \theta}+\tilde{C}_{\mu}\tilde{C}_{\nu}x^{\mu}x^{\nu}$ e a forma explícita de $\Omega_{\sigma}$ é dada no apêndice. 
Os primeiros termos dessa expressão são
\begin{align}
\Omega_{\sigma} & =\chi^{\dagger}\gamma_{0}v\Gamma\chi+B_{\mu}\chi^{\dagger}\gamma_{0}\partial^{\mu}\chi+\tilde{B}_{\mu}\chi^{\dagger}\gamma_{0}\gamma_{5}\partial^{\mu}\chi+\cdots
\end{align}
\indent Para os invariantes remanescentes obtemos expressões similares,
\begin{subequations}\label{eq:identifications}
\begin{align}
J_{\alpha}^{\psi} & =\Delta J_{\alpha}^{\chi}\ +\Omega_{J_{\alpha}},\,(\gamma_{0}\mapsto\gamma_{0}\gamma_{\alpha})\\
S_{\alpha\beta}^{\psi} & =\Delta S_{\alpha\beta}^{\chi}+\Omega_{S_{\alpha\beta}},\,(\gamma_{0}\mapsto\gamma_{0}\gamma_{\alpha\beta})\\
K_{\alpha}^{\psi} & =\Delta K_{\alpha}^{\chi}\ +\Omega_{K_{\alpha}},\,(\gamma_{0}\mapsto\gamma_{0}\gamma_{5}\gamma_{\alpha})\\
\omega^{\psi} & =\Delta\omega^{\chi}\ +\Omega_{\omega},\,(\gamma_{0}\mapsto\gamma_{0}\gamma_{5})
\end{align}
\end{subequations}
onde as identificações entre parênteses significam que, por exemplo, $\Omega_{J_{\alpha}}$ é obtido de $\Omega_{\sigma}$ pela substituição $\gamma_{0}\mapsto\gamma_{0}\gamma_{\alpha}$,
e similarmente para os invariantes remanescentes.\\
\indent Utilizando estas expressões, podemos aplicar a classificação de Lounesto para os espinores transformados (LI). Considere o exemplo $\sigma_{\psi}\neq0,\;\omega_{\psi}\neq0$,
correspondendo ao espinor de tipo-I na teoria LI. Como $\sigma_{\psi}=\Delta\sigma_{\chi}+\Omega_{\sigma}$ e $\omega_{\psi}=\Delta\omega_{\chi}\ +\Omega_{\omega}$, dependendo
do valor de $\Omega_{\sigma}$ e $\Omega_{\omega}$, podemos ter ou $\sigma_{\chi}$ e $\sigma_{\omega}$ iguais a zero ou não e portanto a redefinição do campo pode relacionar um
espinor invariante de Lorentz do tipo 1 com vários tipos de espinores com violação de Lorentz, tais como $\sigma_{\chi}=0=\omega_{\chi}\ $ (tipos 4, 5, e 6) e outros. \\
\indent Uma ampla lista de possibilidades é dada abaixo, onde assumiremos que as funções $\Delta$ e $\Omega$ são não nulas: 
\begin{itemize}
\item[$1_{\psi}$)] $\sigma_{\psi}\neq0,\;\;\;\omega_{\psi}\neq0$.
Como  $\sigma_{\psi}\neq0$ e $\sigma_{\psi}=\Delta\sigma_{\chi}+\Omega_{\sigma}$,
listamos abaixo todas as possibilidades,  dependendo se  $\sigma_{\chi}$
é ou não igual a zero, assim como  $\omega_{\chi}$: 
\begin{enumerate}
\item[$i)$] $\sigma_{\chi}=0=\omega_{\chi}\ $. Estas condições são correspondentes aos campos espinoriais do tipo-(4), tipo-(5) e tipo-(6)  \textemdash{}
respectivamente  {\it flag-dipoles}, {\it flagpole} e {\it dipoles}. 
\item[$ii)$] $\sigma_{\chi}=0$ e $\omega_{\chi}\neq0$, sendo compatíveis com campos espinoriais regulares do tipo-(3). 
A condição  $\sigma_{\chi}=0$ é consistente com  $\sigma_{\psi}\neq0$. 
\item[$iii)$] $\sigma_{\chi}\neq0$ e $\omega_{\chi}=0$. Este caso diz respeito a campos espinoriais regulares do  tipo-(2). 
A condição  $\omega_{\chi}=0$ é consistente com $\omega_{\sigma}\neq0$.
\item[$iv)$] $\sigma_{\chi}\neq0$ e $\omega_{\chi}\neq0$, correspondendo aos campos espinoriais regulares  do tipo-(1), em particular os de Dirac.
\end{enumerate}
\item[$2_{\psi}$)] $\sigma_{\psi}\neq0,\;\;\;\omega_{\psi}=0$.\label{dirac1b}
Embora a condição $\sigma_{\psi}\neq0$ seja consistente com ambas as possibilidades
 $\sigma_{\chi}=0$ e $\sigma_{\chi}\neq0$ (claramente a condição $\sigma_{\chi}\neq0$ é consistente com  $\sigma_{\psi}\neq0$
se $\Delta\sigma_{\chi}\neq-\Omega_{\sigma}$), a condição  $\omega_{\psi}=0$
fornece $\Delta\omega_{\chi}=-\Omega_{\sigma}$, o qual não se anula.
Para resumir: 
\begin{enumerate}
\item[$i)$] $\sigma_{\chi}=0$ e $\omega_{\chi}\neq0$. Este caso corresponde ao tipo-(3), campos espinoriais regulares
. A condição $\sigma_{\chi}=0$ é consistente com  $\sigma_{\psi}\neq0$, contudo como  $\omega_{\chi}\neq0$, a condição adicional
$\Delta\omega_{\chi}=-\Omega_{\sigma}$ deve ser imposta.
\item[$ii)$] $\sigma_{\chi}\neq0$ e $\omega_{\chi}\neq0$. Este caso diz respeito ao tipo-(1), campos espinoriais de Dirac.
\end{enumerate}
\item[$3_{\psi}$)] $\sigma_{\psi}=0,\;\;\;\omega_{\psi}\neq0$.\label{dirac2b} A condição  $\omega_{\psi}\neq0$
é consistente com ambas e complementar a  $\omega_{\chi}=0$ e $\omega_{\chi}\neq0$. Para resumir:
\begin{enumerate}
\item[$i)$] $\omega_{\chi}=0$ e $\sigma_{\chi}\neq0$, correspondendo ao tipo-(2), campos espinoriais regulares. 
A condição $\omega_{\chi}=0$ é consistente com $\omega_{\psi}\neq0$, contudo como $\sigma_{\chi}\neq0$, a condição adicional 
$\sigma_{\psi}=\Delta\sigma_{\chi}+\Omega_{\sigma}\neq0$ deve ser imposta.
\item[$ii)$] $\sigma_{\chi}\neq0$ e $\omega_{\chi}\neq0$. Este caso diz respeito ao tipo-(1), campos espinoriais regulares. 
\end{enumerate}
\item[$4_{\psi}$)] $\sigma_{\psi}=0=\omega_{\psi},\;\;\;\mathbf{K}_{\psi}\neq0,\;\;\;\mathbf{S}_{\psi}\neq0$.\label{tipo4b} 
\item[$5_{\psi}$)] $\sigma_{\psi}=0=\omega_{\psi},\;\;\;\mathbf{K}_{\psi}=0,\;\;\;\mathbf{S}_{\psi}\neq0$.\label{tipo-(5)1b} 
\item[$6_{\psi}$)] $\sigma_{\psi}=0=\omega_{\psi},\;\;\;\mathbf{K}_{\psi}\neq0,\;\;\;\mathbf{S}_{\psi}=0$. 
\end{itemize}
$\qquad$ Todos os campos espinoriais singulares $4_{\psi}$), $5_{\psi}$) e $6_{\psi}$) são definidos pela condição $\sigma_{\psi}=0=\omega_{\psi}$, 
$6_{\psi}$) são definidos pela condição  $\sigma_{\psi}=0=\omega_{\psi}$,
{}{a qual implica} \linebreak $\Delta\sigma_{\chi}=-\Omega_{\sigma}(\neq0)$
e que  $\Delta\omega_{\chi}=-\Omega_{\omega}(\neq0)$. Portanto, espinores singulares na teoria com violação de Lorentz são sempre relacionados a espinores regulares
no modelo com invariância de Lorentz correspondente.
Esta discussão pode ser resumida na Tabela \,\eqref{table}, que descreve a possibilidade de {}{mapear observáveis da estrutura} com violação de Lorentz em observáveis 
nos modelos com covariância Lorentz. 
\begin{table}
\centering{}%
\begin{tabular}{||r|r||r|r||}
\hline 
 & Campos Espinoriais VSL  & Campos Espinoriais Covariantes  & \tabularnewline
\hline 
\hline 
classe ($1_{\psi}$)  & $\psi$-regular  & Regular\qquad{}  & classe (1)\tabularnewline
 &  & Regular\qquad{}  & classe (2)\tabularnewline
 &  & Regular\qquad{}  & classe (3)\tabularnewline
 &  & Flag-dipole \,  & classe (4)\tabularnewline
 &  & Flagpole \,\,\quad{}  & classe (5)\tabularnewline
 &  & Dipole\,\,\,\qquad{}  & classe (6)\tabularnewline
\hline 
classe ($2_{\psi}$)  & $\psi$-regular  & Regular\qquad{}  & classe (3)\tabularnewline
 &  & Regular\qquad{}  & classe (1)\tabularnewline
\hline 
classe ($3_{\psi}$)  & $\psi$-regular  & Regular\qquad{}  & classe (2)\tabularnewline
 &  & Regular\qquad{}  & classe (1)\tabularnewline
\hline 
classe ($4_{\psi}$)  & $\psi$-{\it flag-dipole}  & Regular\qquad{}  & classe (1)\tabularnewline
\hline 
classe ($5_{\psi}$)  & $\psi$-{\it flagpole}  & Regular\qquad{}  & classe (1)\tabularnewline
\hline 
classe ($6_{\psi}$)  & $\psi$-{\it dipole}  & Regular\qquad{}  & classe (1)\tabularnewline
\hline 
\end{tabular}\caption{\label{table}Correspondência entre campos espinoriais VSL ($\chi$) e campos espinoriais LI $\psi$, sob a classificação
de campos espinoriais de Lounesto.}
\end{table}
\\ \indent {}{No contexto LI, a} possibilidade de mapeamento entre diferentes classes de espinores tem sido apontado na literatura. De fato, é conhecido que um campo espinorial regular pode ser
mapeado em qualquer campo espinorial covariante, incluindo mapeamentos exóticos e regimes relacionados com a gravidade quântica \,\cite{daRocha:2011yr,Ablamowicz:2014rpa}.
Tais aplicações são formalmente consistentes e foram realizadas em contextos cinemáticos e dinâmicos. Não obstante, todos estes estudos foram desenvolvidos sobre uma estrutura
com invariância de Lorentz, que é generalizada neste trabalho para incluir a possibilidade de VSL, portanto ampliando consideravelmente a classe de modelos que podem ser
relacionados por estas transformações. {}{Um ponto muito importante são as identidades de Fierz nesse contexto: impondo que as mesmas continuem válidas nos espinores transformados, serão gerados vínculos.} \\
\indent Como um exemplo deste tipo de relação discutida aqui, vamos construir uma transformação que mapeia um espinor de Dirac $\chi$ na estrutura VSL num 
espinor singular $\psi$ (do tipo Majorana, {\it flagpole}) na estrutura LI. Sem perda de generalidade, na representação de Weyl estes espinores são autoespinores do operador de
conjugação de carga e podem ser parametrizados como segue,
\begin{equation}
\chi=\begin{pmatrix}a_{0}\\
a_{1}\\
a_{2}\\
a_{3}
\end{pmatrix}=\begin{pmatrix}\chi_{1}\\
\chi_{2}
\end{pmatrix}\in\mathbb{C}^{4},\quad\qquad\psi=\begin{pmatrix}-i\beta^{*}\\
i\alpha^{*}\\
\alpha\\
\beta
\end{pmatrix}=\begin{pmatrix}\psi_{1}\\
\psi_{2}
\end{pmatrix}\in\mathbb{C}^{4}\,,\label{eq:example1}
\end{equation}
e podemos posteriormente escolher $a_{0}=a_{2}=0$\,\cite{Cavalcanti:2014wia}. A  transformação necessária é da forma 
\begin{eqnarray}\label{spinor_transform}
\psi=(1+v_{\mu}\gamma_{5}\gamma^{\mu})\chi,
\end{eqnarray}
\noindent a qual é equivalente a $\psi-\chi=v_{\mu}\gamma_{5}\gamma^{\mu}\chi$.
Definindo  
\begin{align}
T & \equiv v_{\mu}\gamma_{5}\gamma^{\mu}=\begin{pmatrix}0 & -v_{\mu}\sigma^{\mu}\\
v_{0}I-v_{k}\sigma^{k} & 0
\end{pmatrix}\thinspace,
\end{align}
temos $T\chi=\binom{-v_{\mu}\sigma^{\mu}(\chi_{2})}{v_{0}I-v_{k}\sigma^{k}(\chi_{1})}=\binom{\psi_{1}-\chi_{1}}{\psi_{2}-\chi_{2}}$.
Portanto podemos obter condições de consistência entre as componentes dos espinores, isto é, $i\beta^{*}/a_{3}=-\al/a_{1}\equiv\delta\in\mathbb{C}\thinspace,$
junto com $v_{1}-iv_{2}=\delta$, {}{o que} caracteriza as transformações desejadas. O resultado final é que, no nível clássico o modelo VSL  de um espinor de Dirac
 $\chi$ governado pela Lagrangiana
\begin{equation}
\mathcal{L}_{0}=\frac{i}{2}\bar{\chi}\gamma^{\nu}\stackrel{\leftrightarrow}{\partial_{\nu}}\chi-m\bar{\chi}\chi-iv_{\mu}
\bar{\chi}\gamma_{5}\stackrel{\leftrightarrow}{\partial^{\mu}}\chi,
\end{equation}
\noindent é realmente fisicamente idêntico ao modelo padrão LI de um campo espinorial de Majorana livre. Pode-se recordar que campos espinoriais de Dirac possuem sete graus de 
liberdade, enquanto que espinores de Majorana possuem quatro graus de liberdade. A equação\,\eqref{spintrans} é um vínculo que reduz dois graus de liberdade no espinor de Dirac.
Além disso, a falta do grau de liberdade é {}{contada} quando o calibre $U(1)$ é levado em conta para o espinor de Dirac, mas não para o de Majorana\,\cite{lounesto}.\\
\indent Uma aplicação mais natural entre espinores, {}{alternativamente à equação}\,\eqref{spintrans} seria uma levando espinores de Weyl, que são uma classe particular de espinores
{\it dipole} com simetria de calibre $U(1)$, em espinores regulares de Dirac, o que permite, além disso, bandeiras de Penrose para serem anexadas a espinores de Weyl.
Contudo, a relação entre campos espinoriais de Dirac e Weyl é amplamente explorado ao longo da literatura de TQC. Portanto, optamos por obter um mapeamento entre
campos espinoriais regulares e {\it flag-dipole}, uma vez que espinores {\it flag-dipole} são soluções importantes da equação de Dirac em gravidade $f(R)$ com torção\,\cite{fabbri_ESK}.
De fato considere espinores regulares e do tipo-($4$) respectivamente\,\cite{Cavalcanti:2014wia},
\begin{align}
\chi= & \begin{pmatrix}a_{0}\\
a_{1}\\
a_{2}\\
a_{3}
\end{pmatrix}\in\mathbb{C}^{4}\,,\text{ com }
\begin{matrix}
a_0 \neq -\frac{a_1a_2a_3^*}{||a_2||^ 2},
\end{matrix}\\
\psi= & \begin{pmatrix}c_{0}\\
c_{1}\\
c_{2}\\
c_{3}
\end{pmatrix}=\begin{pmatrix}-\frac{c_1c_2c_3^ *}{||c_2||^ 2}\\
c_1\\
c_2\\
c_3
\end{pmatrix}\in\mathbb{C}^{4}\,,\text{ com }\begin{matrix}
||c_1||^ 2 \neq ||c_3||^2
\end{matrix}.
\end{align}
\indent A transformação que relaciona estes campos espinoriais é dada tomando \linebreak
$v\cdot\Gamma=\lambda\gamma^{5}$ \textendash{} $\psi=(I+\lambda\gamma^{5})\chi$
na equação\,\eqref{spintrans}\,\cite{Cavalcanti:2014wia}. Então  
\begin{eqnarray}
 & & \psi-\chi = \lambda\gamma^{5}\chi, \nonumber \\
 & & \beta_1 = 1 + \lambda = \frac{c_2}{a_2} = \frac{c_3}{a_3}, \nonumber \\
 & & \beta_2 = 1 - \lambda = \frac{c_0}{a_0} = \frac{c_1}{a_1}, \nonumber 
\end{eqnarray}
e o mesmo fornece $\lambda=\frac{\beta_{1}-\beta_{2}}{2}$. \\
\indent Outro exemplo interessante de mapeamento de espinores via essas transformações, é novamente usando espinores regulares e {\it flagpoles }(Majorana), mas dessa vez, utilizando uma transformação do
tipo $\psi - \chi = a_{\mu}\gamma^{\mu}\chi$, a qual, junto com a transformação anterior, serve como base para uma apresentação simplificada (fundo  Minkowski) da equação de Dirac e do respectivo propagador.
Agora obtivemos a condição de consistência $\frac{i\beta^{*}}{a_3} = \frac{\alpha}{a_1} = \delta \in \mathbb{C}$ e $ v_1 - iv_2 = -\delta$. \\
\indent Em suma, mostramos como uma classe geral de redefinições de campos podem relacionar classes diferentes de campos espinoriais na estruturas VSL e LI.
A equivalência física entre os modelos aqui descritos valem nos aspectos cinemáticos e dinâmicos {}{clássicos} destas teorias, deixando em aberto a questão interessante da equivalência
quântica. A segunda quantização de diferentes classes de campos espinoriais pode ser realizada, portanto acreditamos que a investiga\c c\~ao de aspectos quânticos das relações que 
apresentamos é um tópico viável e interessante a ser buscado.

\subsection{A equação de Dirac e o respectivo propagador fermiônico com violação de Lorentz}

\indent Nessa seção, apresentamos um exemplo simplificado (fundo Minkowski) de Lagrangiana com violação de Lorentz usando a transformação \eqref{spinor_transform}. \\
\indent A Lagrangiana de férmions de massa $m$ com os termos de quebra de simetria $CPT$ é dada por \cite{DeOliveira:2010ig}:
\begin{equation}\label{lagranviolation}
{\cal
L}=\bar{\psi}(\not\!p -\not\!b\gamma_5-m)\psi\,,
\end{equation}
onde $b^\mu=(b_0,{\bf b})$ é um quadrivetor constante e variáveis do tipo $\not\!y$ são por definição $\not\!y = \gamma^{\mu}y_{\mu}$.
\smallskip

\indent Procuramos por soluções do tipo ondas-planas, em que $\psi^{(\alpha)}=N_u^{(\alpha)}\,u^{(\alpha)}(p)e^{-ip\cdot x}$
($\alpha=1,2)$ é um espinor de Dirac modificado com quatro componentes, onde $N_u^{(\alpha)}$ é uma constante de normalização a ser determinada.
\smallskip

\indent Da equação \eqref{lagranviolation}, obtemos a equação de Dirac modificada
\begin{equation}\label{diracLV}
(\not\!p -\not\!b\gamma_5-m)\psi=0\,.
\end{equation}

\indent Aplicando o {\it ansatz} para $\psi$ e multiplicando a equação acima pela esquerda por \linebreak $(\not\!p -\not\!b\gamma_5+m)$, obtemos a seguinte relação:
\begin{equation}\label{nondiagonalrelation}
\left\{p^2-b^2-m^2-[\not\! p,\not
b]\gamma_5\right\}u(p)=0\,.
\end{equation}

\indent A expressão acima ainda não é diagonal, uma vez que na mesma aparecem termos com matrizes não diagonais. 
Para obter a equação algébrica para este modelo, \'e necess\'ario multiplicar a equa\c{c}\~ao 
acima por $p^2-b^2-m^2+[\not\! p,\not b]\gamma_5$ pela esquerda:
\begin{equation}
\{[p^2-b^2-m^2]^2-\left([\not\! p,\not b]\gamma_5\right)^2\}u(p)=0\,.
\end{equation}

\indent Vamos trabalhar o comutador acima: 
\begin{eqnarray}
\left([\not\! p,\not b]\gamma_5\right)^2&=&(\not\!p\not\!b\,-\not\!b\not\!p)\gamma_5(\not\!p\not\!b\,-\not\!b\not\!p)\gamma_5\nonumber\\
&=&\not\!p\not\!b\not\!p\not\!b\,-\not\!p\not\!b\not\!b\not\!p\,-\not\!b\not\!p\not\!p\not\!b\,+\not\!b\not\!p\not\!b\not\!p\nonumber\\
&=&\not\!p\not\!b\left[-\not\!b\not\!p+2(p\cdot b)\right]-2p^2b^2+\not\!b\not\!p\left[-\not\!p\not\!b+2(p\cdot b)\right]\nonumber\\
&=&-4p^2b^2+2(p\cdot b)\not\!p\not\!b+2(p\cdot b)\not\!b\not\!p\nonumber\\
&=&-4p^2b^2+2(p\cdot b)\left[-\not\!b\not\!p+2(p\cdot b)\right]\,+2(p\cdot b)\not\!b\not\!p\nonumber\\
&=&-4p^2b^2+4(p\cdot b)^2\,,
\end{eqnarray}
onde foi utilizada a identidade $\not\!c\not\!d=-\not\!d\not\!c+2(c\cdot d)$.
\smallskip

\indent Portanto, a rela\c{c}\~ao de dispers\~ao para o modelo \'e
\begin{equation}\label{dispersionrelation}
[p^2-b^2-m^2]^2-4[b\cdot p]^2+4b^2p^2=0\,.
\end{equation}

\indent Esta relação de dispersão é quártica na variável $p^0({\bf p})$. Ela possui duas raízes positivas $E_u^{(\alpha)}$ e duas negativas $E_v^{(\alpha)}$, onde $\alpha=1,2$.
\smallskip

\indent A equação \eqref{dispersionrelation} é facilmente resolvida para os casos em que $b^\mu$ é estritamente temporal ou espacial. Para o caso $b^\mu=(b_0,{\bf 0})$, a relação de dispersão fornece
\begin{eqnarray}
E_u^{(\alpha)}&=&\sqrt{(|{\bf p}|+(-1)^{\alpha}b_0)^2+m^2}\,,\nonumber\\
&&\\
E_\upsilon^{(\alpha)}&=&\sqrt{(|{\bf
p}|-(-1)^{\alpha}b_0)^2+m^2}\,,\nonumber
\end{eqnarray}
em que $E_{u,\upsilon}^{(\alpha)}$ indicam as energias para as partículas e suas antipartículas, respectivamente. 
\smallskip

\indent Para o caso $b^\mu=(0,{\bf b})$, as solu\c{c}\~oes s\~ao
\begin{eqnarray}\label{energies}
E_u^{(\alpha)}&=&\sqrt{{\mbox{\boldmath{$p$}}}^2+m^2+{\mbox{\boldmath{$b$}}}^2+(-1)^{\alpha}2\sqrt{({\mbox{\boldmath{$b$}}}\cdot{\mbox{\boldmath{$p$}}})^2+m^2}}\,,\nonumber\\
&&\\
E_\upsilon^{(\alpha)}&=&\sqrt{{\mbox{\boldmath{$p$}}}^2+m^2+{\mbox{\boldmath{$b$}}}^2-(-1)^{\alpha}2\sqrt{({\mbox{\boldmath{$b$}}}\cdot{\mbox{\boldmath{$p$}}})^2+m^2}}\,.\nonumber
\end{eqnarray}

\indent Quando multiplicamos a equação \eqref{diracLV} pela esquerda por $\gamma^0$, esta equação de movimento pode ser escrita na forma hamiltoniana, onde $i\dfrac{\partial\psi}{\partial t}=H\psi$. 
Assim, temos que\footnote{$\displaystyle {\bf \Sigma}={\mbox{\boldmath{$
\alpha$}}}\gamma^5=\left(\begin{array}{cc}{\mbox{\boldmath{$
\sigma$}}}&0\\0&{\mbox{\boldmath{$ \sigma$}}}\end{array}\right)\,$, onde $\boldmath{\sigma} = \sigma^i\rm{e}_i$.}
\begin{eqnarray}
H = {\mbox{\boldmath{$\alpha$}}}\cdot{\bf p} +m\gamma^0 + \gamma_5b_0+{\bf
\Sigma}\cdot{\bf b}\,.
\end{eqnarray}
\indent Vamos construir os espinores para o caso $b^\mu$ puramente temporal. O Hamiltoniano é então dado por:
\begin{equation}
H={\mbox{\boldmath{$\alpha$}}}\cdot\mathbf{p} + m\gamma^0 + b_0\gamma_5\,.
\end{equation}
\indent Na representação padrão das matrizes $\gamma$ de Dirac, como \'e usual, obtemos os seguintes espinores
\begin{equation}
u^{(\alpha)}(p)=N_u^{(\alpha)}\,\left(\begin{array}{c}\chi^{(\alpha)}\\\xi_u^{(\alpha)}\,\chi^{(\alpha)}\end{array}\right)
\end{equation}
para os estados de energia positiva e, para os estados de energia negativa,
\begin{equation}\label{negative_energy}
\upsilon^{(\alpha)}(p)=N_\upsilon^{(\alpha)}\,\left(\begin{array}{c}\xi_\upsilon^{(\alpha)}\,\eta^{(\alpha)}\\\eta^{(\alpha)}\end{array}\right)\,,
\end{equation}
onde
\begin{equation}
\xi_u^{(\alpha)}=\frac{{\mbox{\boldmath{$\sigma$}}}\cdot {\mbox{\boldmath{$p$}}}-b_0}{E_u^{(\alpha)} + m}\,.
\end{equation}
\indent A solução para $\xi_\upsilon^{(\alpha)}$ é obtida pela troca, na expressão acima,  $b_\mu\mapsto - b_\mu$ e \linebreak $E_u^{(\alpha)}\mapsto E_\upsilon^{(\alpha)}$.
\\
\indent O espinor \eqref{negative_energy} pode ser normalizado, se escolhermos a mesma condi\c{c}\~ao de normaliza\c{c}\~ao do caso da teoria LI:
\begin{equation}
\bar{u}^{(\alpha)}(p)u^{(\alpha')}(p)=\delta^{\alpha\alpha'}\,.
\end{equation}
\indent Utilizando a definição $\bar{u}=u^\dagger\gamma^0$ e a auto-energia positiva \eqref{energies} para a partícula, encontramos a constante $N_u^{(\alpha)}$ de normalização:
\begin{equation}
N_u^{(\alpha)}=\sqrt{\frac{E_u^{(\alpha)}+ m}{2M}}\,.
\end{equation}
\indent Se o espinor de duas componentes $\chi^{(\alpha)}$ for escolhido ser autovetor do operador 
${\mbox{\boldmath{$\sigma$}}}\cdot\dfrac{{\mbox{\boldmath{$p$}}}}{|{\mbox{\boldmath{$p$}}}|}$ 
com autovalor $-(-1)^\alpha$, o espinor de Dirac modificado e normalizado fica
\begin{equation}
u^{(\alpha)}(p)=\sqrt{\frac{E_u^{(\alpha)} + m}{2m}}\left(\begin{array}{c}\chi^{(\alpha)}\\\displaystyle{\frac{-(-1)^{(\alpha)}|{\mbox{\boldmath{$p$}}}
|-b_0}{E_u^{(\alpha)} + m}}\,\chi^{(\alpha)}\end{array}\right)\,.
\end{equation}

\indent Procedendo de modo análogo, obtemos o espinor para os anti-férmions:

\begin{equation}
\upsilon^{(\alpha)}(p)=\sqrt{\frac{E_\upsilon^{(\alpha)} + m}{2m}}\left(\begin{array}{c}\displaystyle{\frac{-(-1)^{(\alpha)}|{\mbox{\boldmath{$p$}}}| 
+ b_0}{E_\upsilon^{(\alpha)} + m}}\,\chi^{(\alpha)}\\\chi^{(\alpha)}\end{array}\right)\,.
\end{equation}

\indent Ainda podemos obter outro elemento de grande importância na teoria da EDQ estendida: o {\it propagador fermiônico modificado}. O propagador de Feynman escolhido deve satisfazer
\begin{equation}
(i\!\not\!\partial -\not\!b\gamma_5-m)S_{b}(x-y)=i\delta^{(4)}(x-y)\,,
\end{equation}
{}{que, }escrita no espaço de Fourier,
\begin{equation}
\int\frac{d^4p}{(2\pi)^4}(\not\!p
-\not\!b\gamma_5-m)e^{-ip\cdot(x-y)}S_{b}(p)=i\delta^{(4)}(x-y)
\end{equation}
fornece, através da representação de Fourier da Delta de Dirac,
\begin{equation}
S_{b}(p)=\frac{i}{\not\!p - \not\!b\gamma_5-m}\,.
\end{equation}
\indent Podemos ainda, usar as referências \cite{DeOliveira:2010ig, Oliveira:2010wa} para {}{escrever} o propagador da seguinte forma
\begin{eqnarray}
 S_{b}(p)&=&\frac{i}{\not\!p -\not\!b\gamma_5-m} = 
 \frac{i(\not\!p -\not\!b\gamma_5+m)}{(\not\!p -\not\!b\gamma_5-m)(\not\!p -\not\!b\gamma_5+m)}\nonumber\\
&&\nonumber\\
&=&  \frac{i(\not\!p -\not\!b\gamma_5+m)}  { \left\{p^2-b^2-m^2-[\not\! p,\not
b]\gamma_5\right\}}\,,
\end{eqnarray}
onde foi utilizada \eqref{nondiagonalrelation}. Utilizando também \eqref{dispersionrelation}, segue que
\begin{equation}
 S_{b}(p)=\frac{i(\not\!p -\not\!b\gamma_5+m)\{ p^2-b^2-m^2+[\not\! p,\not
b]\gamma_5 \}}{[p^2-b^2-m^2]^2-4[p\cdot b]^2+4p^2b^2}\,.
\end{equation}
Portanto, vemos que a quebra da simetria de Lorentz, adotada em \eqref{spinor_transform} modifica as relações de dispersão, as autoenergias e os espinores da teoria de Dirac LI, além de gerar uma perturbação no Hamiltoniano de Dirac livre.

%% file: conclusoes.tex
\chapter{Conclusões e desenvolvimentos futuros}

$\qquad$ Nessa tese, investigamos como a classificação de espinores de Lounesto ilumina várias áreas da física teórica de ponta. \\
\indent No capítulo 1, revimos brevemente o conceito de grupo de Lorentz e construímos explicitamente as representações irredutíveis do seu recobrimento duplo, o chamado grupo spin, 
que é identificado com o grupo $SL(2,\mathbb{C})$, após isso, apresentamos os bilineares covariantes e discutimos o seu significado físico. Terminamos o capítulo mostrando via
transformações de Fierz, como o grupo de Lorentz $SO(1,3)$ age na álgebra $M(4,\mathbb{C})$ (a álgebra de Clifford do espaço-tempo). \\
\indent Devemos frisar que a classificação de Lounesto não se resume ao espaço-tempo de Minkowski.
Recentemente, novas classes de espinores em espaços de Lorentz de cinco dimensões foram construídas \cite{deBrito:2016qzl}, o que tem grande 
utilidade na busca por novas soluções espinoriais de equações de primeira ordem que podem ser acopladas com equações de Einstein. Tal classificação é análoga à 
classificação de Lounesto para o espa\c co $\mathbb{R}^{4,1}$ Tais novas classes emulam espinores em \emph{bulk} do tipo anti-de Sitter (AdS$_5$) e sua 
subsequente localização no que chamamos de brana, que descreve nosso Universo, ainda é um problema em aberto e deve ser explorado \cite{Bernardini:2016uhj,German:2013sk}. 
Para espinores singulares no \emph{bulk}, somente o campo Elko foi usado para tal procedimento \cite{Liu:2011nb,Jardim:2014cya,Jardim:2014xla} e novos espinores do tipo 
\emph{flagpole} e \emph{flag-dipole} foram descobertos em \cite{daRocha:2016bil}. Outras classes preditas em  \cite{deBrito:2016qzl} e ainda inexploradas serão palco de investigação. 
O caso de seis dimensões, neste contexto, é ainda aberto, tendo sido implementado para o Elko em \cite{Dantas:2015mfi}. Além disso, recentemente o análogo da classificação 
de Lounesto para sete dimensões (tanto o caso Lorentziano quanto o Riemanniano) foi construído, com aplicações nas soluções da equação de Dirac no espa\c co AdS$_4$ $\times S^7$ 
que são espinores singulares \cite{Bonora:2014dfa,Bonora:2015ppa}. Visamos, ainda usar o formalismo desenvolvido no regime da escala de Planck em \cite{Ablamowicz:2014rpa} para
refinar o capítulo 4, bem como implementar sua versão exótica \cite{Bernardini:2012sc,daRocha:2011yr,daRocha:2011xb,Rocha:2014gqa,daSilva:2016htz}. \\
\indent No capítulo 2 revisamos as ferramentas geométricas necessárias para o tratamento de teorias de gravitação: variedades diferenciáveis, campos vetoriais e tensoriais, focando principalmente 
no conceito de tensor métrico com sua conexão canônica (de Levi-Civita). Por questões de didatismo, apresentamos com detalhes a dedução da equação de campo da teoria da gravitação de Einstein, partindo da
ação de Einstein-Hilbert.\\
\indent No capítulo 3, num contexto de espaço-tempo do 
tipo Riemann-Cartan, mostramos que os bilineares covariantes também surgem na dinâmica do sistema, estando presentes na equação de movimento do campo fermiônico num fundo 
ECSK, vinculando explicitamente o tipo de espinor \eqref{2.16}. Ainda no capítulo 2, com uma análise posterior, mostramos que o {\it ansatz} \eqref{generalspinor} é solução para 
a equação do campo fermiônico e identificamos como sendo um espinor do tipo-$4$, que pela primeira vez, surge naturalmente num modelo teórico. \\
\indent No capítulo 4, exploramos com grande generalidade, a quebra de simetria de Lorentz tomando um fundo do tipo $f(\mathit{R})$ com torção e adicionando um 
campo fermiônico ao mesmo, mostramos que para um tipo específico de espinor singular, o acoplamento com a torção é o menos sensível. 
Além disso, apresentamos exemplos concretos de como a transformação \eqref{spintrans} afeta os bilineares covariantes \eqref{eq:example1}. {}{O fundo $f(\mathit{R})$ tanto quanto espaços Riemann-Cartan 
são imprescindíveis para {\it flag-dipoles}.}\\
\indent Para desenvolvimentos futuros adicionais, temos duas propostas: buscar de forma natural a quebra de simetria de Lorentz usando um fundo como o da gravitação de 
Einstein é difícil 
(artificial na verdade), pois usando coordenadas locais, o tensor métrico se torna o de Minkowski em torno de um ponto. Acreditamos que partindo de um modelo de gravitação 
que não seja Riemanniano é o natural, pois se a fonte do campo gravitacional não é um tensor métrico, a linearização local não fornecerá imediatamente o modelo da relatividade especial.
Uma teoria geométrica naturalmente não Riemmaniana é a teoria dos espaços de Finsler \cite{xiaohuan}. Em \cite{Kostelecky:2016ufw} essa questão é abordada, sugerindo indícios 
teóricos recentes que apoiam tal enfoque \cite{Colladay:2012rv}. Um tópico totalmente inovador, seria estender a classificação de Lounesto para campos fermiônicos no 
âmbito da segunda quantização, o que implicaria vínculos nas teorias quantizadas. Ainda, a classificação de espinores em segunda quantização é um problema
em aberto, cuja solução está próxima \cite{roldaosecondquant}.\\
\indent Outra questão interessante, é o setor gravitacional do modelo padrão estendido (SME), cujo espaço de coeficientes carecem de ser investigados \cite{Kostelecky:2015dpa}.

%% file: apendice_omega.tex
\chapter*{Apêndice: a expressão completa de $\Omega_{\sigma}$}

$\qquad$ Por uma questão de exaustividade, citamos aqui a expressão completa da função $\Omega_{\sigma}$ aparecendo em\,\eqref{eq:upsigma},
\begin{eqnarray*}
\Omega_{\sigma} & = & \Delta\{\tilde{B}_{\mu}\partial^{\mu}\chi^{\dagger}+i{\rm \theta}\tilde{B}_{\mu}\partial^{\mu}\chi^{\dagger}+i\tilde{B}_{\mu}\tilde{C}_{\alpha}x^{\alpha}\partial^{\mu}\chi^{\dagger}\}\gamma_{5}\gamma_{0}\chi\\
 & + & \{\chi^{\dagger}+\chi^{\dagger}(v\Gamma)^{\dagger}-i\tilde{C}_{\mu}x^{\mu}\chi^{\dagger}+\Delta(B_{\mu}\partial^{\mu}\chi^{\dagger}+\tilde{B}_{\mu}\partial^{\mu}\chi^{\dagger}\gamma_{5}+C_{\mu\nu}x^{\mu}\partial^{\nu}\chi^{\dagger})-i{\rm \theta}^{*}\chi^{\dagger}\}\gamma_{0}v\Gamma\chi\\
 & + & \{\chi^{\dagger}(v\Gamma)^{\dagger}+\Delta(B_{\mu}\partial^{\mu}\chi^{\dagger}+C_{\mu\nu}x^{\mu}\partial^{\nu}\chi^{\dagger})+i{\rm \theta}\chi^{\dagger}(v\Gamma)^{\dagger}+i\tilde{C}_{\alpha}x^{\alpha}\chi^{\dagger}(v\Gamma)^{\dagger}\\
 & + & i{\rm \theta}\Delta(B_{\mu}\partial^{\mu}\chi^{\dagger}+i\tilde{C}_{\alpha}B_{\mu}x^{\alpha}\partial^{\mu}\chi^{\dagger}+i{\rm \theta}C_{\mu\nu}x^{\mu}\partial^{\nu}\chi^{\dagger}+iC_{\mu\nu}x^{\mu}\tilde{C}_{\alpha}x^{\alpha}\partial^{\nu}\chi^{\dagger})\}\gamma_{0}\chi\\
 & + & \{\tilde{B}_{\alpha}\chi^{\dagger}+\Delta(\tilde{B}_{\alpha}\chi^{\dagger}(v\Gamma)^{\dagger}-\tilde{C}_{\mu}x^{\mu}\tilde{B}_{\alpha}\chi^{\dagger})+B_{\mu}\tilde{B}_{\alpha}\partial^{\mu}\chi^{\dagger}-\tilde{B}_{\mu}B_{\alpha}\partial^{\mu}\chi^{\dagger}\\
 & + & \tilde{B}_{\mu}\tilde{B}_{\alpha}\partial^{\mu}\chi^{\dagger}\gamma_{5}-\Delta\tilde{B}_{\mu}x^{\beta}C_{\beta\alpha}\partial^{\mu}\chi^{\dagger}+\tilde{B}_{\alpha}C_{\mu\nu}x^{\mu}\partial^{\nu}\chi^{\dagger}-i{\rm \theta}^{*}\Delta\tilde{B}_{\alpha}\chi^{\dagger}\}\gamma_{0}\gamma_{5}\partial^{\alpha}\chi\\
 & + & \{B_{\alpha}\chi^{\dagger}+C_{\beta\alpha}x^{\beta}\chi^{\dagger}-i\tilde{C}_{\mu}\Delta C_{\beta\alpha}x^{\mu}x^{\beta}\chi^{\dagger}+B_{\mu}B_{\alpha}\partial^{\mu}\chi^{\dagger}+B_{\mu}C_{\beta\alpha}x^{\beta}\partial^{\mu}\chi^{\dagger}\\
 & + & B_{\alpha}C_{\mu\nu}x^{\mu}\partial^{\nu}\chi^{\dagger}+C_{\mu\nu}C_{\beta\alpha}x^{\mu}x^{\beta}\partial^{\nu}\chi^{\dagger}-i\Delta({\rm \theta}^{*}B_{\alpha}\chi^{\dagger}-i{\rm \theta}^{*}C_{\beta\alpha}x^{\beta}\chi^{\dagger}\\
 & + & B_{\alpha}\chi^{\dagger}(v\Gamma)^{\dagger}+C_{\beta\alpha}x^{\beta}\chi^{\dagger}(v\Gamma)^{\dagger}-i\tilde{C}_{\mu}B_{\alpha}x^{\mu}\chi^{\dagger})\}\gamma_{0}\partial^{\alpha}\chi,
\end{eqnarray*}
onde $\Delta$ engloba o sinal em relação a conjugação Hermitiana.
As funções $\Omega$ aparecendo em \eqref{eq:identifications} são obtidas das equações anteriores após as identificações indicadas em \eqref{eq:identifications}. 